\documentclass[a4paper,11pt]{article}

\pdfoutput=1

\usepackage{jheppub}
\usepackage[T1]{fontenc} 
\usepackage{graphicx}
\usepackage{epsfig}
\usepackage{rotating}
\usepackage{amssymb}
\usepackage{dsfont}
\usepackage{psfrag}
\usepackage{amsmath,euscript,array,mathrsfs}
\usepackage{bbold}
\usepackage{epsf}
\usepackage{slashed}

\usepackage{caption,subcaption,wrapfig}
\usepackage[colorlinks=true,linktocpage=true,linkcolor=blue,citecolor=blue]{hyperref}

\usepackage{tikz}
\usetikzlibrary{decorations.markings,decorations.pathmorphing}

\newcommand{\be}{\begin{equation}}
\newcommand{\ee}{\end{equation}}
\newcommand{\beqs}{\begin{eqnarray}}
\newcommand{\eeqs}{\end{eqnarray}}

\newcommand{\dd}{\mathrm{d}}

\newcommand{\AAA}{\mathcal{A}}
\newcommand{\FF}{\mathcal{F}}
\newcommand{\OO}{\mathcal{O}}
\newcommand{\GG}{\mathcal{G}}
\newcommand{\MM}{\mathcal{M}}
\newcommand{\LL}{\mathcal{L}}
\newcommand{\BB}{\mathcal{B}}
\newcommand{\B}{\mathbb{B}}

\newcommand{\Bconf}{\mathbb{B}_8^{\rm{conf}}}

\preprint{ICCUB-20-005}
\title{Phase transitions in a three-dimensional analogue of Klebanov--Strassler}

\author[1]{Daniel Elander,}
\author[2,3,4]{Ant\'on F. Faedo,}
\author[2,5]{David Mateos,}
\author[2]{Javier G. Subils}

\affiliation[1]{Laboratoire Charles Coulomb (L2C), University of Montpellier, CNRS, Montpellier, France.}
\affiliation[2]{Departament de F\'isica Qu\`antica i Astrof\'isica \& Institut de Ci\`encies del Cosmos (ICC), \\ Universitat de Barcelona, Mart\'i Franqu\`es 1, ES-08028, Barcelona, Spain.}
\affiliation[3]{Departamento de F\'{i}sica, Universidad de Oviedo, Federico Garc\'{i}a Lorca 18, ES-33007, Oviedo, Spain.}
\affiliation[4]{Instituto Universitario de Ciencias y Tecnolog\'{i}as Espaciales de Asturias (ICTEA),\\ Calle de la Independencia 13, ES-33004, Oviedo, Spain.}
\affiliation[5]{Instituci\'o Catalana de Recerca i Estudis Avan\c cats (ICREA), Passeig Llu\'\i s Companys 23, \\ ES-08010, Barcelona, Spain.}

\date{\today}

\abstract{
We use top-down holography to study the thermodynamics of a one-parameter family of three-dimensional, strongly coupled Yang--Mills--Chern--Simons theories with M-theory duals.  
For generic values of the parameter, the theories exhibit a mass gap but no confinement, meaning no linear quark-antiquark potential. For two specific values of the parameter they flow to an infrared fixed point or to a confining vacuum, respectively.  As in the Klebanov--Strassler solution, on the gravity side the mass gap is generated by the smooth collapse to zero size of a cycle in the internal geometry.  We uncover a rich phase diagram with thermal phase transitions of first and second order, a triple point and a critical point.}

\begin{document}
\maketitle
\flushbottom

\section{Introduction}

Three-dimensional gauge theories enjoy properties that can challenge our four-dimensional intuition. A prominent example is the existence of the Chern--Simons (CS) term, which is a relevant deformation of the Yang--Mills action. Despite being topological, it has dramatic effects on the dynamics.  Besides, since the gauge coupling has positive mass dimension, three-dimensional Yang--Mills theories  admit a free ultraviolet (UV) fixed point. Conversely, they are generically strongly coupled in the infrared (IR). This gives rise to interesting nonperturbative phenomena, including the appearance of novel quantum phases, which may play a role in condensed matter systems where the gauge symmetry is typically emergent. These new phases cannot be uncovered using semiclassical approximations such as the large-CS level, large-number of flavours or large-mass limits, but their existence can be inferred employing dualities, anomaly matching or other arguments, as in \cite{Braun:2014wja,Gomis:2017ixy,Komargodski:2017keh,Choi:2018ohn}. As usual, supersymmetry is a valuable tool in this type of analyses \cite{Witten:1999ds}, and lattice techniques can also be adapted to these kinds of models \cite{Diamantini:1994xr,Damgaard:1998yv,Karthik:2016bmf}.

Holography also provides a useful tool to study strongly coupled gauge theories, especially when the approaches above are not applicable. In this paper we will use this tool to study a one-parameter family of three-dimensional gauge theories at non-zero temperature by means of their gravitational duals. The interactions consist of Yang--Mills interactions associated to a two-sites quiver, similarly to the Klebanov--Witten theory \cite{Klebanov:1998hh}, as well as CS terms for both gauge groups. The parameter distinguishing one theory from another, which we call $b_0$,  is related to the difference between the inverse squared couplings of each of the two gauge groups. In our conventions $b_0$ takes values in the interval 
$[0,1]$. The gravitational duals at zero temperature can be described in ten- or eleven-dimensional supergravity and are therefore firmly embedded in string or M-theory. They preserve $\mathcal{N}=1$ supersymmetry and were studied in detail in \cite{Faedo:2017fbv}, based on the results of \cite{Cvetic:2001ye,Cvetic:2001pga}. For $0< b_0<1$  the theories exhibit a mass gap but no confinement in the sense of a linear quark-antiquark potential at large distances \cite{Faedo:2017fbv}. For $b_0=0$  there is no mass gap and the theory flows  to a fixed point in the IR. For $b_0=1$ it exhibits not just a mass gap but also confinement. These properties are pictorially represented in Fig.~\ref{fig:triangle}. 

The aim of this work is to study the thermodynamics of the family of theories above. The results are summarised in Fig.~\ref{fig:PhaseDiagram}. Generically, as the temperature increases gradually from zero, the system undergoes a phase transition from the gapped state to an ungapped phase, where the latter is  characterized by the presence of a black brane horizon. We will refer to this as a ``degapping transition.'' The details of this transition and the behaviour of the system at even higher temperatures depends on the value of $b_0$. 

For $b_0 \in (b_0^\textrm{triple},1]$, with 
$b_0^\textrm{triple}\approx 0.6847$, there is a  first-order, Hawking--Page-like phase transition between the gapped geometry and the black brane geometry. As expected, the entropy density jumps discontinuously from zero to a positive value across this phase transition. The critical temperature  depends on $b_0$, as indicated by the dashed, red curve in Fig.~\ref{fig:PhaseDiagram}.  In the particular case $b_0=1$ the phase transition is a deconfinement phase transition. For these values of $b_0$ no other phase transitions occur as the temperature is further increased. 

For $b_0 \in (0, b_0^\textrm{triple})$, the system also undergoes a phase transition from the gapped to the black brane geometry, as indicated by the 
solid, magenta curve in Fig.~\ref{fig:PhaseDiagram}. However, in this case the entropy density increases continuously from zero. In other words, from the viewpoint of the entropy density alone the degapping transition for $b_0 \in (0, b_0^\textrm{triple})$ looks like a second-order phase transition. However, other quantities change discontinuously, so the transition is still first-order. For $b_0 \in (0, b_0^\textrm{critical})$, with $b_0^\textrm{critical}\approx 0.6815$, no other  phase transitions take place as the temperature is further increased. In contrast, for 
\mbox{$b_0 \in (b_0^\textrm{critical},b_0^\textrm{triple})$}, 
a second phase transition occurs as the temperature is further increased, as indicated by the dashed, black line in Fig.~\ref{fig:PhaseDiagram}. In this case this is a first-order transition between two black-brane solutions with a discontinuous jump in the entropy density. By continuity, it follows that this line of phase transitions ends at a critical point at 
$b_0=b_0^\textrm{critical}$, at which the phase transition is second-order. The three curves of phase transitions in Fig.~\ref{fig:PhaseDiagram} meet at a triple point at $b_0=b_0^\textrm{triple}$, at which the three phases can coexist. For $b_0=0$ there are no phase transitions at any temperature. 

Some properties of our system are similar to those of the Klebanov--Strassler (KS) solution  \cite{Klebanov:2000hb}, which is dual to a four-dimensional gauge theory. For example, the mass gap at zero temperature arises from the shrinking to zero size of a cycle in the internal part of the geometry. This means that our theories are truly three-dimensional at all energy scales, as opposed to Witten-like  models \cite{Witten:1998zw} in which the gap arises from a compact  gauge theory direction that becomes important at high energies. A further similarity with the KS case \cite{Aharony,Buchel} is the presence of a phase transition from a gapped to an ungapped phase. There are two main differences between our system and the KS model. The first one is that our theories have a simple, weakly coupled UV completion, since they are asymptotically free. The second one is that these theories come in a one-parameter family, which gives rise to a  rich phase diagram including a critical point and a triple point. 

The organization of the paper is as follows. First, in Section~\ref{LowT} we discuss the low-temperature phases. Since these are constructed directly from the zero-temperature solutions, we review their properties and give the details of the expected gauge theory duals. In Section~\ref{HighT} we construct numerically the competing phases, which take the form of black brane solutions. Equipped with all these solutions, we explore their thermodynamic properties and present the relevant phase diagram in Section~\ref{sec:thermodynamics}. We finally discuss the results and conclude in Section~\ref{sec:conclusions}. Most technical aspects are relegated to several appendices.

\section{Low-temperature phases}
\label{LowT}

The ground state of the system corresponds to the regular supersymmetric solutions of \cite{Faedo:2017fbv}. These take the form of a stack of $N$ coincident M2-branes where the transverse space is one of the eight-dimensional manifolds pertaining to the $\mathbb{B}_8$ class, found originally in \cite{Cvetic:2001ye,Cvetic:2001pga}. These metrics have Spin(7) holonomy in order to preserve $\mathcal{N}=1$ supersymmetry. They come in a one-parameter family, characterised for instance by the value of the radial coordinate at which the geometry ends smoothly. Moreover, they posses a non-trivial four-cycle that does not collapse at the end-of-space and whose size provides a scale and consequently a mass gap in the dual theory. Nevertheless, there is no confinement in the sense explained in \cite{Faedo:2017fbv}.  

Regularity of the solution is ensured by the presence of a complicated four-form, with different internal components. We should mention that, once we pick one of the $\mathbb{B}_8$ metrics, specified by the scale of the mass gap or, equivalently, by the position of the end-of-space, the four-form is completely fixed by the requirement that the entire geometry be regular. 

The UV of the model is best understood by reducing on the appropriate internal circle and considering the ten-dimensional version of the solution. Since the circle is internal, the reduced type IIA string-frame metric takes the form of a stack of D2-branes
\begin{eqnarray}
\label{D2brane}
\dd s_{\rm st}^2 &=&h^{-\frac12}\, \dd x_{1,2}^2 +h^{\frac12}\bigg(\dd r^2+e^{2f}\dd\Omega_4^2+e^{2g}\left[\left(E^1\right)^2+\left(E^2\right)^2\right] \bigg)\,,
\end{eqnarray}
together with a slightly modified dilaton $e^{\Phi}=h^{\frac14}e^{\Lambda}$. The vielbeins $E^1$ and $E^2$ describe a ${\rm S}^2$ fibration over ${\rm S}^4$, the latter with metric $\dd\Omega_4^2$ (see Appendix~\ref{sec:10Dansatz} for the details). For fixed values of $e^f$ and $e^g$, the internal metric is that of $\mathbb{CP}^3\simeq{\rm Sp}(2)/{\rm U}(2)$, which is squashed with respect to the Fubini--Study metric whenever $e^f\ne e^g$. In the complete solution, this squashing changes with the radial direction and thus along the RG flow. The four-sphere coincides with the four-cycle that does not contract in the IR of the M-theory realisation. However, in ten dimensions the presence of the mass gap is obscured by the fact that the warp factor $h$ has an IR singularity. 

\begin{figure}[t]
	\centering
	\begin{tikzpicture}[scale=3.3,very thick,decoration={markings,mark=at position .5 with {\arrow{stealth}}}]
	
	\node[above] at (0,0) {\footnotesize{SYM-CSM $|$ D2}};
	\node[below] at (0,-2.2) {\footnotesize{Mass gap $|$ $\mathbb{R}^4\times {\rm S}^4$}};
	\node[left] at (-1,-1) {\footnotesize{CFT $|$ AdS}};
	\node[right] at (2,-2) {\footnotesize{Confinement $|$ $\mathbb{R}^3\times{\rm S}^1\times {\rm S}^4$}};
	
	\draw [black, ultra thick] (0,0) circle [radius=0.015];
	\draw [black, ultra thick] (-1,-1) circle [radius=0.015]; 
	
	\node at (-.6,-1) {$\mathbb{B}_8^+$};
	\node at (.63,-1) {$\mathbb{B}_8^-$};
	
	\draw[postaction={decorate},very thick] (0,0) --  (0,-2) node[left,midway]{$\mathbb{B}_8$};
	\draw[postaction={decorate},ultra thick] (0,0) -- (-1,-1) node[left,midway]{$\mathbb{B}_8^\infty$\,};
	\draw[postaction={decorate},ultra thick] (0,0) -- (2,-2) node[right,midway]{\, $\mathbb{B}_8^{\rm{conf}}$};
	\draw[postaction={decorate},ultra thick] (-1,-1) -- (-1,-2) node[left,midway]{$\mathbb{B}_8^{\textrm{\tiny OP}}$\,\,};
	
	\draw[postaction={decorate},thick, red] (0,0) .. controls (-.9,-.9) and (-1,-1) .. (-0.98,-2);
	\draw[postaction={decorate},thick, red] (0,0) .. controls (-.5,-.5) and (-.5,-1.5) .. (-.5,-2);
	\draw[postaction={decorate},thick, blue] (0,0) .. controls (.5,-.5) and (.5,-1.5) .. (.5,-2);
	\draw[postaction={decorate},thick, blue] (0,0) .. controls (.9,-.9) and (.95,-1.05) .. (1,-2);
	\draw[postaction={decorate},thick, blue] (0,0) .. controls (.9,-.9) and (1.5,-1.6) .. (1.5,-2);
	\draw[postaction={decorate},thick, blue] (0,0) .. controls (.95,-.95) and (1.8,-1.8) .. (1.9,-2);
	
	\draw[|-|] (-1,-2) -- (0,-2);
	\draw[-stealth] (0,-2) -- (2,-2);
	
	\node[left=5] at (-1,-2) {$b_0$};
	\node[below=5] at (-1,-2) {$0$};
	\node[below=5] at (0,-2) {$2/5$};
	\node[below=5] at (2,-2) {$1$};
	
	\end{tikzpicture}
	\caption{\small Pictorial representation of the different ground state solutions (see \cite{Faedo:2017fbv} for further details).
	}\label{fig:triangle}
\end{figure}
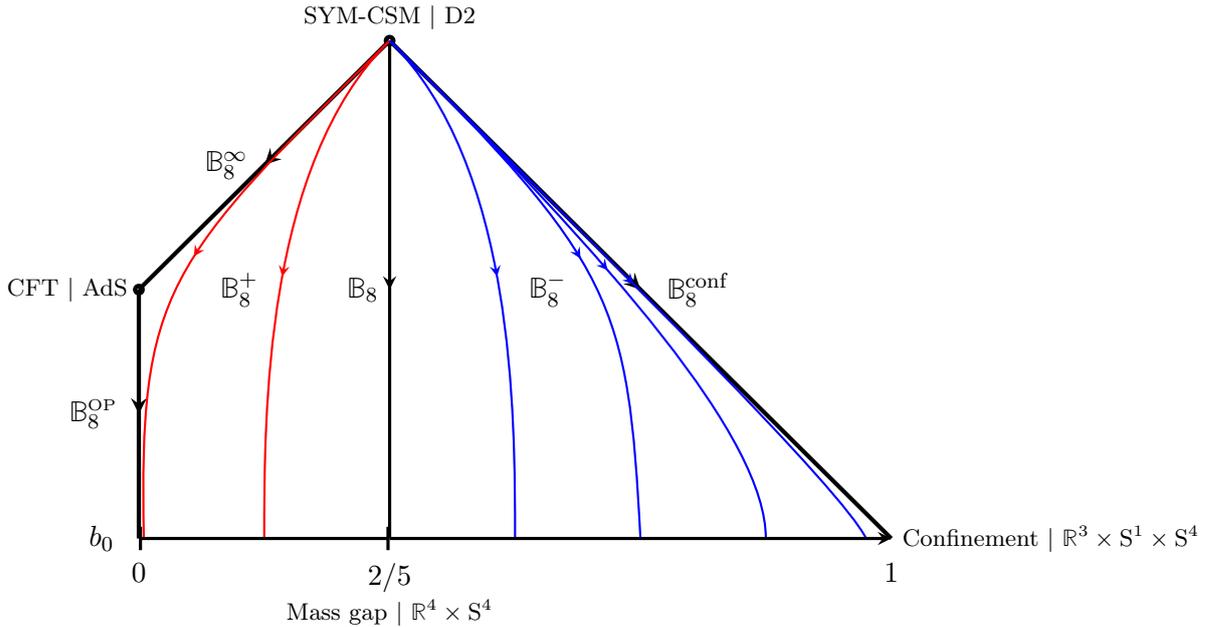

The exact form of the metric functions can be found in \cite{Faedo:2017fbv} and will not be needed here. It can be seen that the UV of the entire family of geometries is that of $N$ coincident D2-branes in the decoupling limit
\begin{equation}
e^{2f}\,=\,2\,e^{2g}\,\sim\, r^2\,,\qquad\qquad e^{\Phi}\,\sim\,h^\frac14\,,\qquad\qquad h\,\sim\,N\,r^{-5}\,.
\end{equation}
The asymptotic value of the squashing, $e^{2f-2g}=2$, is fixed for all the solutions and corresponds to the nearly K\"ahler point of the $\mathbb{CP}^3$. The gauge theory dual for such metric was proposed in \cite{Loewy:2002hu} and consists of a two-sites Yang--Mills quiver with ${\rm U}(N)\times{\rm U}(N)$ gauge group and bifundamental matter, similar to the Klebanov--Witten quiver in four dimensions. Additionally, since the circle on which we reduce from eleven dimensions is non-trivially fibered, we get a non-vanishing internal two-form, which generates Chern--Simons interactions in the gauge theory dual. Finally, there are also internal three- and four-form fluxes signalling the presence of fractional D2-branes. We expect these to correspond to a shift $M$ in the rank of one of the gauge groups. Thus, the conjecture is that these solutions are dual to RG flows in a 
\begin{equation}\label{quiver}
{\rm U}(N)_k\times{\rm U}(N+M)_{-k}
\end{equation}
quiver gauge theory with CS interactions at level $k$ and preserving $\mathcal{N}=1$ supersymmetry.
Given that the ten-dimensional metric is IR singular, in this setup the correct label for the different solutions is not the position of the end-of-space (equivalently the radius of the four-sphere at that point) but the asymptotic (UV) value of the NS two-form through a two-cycle inside $\mathbb{CP}^3$, which is in one-to-one correspondence with the former \cite{Faedo:2017fbv}. This parameter, that we call $b_0$, has the advantage of having a direct interpretation in the gauge theory dual as controlling the asymptotic difference between the microscopic Yang--Mills couplings of each of the two factors in the gauge group \cite{Hashimoto:2010bq}
\begin{equation}
b_0\,\sim\,\frac{1}{g_1^2}-\frac{1}{g_2^2}\,.
\end{equation}
In our conventions $b_0\in[0,1]$. The two endpoints of the interval are special in that they lead to IR physics qualitatively different from the rest of the family. On the one hand, the solution with $b_0=1$, constructed originally in \cite{Cvetic:2001ma}, is not only gapped but confining. This was attributed to the vanishing of the CS level in \cite{Faedo:2017fbv}, in agreement with the arguments in \cite{Herzog}. We denote this solution $\mathbb{B}_8^{\rm{conf}}$. On the other hand, when the difference between the couplings vanishes, $b_0=0$, the mass gap is lost and the theory flows to an IR fixed point described by the Ooguri--Park CFT \cite{Ooguri:2008dk}. This flow is denoted $\mathbb{B}_8^\infty$. Notice that for arbitrarily-small but non-vanishing values of $b_0$ the RG flow will pass arbitrarily close to the fixed point before reaching the gapped phase, giving rise to quasi-conformal dynamics in a certain range of energies. The setup is summarised in Fig.~\ref{fig:triangle}, where we depict the different transverse geometries as a function of $b_0$, together with their IR limit and the associated gauge theory dual interpretation. A more detailed explanation can be found in \cite{Faedo:2017fbv}. 

This family of solutions can be straightforwardly heated up by going to Euclidean space and declaring the time direction to be compact. As usual, the period $\beta$ of the Euclidean time is related to the temperature $T$ in the dual gauge theory as
\begin{equation}\label{beta}
\beta\,=\,\frac{1}{T}\,.
\end{equation}
 
The thermodynamic properties are very simple. The ground states are supersymmetric, so in the appropriate renormalisation scheme (see Appendix~\ref{sec:4Deffectivetheory}) the free energy $F$, computed as the bulk on-shell action,  vanishes. Since formally these 
finite-temperature solutions coincide with the supersymmetric ones (they only differ globally in the time direction), their free energy will also vanish in the same scheme. This happens independently of the temperature, and therefore the entropy $S$ is also zero. 

These thermal states are continuously connected to the ground state, so we expect them to correspond to a phase still exhibiting a mass gap. On the other hand, we know that replacing the regular IR with a horizon also introduces temperature into the system while removing the gap. These solutions, describing black branes, can have the same UV asymptotics and therefore correspond to other finite-temperature states of the same gauge theories. In the following we construct the black brane solutions and show the existence of several phase transitions.

\section{High-temperature phases}
\label{HighT}

In this section we construct high-temperature phases of the gauge theories we have discussed. On the gravity side of the duality this corresponds to the replacement of the regular IR by a horizon, giving rise to a black brane. These black branes need to satisfy, at leading order, identical D2-brane UV boundary conditions as the zero-temperature solutions in order to correspond to states in the same gauge theory duals. Therefore we will pick the following ansatz for the type-IIA string frame metric and dilaton 
\begin{eqnarray}
\label{10Dansatz}
\dd s_{\rm st}^2 &=&h^{-\frac12}\left(-\mathsf{b}\,\dd t^2 + \dd x_1^2 + \dd x_ 2^2\right)+h^{\frac12} \left(\frac{\dd r^2}{\mathsf{b}}+e^{2f}\dd\Omega_4^2+e^{2g}\left[\left(E^1\right)^2+\left(E^2\right)^2\right] \right)\,,\nonumber\\[2mm]
e^\Phi&=&h^{\frac14} \, e^\Lambda \,,
\end{eqnarray}
which differs from that in Eq.~\eqref{D2brane} only in the presence of a blackening factor $\mathsf{b}$. This function must have a simple zero at the position of the horizon for the solution to describe a black brane. 

The ansatz for the different fluxes takes the exact same form as for the supersymmetric solutions, as given in Eq.~\eqref{fluxesansatz}. They are parametrised by three functions of the radial coordinate, $b_J$, $b_X$ and $a_J$, together with three constants $Q_c$, $q_c$ and $Q_k$. These constants are Page charges for D2, D4 and D6-branes and are thus quantised. Moreover, they are related to the gauge theory parameters appearing in Eq.~\eqref{quiver} as \cite{Faedo:2017fbv} 
\begin{equation}\label{gaugeparam}
Q_c\,=\,3\pi^2\ell_s^5 g_s\,N\,,\qquad\qquad q_c\,=\,\frac{3\pi\ell_s^3g_s}{4}\,\overline{M}\,,\qquad\qquad Q_k\,=\,\frac{\ell_sg_s}{2}\,k\,,
\end{equation}
where we have defined the combination
\begin{equation}
\overline{M}\,=\,\left(M-\frac{k}{2}\right)\,.
\end{equation}
In our conventions we must take $Q_k<0$. 

These equations make explicit the length dimensions of the different charges. We can thus use them to shift and rescale the various functions of the ansatz as  
\begin{equation}\label{dimlessfunctions}
\begin{aligned}
e^f &=  |Q_k|\ e^\FF\, , \qquad \qquad  e^g = |Q_k|\ e^\GG\,,  \qquad\qquad h= \frac{4q_c^2 +3 |Q_k| Q_c}{|Q_k|^6} \ \mathbf{h}\,,\\[2mm] 
a_J &= -\frac{q_c}{6} - \sqrt{4q_c^2 + 3|Q_k| Q_c} \ \AAA_J\,, \qquad\quad b_X =  \frac{2q_c}{3 |Q_k|} + \frac{\sqrt{4 q_c^2 + 3 |Q_k| Q_c}}{3 |Q_k|}\ \BB_X\,,\\[2mm]
b_J &= - \frac{2q_c}{3 |Q_k|} - \frac{\sqrt{4 q_c^2 + 3 |Q_k| Q_c}}{3 |Q_k|}\ \BB_J\,,
\end{aligned}
\end{equation}
so that the redefined functions are dimensionless. Furthermore, working with the dimensionless radial coordinate 
\begin{equation}\label{ucoord}
u\,=\,\frac{|Q_k|}{r}
\end{equation}
all the charges drop from the equations. This means that, up to simple rescalings, the only parameter distinguishing one theory from another is $b_0$. In particular, the mass gap at zero temperature (in units of the 't Hooft couplings) is fixed by this parameter. The thermal phase transitions that we will exhibit will  take place at specific values of the ratio $T/M_\textrm{gap}$ that are determined by $b_0$. 
 
\subsection{Boundary conditions} 

In the radial coordinate defined in Eq.~\eqref{ucoord} the UV is located at $u=0$. The leading behaviour of the functions must coincide with that of the ground state. It is then possible to solve the equations order by order in the radial coordinate around this point, obtaining an expansion of the form 
 \begin{eqnarray}\label{UVexpansions}
e^\FF&=& \frac{1}{u\sqrt{2}}  \Big[1+{ f_1}{u}+\cdots+{ f_4}{u^4}+ { f_5}{u^5}+\cdots+{ f_{10}}{u^{10}}+\OO(u^{11})\Big]\,,\nonumber\\[2mm]
e^\GG&=& \frac{1}{2u}  \Big[1+\OO(u)\Big]\,, \qquad \qquad \qquad\mathbf{h} \,=\,\frac{16}{15}(1-b_0^2)\ u^5\   \Big[1+\OO(u) \Big]\,,\nonumber\\[2mm]
\mathsf{b}&=& 1+{\mathsf{b}_{5}}{u^{5}}+\OO(u^6)\,,\qquad \qquad \,e^\Lambda\,=\, 1+\OO(u)\,,\\[2mm]
\BB_J&=& b_0 +\cdots + { b_4} {u^4} +\cdots + { b_6} {u^6} + \cdots + { b_9} {u^9} +\OO(u^{10})\,, \nonumber\\[2mm]
\quad \BB_X&=&b_0+ \OO(u)\,,\qquad\qquad\qquad\quad\,\AAA_J\,=\,\frac{b_0}{6}+ \OO(u)\,, \nonumber
\end{eqnarray}
where we have made explicit the parameters that remain undetermined by the equations of motion. The first coefficients not explicitly shown in this expansion, written in terms of the ones that are shown, can be found in Appendix \ref{ap:UVexp}. As we already argued, $b_0$ selects a particular member of the family of gauge theories. The radial gauge chosen in Eq.~\eqref{10Dansatz}, $g_{tt} g_{rr}=-1$, still leaves a residual freedom to shift the radial coordinate, $r\to r+ \mbox{constant}$. This allows us to fix the value of $f_1$ without loss of generality. We choose $f_1=-1$ in order to facilitate the comparison with the supersymmetric ground state.  We are thus left with seven subleading, undetermined parameters in the UV. They correspond to normalisable modes and are hence related to the temperature (see Eq.~\eqref{b5_conserved} below) and to vacuum expectation values for different operators in the dual gauge theory, including the energy-momentum tensor. These parameters are fixed dynamically in the complete solution once we impose the presence of a regular horizon.


The existence of a horizon  is encoded in a simple zero of the blackening factor $\mathsf{b}$. At the horizon we demand regularity for the rest of the functions, so that they reach a finite value. We denote the position of the horizon as $u=u_h$ in the dimensionless radial coordinate introduced in Eq.~\eqref{ucoord}. In this way the functions enjoy an expansion of the form
\begin{equation}\label{Horizon_expansions}
\begin{array}{rclrclrcl}
 e^\FF &=&  f_h  +\OO(u-u_h)  \ ,&
\quad e^\GG&=&  g_h + \OO(u-u_h)  \ , &
\quad e^\Lambda&=& \lambda_h + \OO(u-u_h) \ , \\[2mm]
\mathbf{h}& =& h_h +\OO(u-u_h)  \ ,&
\quad\BB_J&=& \xi_h +\OO(u-u_h)  \ ,&
\quad\BB_X&=& \chi_h +\OO(u-u_h)   \ ,\\[2mm]
\AAA_J&=& \alpha_h+\OO(u-u_h)  \ ,&
\quad \mathsf{b}&=&  \mathsf{b}_h (u-u_h)+\OO(u-u_h)^2 \ .
\end{array}
\end{equation}
All the subleading coefficients are determined in terms of these eight leading-order ones (see Appendix \ref{ap:horizonexp} for the first coefficients). The position of the horizon $u_h$ controls the temperature of the black brane, so the phase diagram can be explored by changing its value.

\subsection{Numerical integration}

The perturbative analysis of the equations of motion with the desired boundary conditions in the UV and IR leaves us with the following undetermined parameters 
\begin{eqnarray}
\text{in the UV:}&\qquad&  f_4,\ f_5,\ f_{10},\ b_4,\ b_6,\ b_9,\ \mathsf{b}_5\nonumber\\
\text{at the horizon:}&\qquad& g_h,\ f_h,\ \lambda_h,\ h_h,\ \mathsf{b}_h,\ \alpha_h,\ \xi_h,\ \chi_h \,.
\end{eqnarray}
For each value of $b_0$ and $u_h$ we must find these fifteen parameters dynamically in the complete solution. Since we are solving eight second-order equations subject to one first-order (Hamiltonian) constraint, the problem is well posed. It should be mentioned that, as we note in Eq.~\eqref{conserved_current}, there is a conserved quantity along the radial coordinate. When substituting the expansions, this conserved quantity relates one of the UV parameters with horizon data as
\begin{equation}\label{b5_conserved}
\mathsf{b}_5 = \frac{16}{5}\, \frac{ \mathsf{b}_h\ f_h^4\ g_h^2}{\lambda_h^2}\ u_h^2\,.
\end{equation}
This equation can be used as a check for the numerical solutions. Alternatively, it can be imposed {\it ab initio}, reducing the number of parameters to be found numerically to fourteen.  

The solutions were constructed using a shooting method. The boundary conditions, Eqs.~\eqref{UVexpansions} and \eqref{Horizon_expansions}, were imposed close to the UV and the horizon, respectively, with an initial guess for the unfixed parameters, giving us a seed solution. The full equations were then integrated numerically both from the UV and from IR up to a designated intermediate matching point. For arbitrary values of the parameters the functions would be discontinuous at that point. Starting from this seed, we then used the Newton--Raphson algorithm to find the values of the parameters such that the functions and their derivatives are continuous at the matching point with the desired precision.    
 
The main challenge when solving a system of equations via the shooting method is to find a good initial seed. Ultimately, we are using a Newton--Raphson algorithm on a fourteen-dimensional parameter space (after making use of Eq.~\eqref{b5_conserved}), so we need to start the search close enough to the correct values of the unknown constants for it to converge. A good seed for a solution can be provided by another solution close enough in parameter space since, by continuity, the values of the unknown constants will be similar. In other words, once we have a black brane with given control parameters $b_0$ and $u_h$, we can use the values of the rest of the parameters as the initial guess for another solution with slightly corrected control parameters. The problem thus reduces to finding the first black brane.  

Fortunately the system admits one analytic solution describing a black brane. As we already mentioned, the RG flow corresponding to $b_0=0$, denoted by $\B_8^\infty$ at the left of Fig.~\ref{fig:triangle}, ends at an IR fixed point. This is located at $r=-\pi  Q_k$ (equivalently, $u= \pi^{-1}$) so that near this point the functions in the metric read
\begin{equation}\label{AdS-OP}
\begin{array}{rclcrcl}
e^{2f}&=& \frac{9}{5} \left(r+\pi Q_k\right){}^2\,, &\qquad\qquad\qquad& e^{2g}& =&\frac{9}{25} \left(r+\pi Q_k\right){}^2\,,\\[2mm]
e^\Lambda& =& \frac{3}{5 |Q_k|} \left(r+\pi Q_k\right)\,,&\qquad\qquad\qquad&h&=&\frac{125 \,Q_k^4}{2187\left(r+\pi Q_k\right)^4}\,,
\end{array}
\end{equation}
giving rise to an ${\rm AdS}_4$ geometry, as expected. It is then straightforward to include a horizon by considering the blackening factor
\begin{equation}\label{AdSblackening}
\mathsf{b}(r)=1-\frac{(r_h+\pi Q_k)^3}{(r+\pi Q_k)^3}\,,
\end{equation}
which describes an AdS-Schwarzschild black brane. This is an exact solution to the equations of motion when $b_0=0$. 

Based on this observation, the strategy to find the first black brane with the desired asymptotics is the following. We put a horizon in the region where the geometry is close to AdS-Schwarzschild and we choose as horizon data the values of the functions for $\B_8^\infty$ at that point. Since this should correspond to a solution at low temperature, the UV parameters are then taken as the ones for the metric without a horizon. Also, we use Eq.~\eqref{AdSblackening} as the seed for the blackening factor at both the UV and at the horizon. As a final simplification, we note that for $b_0=0$ the dimensionless functions $\BB_J$, $\BB_X$ and $\AAA_J$ from Eq.~\eqref{dimlessfunctions} vanish exactly and, consequently, there are three equations which are trivially satisfied and six fewer parameters  to be found. In this way we were able to find a black brane solution in the $\B_8^\infty$ RG flow, with the horizon located at $u_h\approx \pi^{-1} -4\times 10^{-4}$. 

Once we have the first black brane we slightly decrease $u_h$ and use as seed the values of the parameters we have just obtained, finding in this way another solution. Thus, we can iteratively find hotter black branes by interpolating the values of the parameters of the previous solutions to obtain a good seed for the next one. The resulting family of black branes has the expected properties. At low temperatures the dependence of the free energy on the temperature is dictated by conformal invariance,
\begin{equation}\label{FreeECFT}
F\,\sim\, \frac{1}{|k|}\left(\frac{\overline{M}^2}{2} +N|k|\right)^{3/2}T^3\,.
\end{equation}
In the opposite, UV-limit, for temperatures above 
\begin{equation}\label{Tconformal}
\lambda\,\frac{k^2}{N}\left(\frac{\overline{M}^2}{2} +N|k|\right)^{-1/2}\,,
\end{equation}
with $\lambda=g_s\ell_s^{-1}N$ denoting 't Hooft's coupling, we  recover the D2-brane solution and the  free energy behaves as in a three-dimensional Yang--Mills theory at strong coupling:
\begin{equation}
F\,\sim\, \lambda^{-1/3}\left(\frac{N}{|k|^5}\right)^{1/3}\left(\frac{\overline{M}^2}{2} +N|k|\right)^{5/3}T^{10/3}\,.
\end{equation}
This UV behaviour will be shared by the entire family of solutions. Note that in the 
$N\to \infty$ limit the free energy scales with the number of colours as $N^2$, as expected on a deconfined phase. 

Taking into account that all the flows share the same UV asymptotics except for the value of $b_0$, we can use the solutions we just constructed as a seed to find black branes with non-vanishing $b_0\gtrsim0$. The seed will be better the higher the temperature is, since the flows differ mostly in the IR. Now $\BB_J$, $\BB_X$ and $\AAA_J$ are also non-vanishing, hence we have to solve the full system of equations and shoot for the complete set of parameters. In this way we obtained the first high-temperature black brane for a small non-vanishing value of $b_0$, whose ground state is gapped. Again, using this first solution as a seed it is possible to gradually increase the value of $u_h$ and find colder black branes. The process ends when some of the functions, namely $e^\FF$, $e^\GG$ and $e^\Lambda$, vanish at the horizon within numerical precision.  We will refer to the value of $u_h$ where this happens as $u_N$,  which is a meaningful quantity because we have fixed the radial gauge completely. In the case of the zero-temperature, regular solutions the vanishing of these metric functions is precisely what produces the end-of-space and the mass gap. Below we will investigate whether this happens in the current context in a smooth or in a singular manner. 

Once we have the entire family of black branes for this particular $b_0$, we slightly increase its value. The hottest black branes would again be similar to the ones we had constructed, that may then be used as a seed. With this strategy it is possible to explore the entire space of parameters in the $\left(b_0,u_h\right)$-plane. Plots for some of the solutions can be found in Appendix~\ref{ap:solutions}. 

\subsection{Zero-entropy limit} \label{sec:zeroEntropyLimit}

Before moving into the  study of the phase diagram, let us discuss in more detail the solutions that we obtain in the limit in which we remove the horizon. This is achieved when the values of the functions $e^\FF$, $e^\GG$ and $e^\Lambda$ become zero at $u_h=u_N$, since the vanishing of these functions implies that the area of the black branes, and therefore their entropy, goes to zero. As we will see, in this limit the temperature is still non-zero. 

\begin{figure}[t]
	\begin{center}
		\includegraphics[width=.58\textwidth]{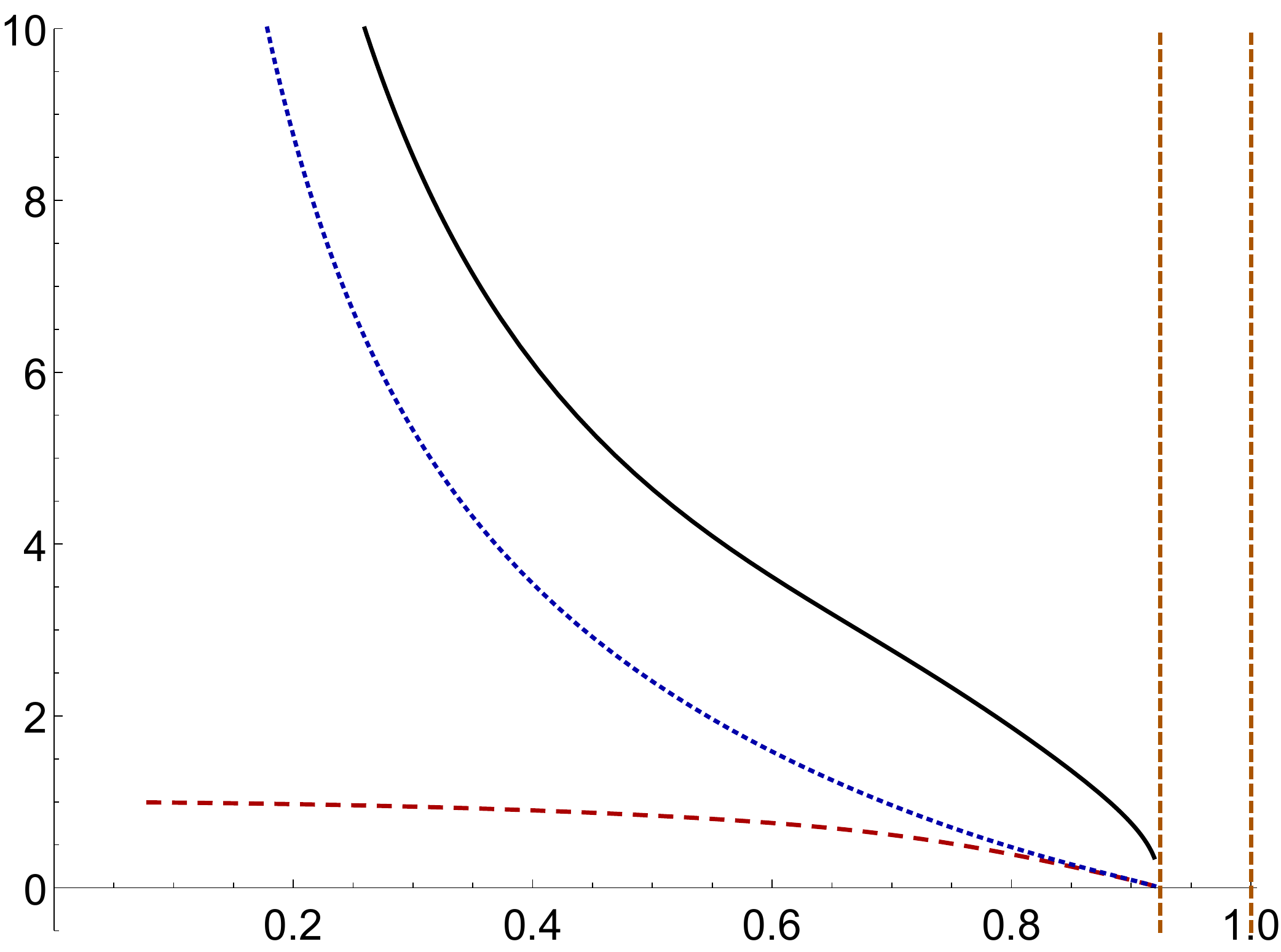} 
		\put(-266,190){\rotatebox{0}{$f_h,\ g_h,\ \lambda_h$}}
		\put(9,9){$u_h/u_s$}
		\caption{\small Horizon values of $e^\FF$, $e^\GG$ and $e^\Lambda$ in solid black, dotted blue and dashed red, respectively, as a function of the position of the horizon normalized to $u_s$, which is the where the supersymmetric regular ground-state solution ends. The dashed orange vertical line to the left corresponds to $u_N$, while the one to the right represents $u_s$. Between those two lines we have not found any black brane solutions. In this plot, we have fixed $b_0=2/5$.}
		\label{fig:HorizonValues}
	\end{center}
\end{figure}

As we have defined it, $u_N$ is the maximum value fo $u_h$ for a given $b_0$. The first observation is that this does not coincide with $u_s(b_0)$, defined as the position of the end-of-space for the supersymmetric, regular, zero-temperature solution with the same $b_0$. Note that it is meaningful to compare these two coordinate values because we have fixed the radial gauge completely. The discrepancy can be seen for the particular example of $b_0=2/5$ in Fig.~\ref{fig:HorizonValues}, where we show the value of $e^\FF$, $e^\GG$ and $e^\Lambda$ at the horizon as a function of the position of the horizon normalized to $u_s$. All three values vanish at the same point $u_N<u_s$. The same phenomenon can be seen for different values of $b_0$ in Appendix~\ref{ap:solutions}.

As a consequence, when we take the area of the horizon to zero, which we refer to as the ``zero-entropy limit,'' we do not recover the  ground state solutions discussed in Section~\ref{LowT}. This is in sharp contrast with well known examples such as AdS-Schwarzschild or black Dp-branes, where the usual AdS and Dp-brane metrics are recovered as one removes the horizon. It is therefore interesting to understand the nature of the solution obtained in this regime.

\begin{figure}[t]
	\begin{center}
		\includegraphics[width=.80\textwidth]{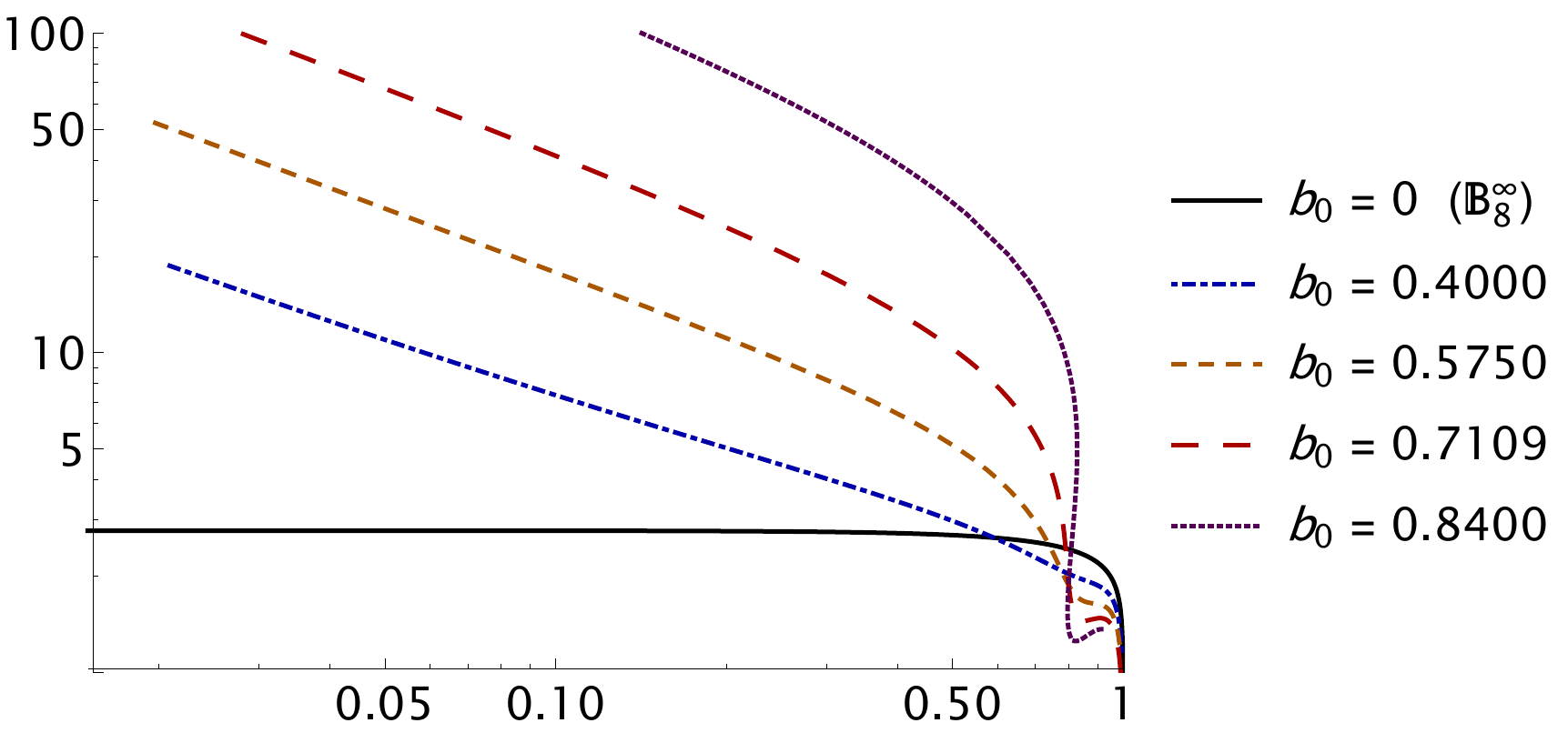} 
		\put(-400,173){\rotatebox{0}{$-\left(4q_c^2+3Q_c|Q_k|\right)^{1/3} R_h$}}
		\put(-85,13){$\lambda_h$}
		\caption{\small Eleven-dimensional Ricci scalar as a function of $e^\Lambda$, both evaluated at the horizon, for different black brane solutions. Notice that as $\lambda_h$ goes to zero (i.e.~in the zero-entropy limit), the Ricci scalar diverges for a generic value of $b_0>0$, whereas for $\B_8^\infty$ it approaches the constant value corresponding to the IR AdS. The curvature invariants obtained by squaring the Ricci and Riemann tensors have an analogous behaviour.}
		\label{fig:Ricci}
	\end{center}
\end{figure}

\begin{figure}[t]
\begin{center}
		\begin{subfigure}{0.45\textwidth}
			\includegraphics[width=\textwidth]{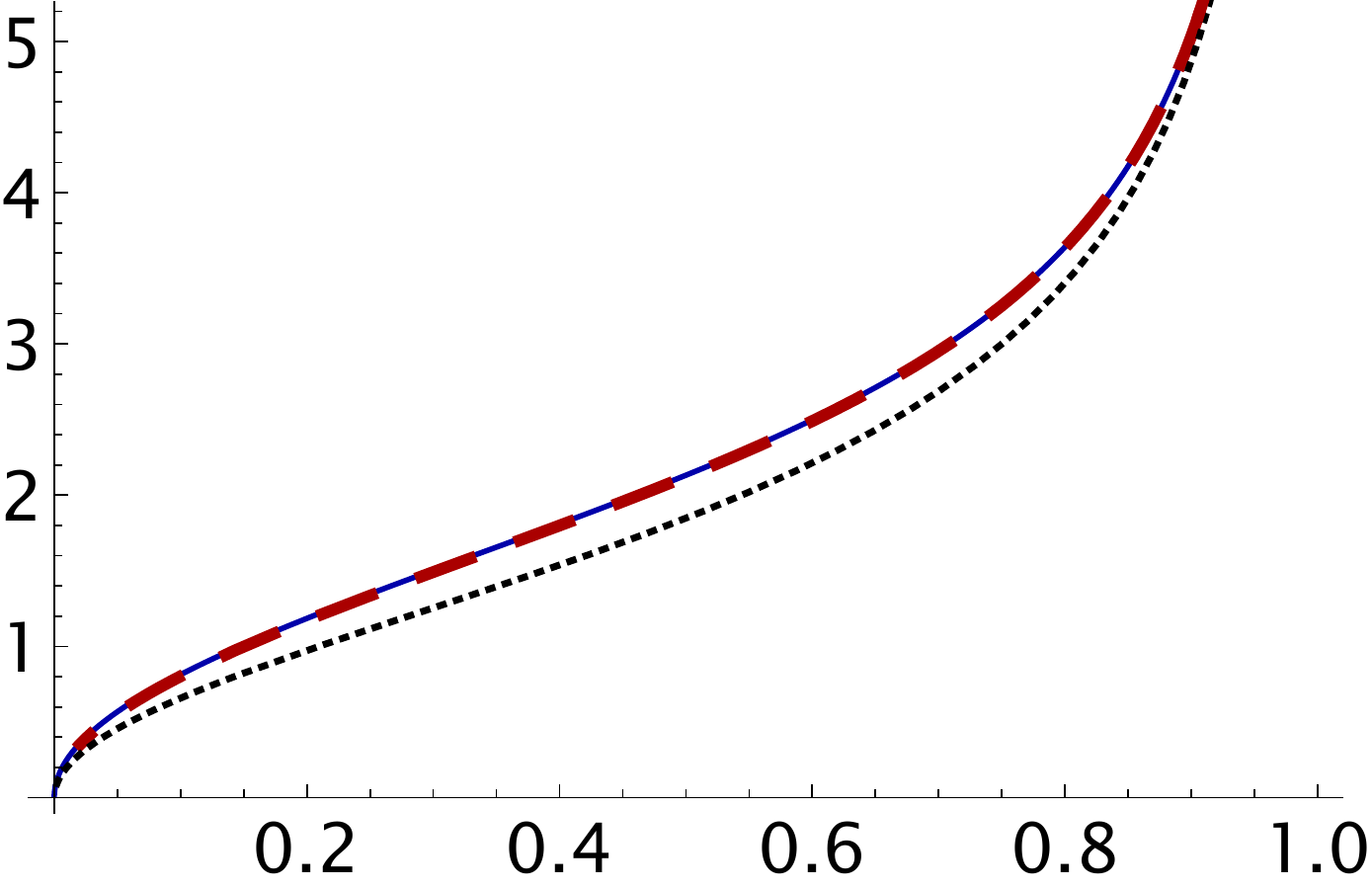} 
			\put(-190,130){$e^{\FF}$}
			\put(-5,20){$e^{\Lambda}$}
		\end{subfigure}\hfill
		\begin{subfigure}{.45\textwidth}
			\includegraphics[width=\textwidth]{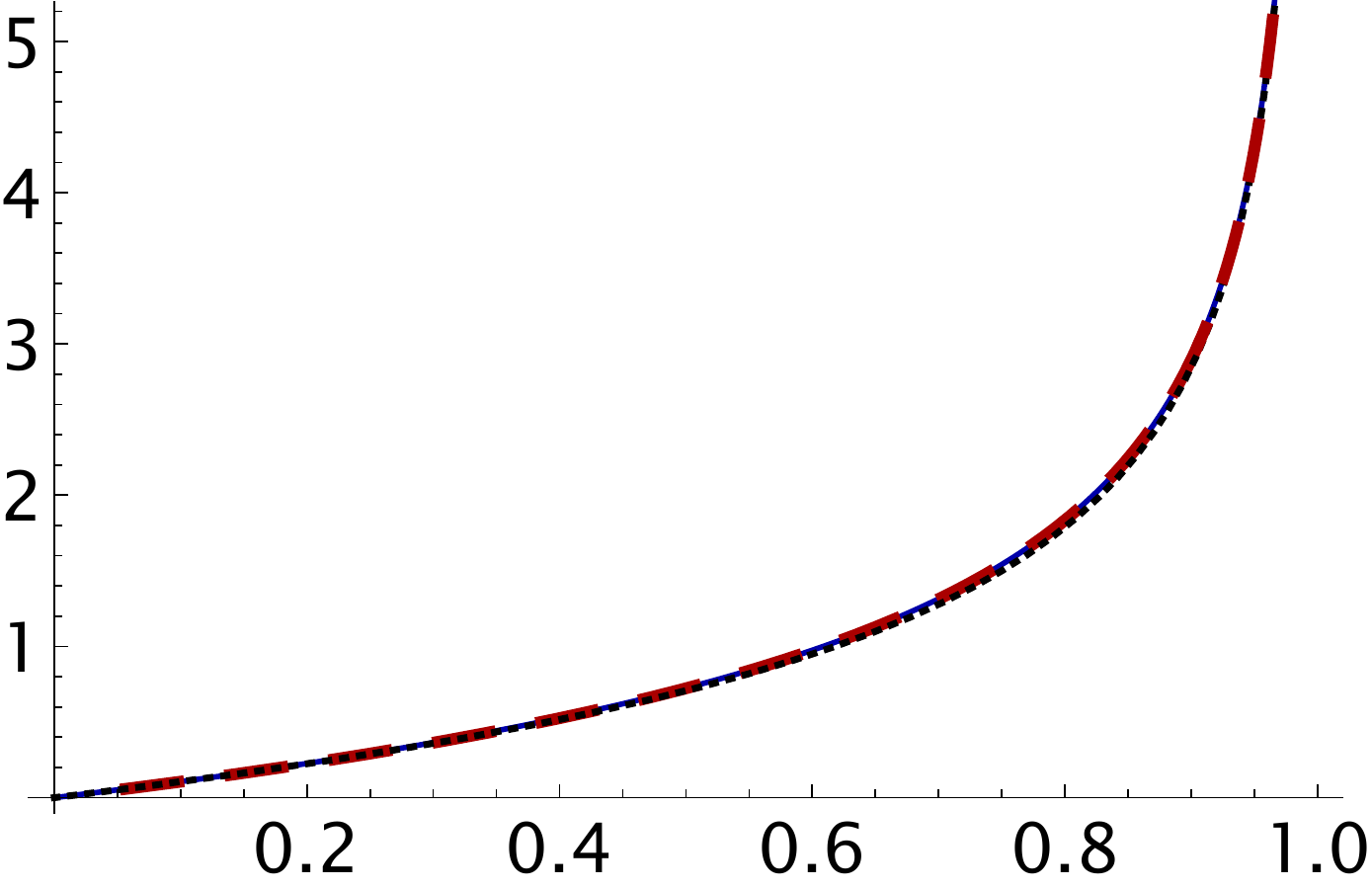} 
			\put(-190,130){$e^{\GG}$}
			\put(-5,20){$e^{\Lambda}$}
		\end{subfigure}\vspace{3mm}
		\begin{subfigure}{0.45\textwidth}
			\includegraphics[width=\textwidth]{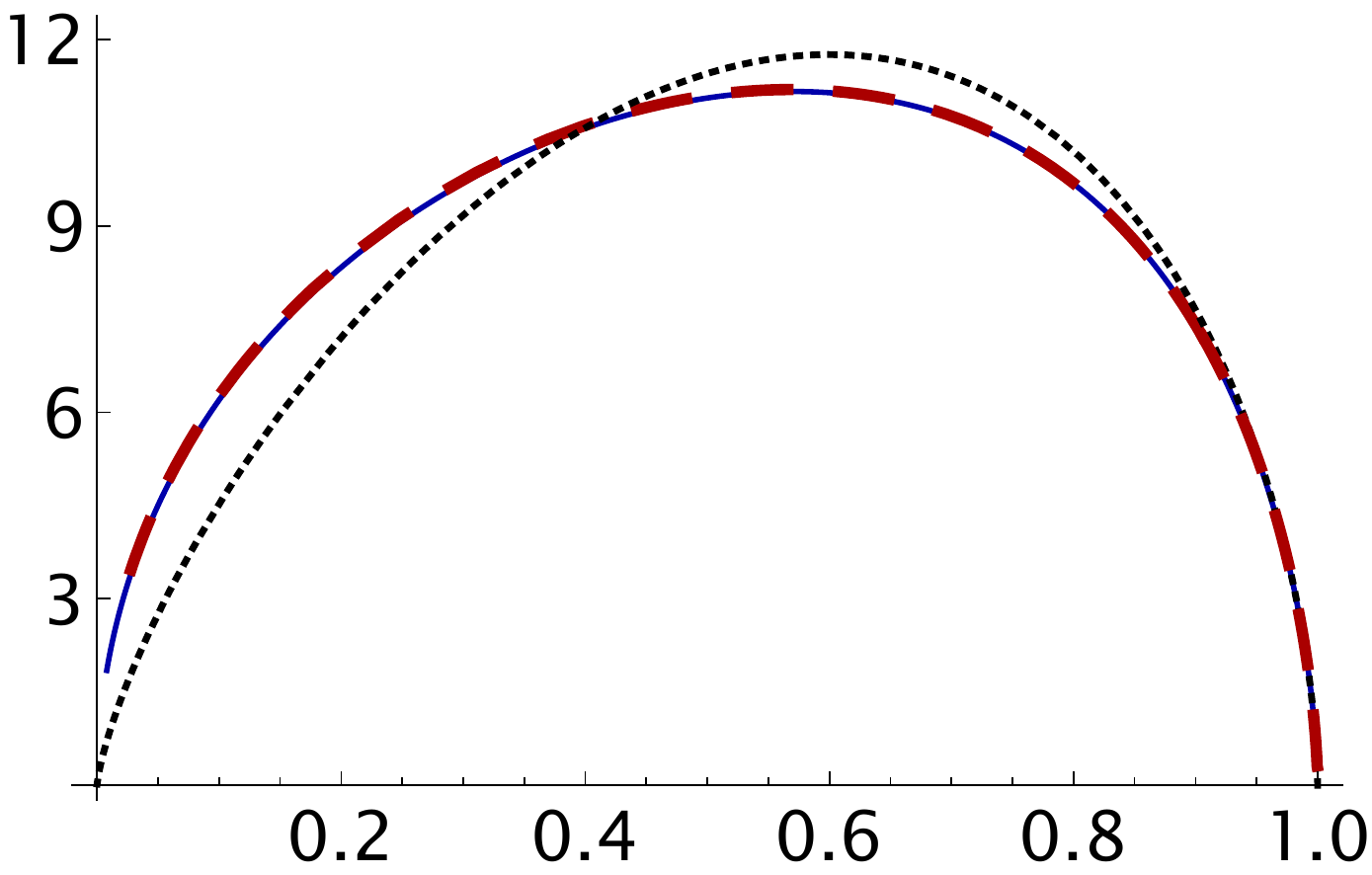} 
			\put(-190,140){$\frac{|Q_k|^{\frac{3}{2}}}{(4 q_c^2 + 3Q_c |Q_k|)^{\frac{1}{4}}}10^2e^\Phi$}
			\put(-5,20){$e^{\Lambda}$}
		\end{subfigure}\hfill
		\begin{subfigure}{.45\textwidth}
			\includegraphics[width=\textwidth]{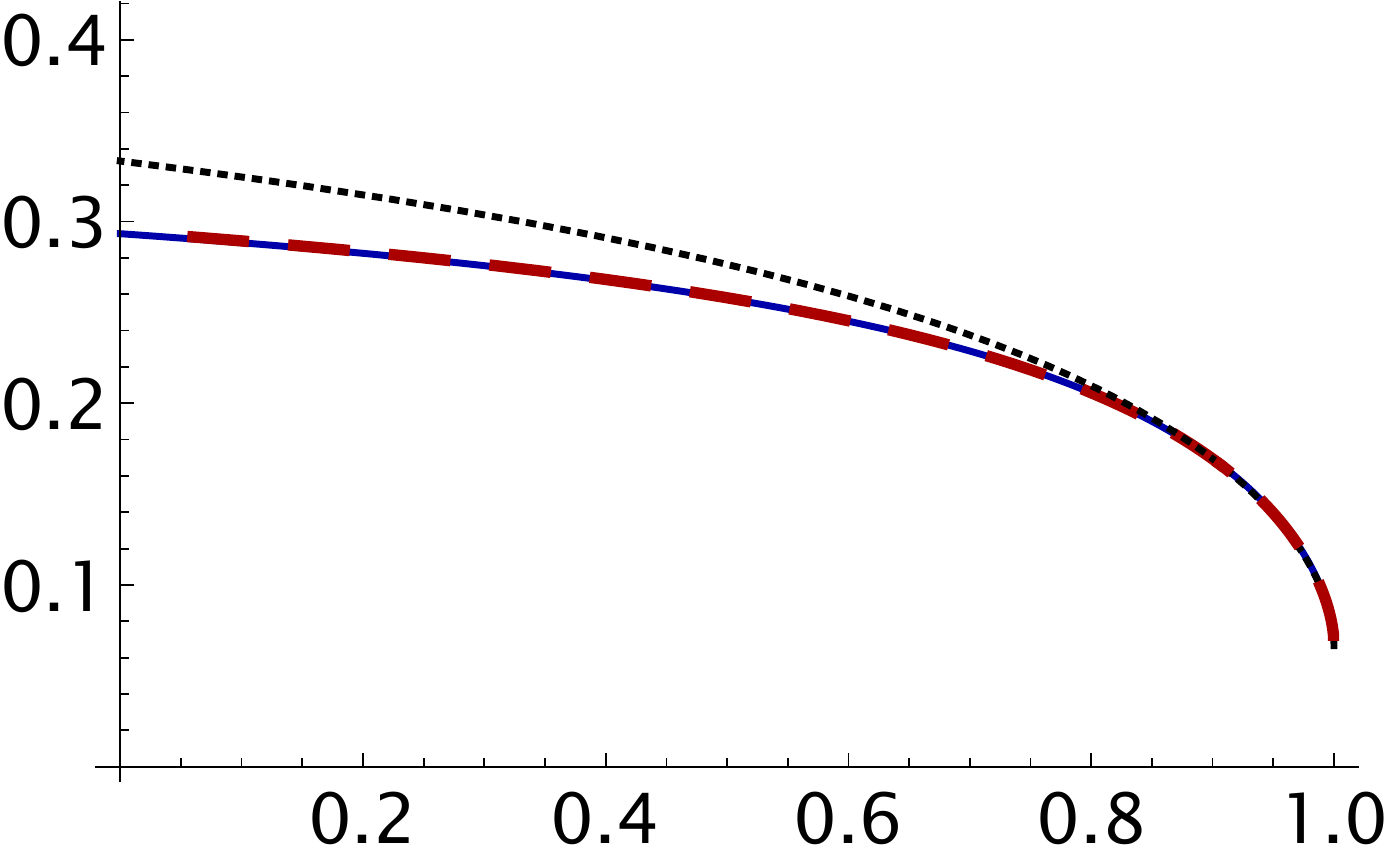} 
			\put(-190,130){$\AAA_J$}
			\put(-5,20){$e^{\Lambda}$}
		\end{subfigure}\vspace{5mm}
		\begin{subfigure}{0.45\textwidth}
			\includegraphics[width=\textwidth]{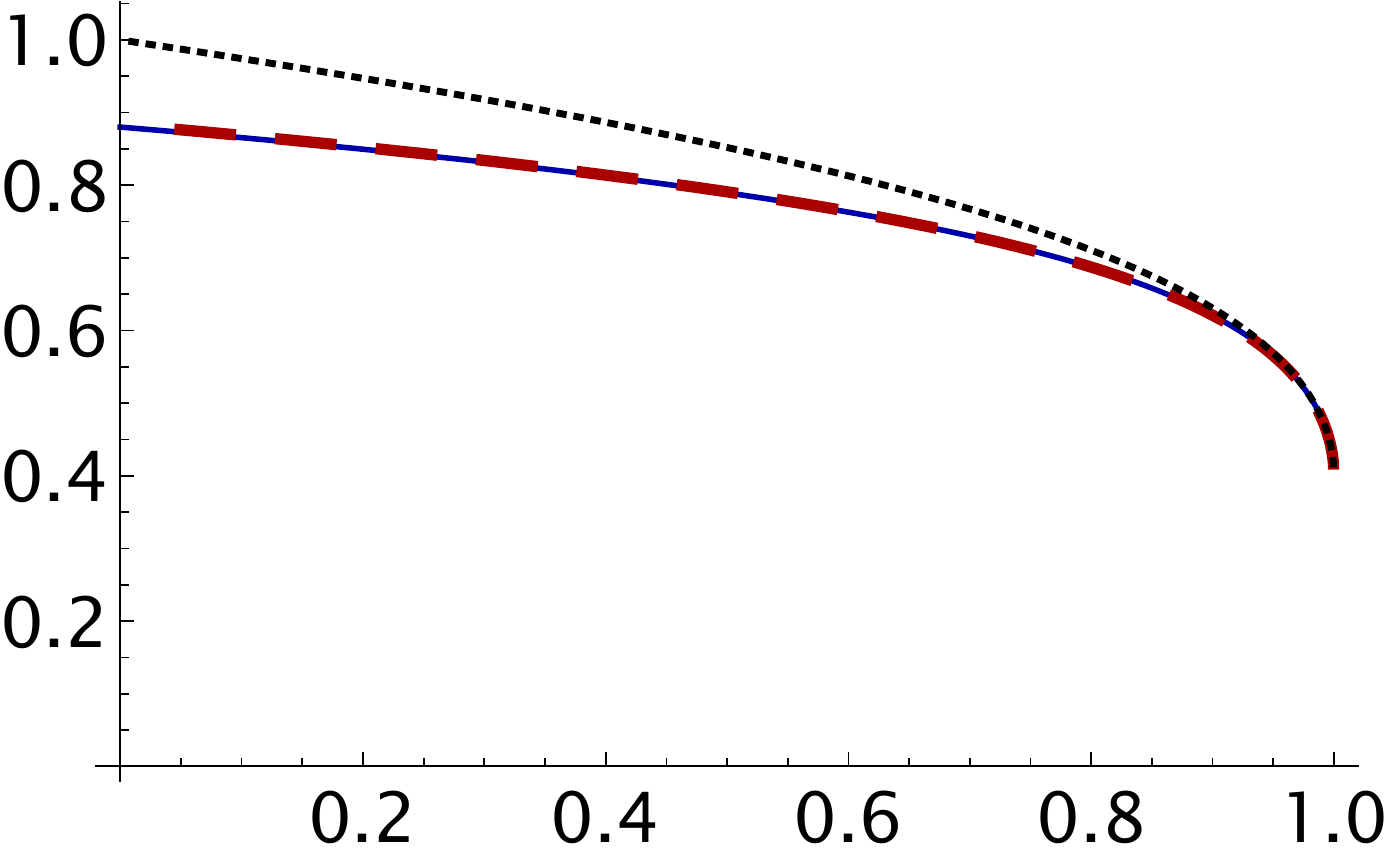} 
			\put(-190,130){$\BB_J$}
			\put(-5,20){$e^{\Lambda}$}
		\end{subfigure}\hfill
		\begin{subfigure}{.45\textwidth}
			\includegraphics[width=\textwidth]{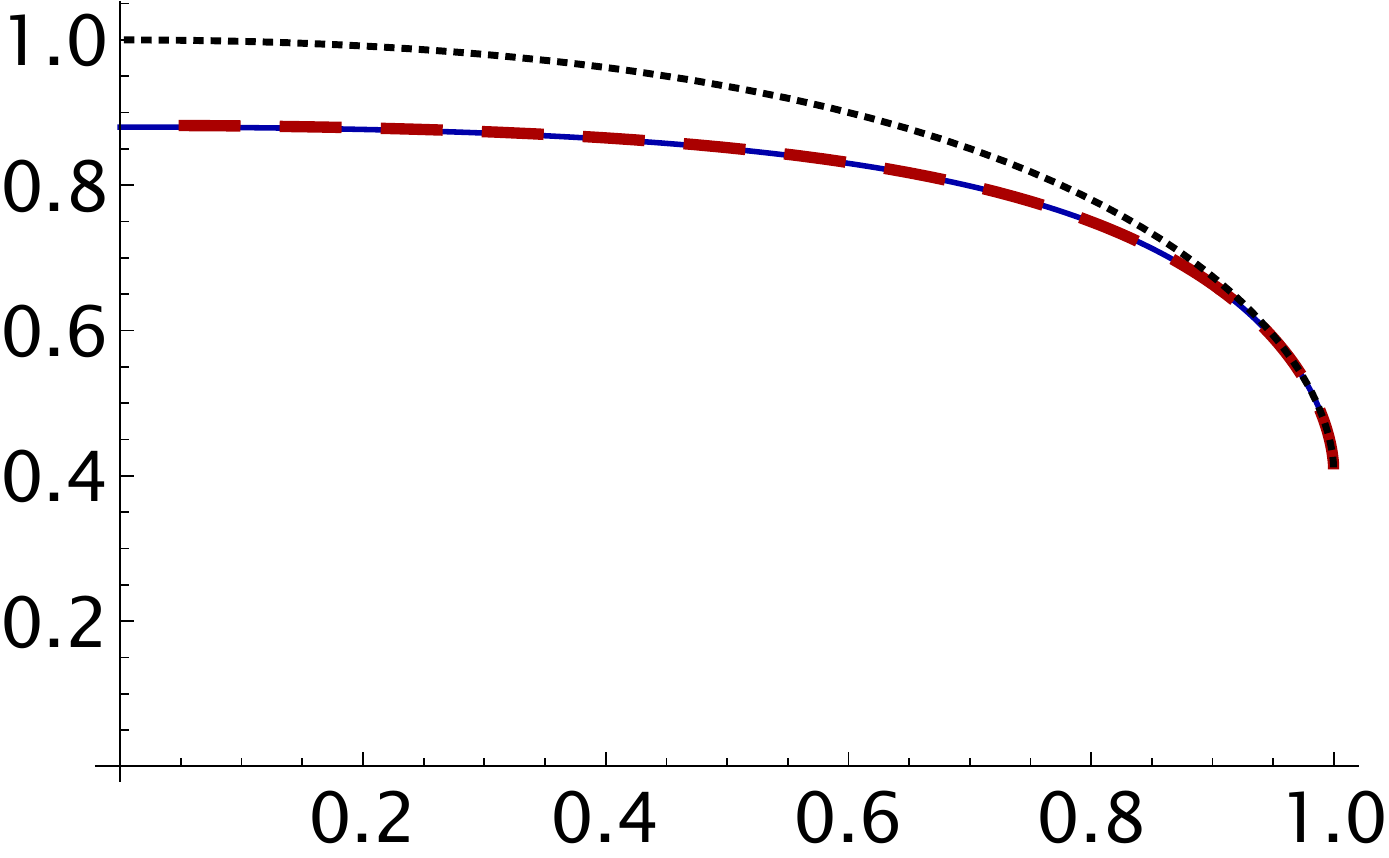} 
			\put(-190,130){$\BB_X$}
			\put(-5,20){$e^{\Lambda}$}
		\end{subfigure}
		\caption{\small Comparison of three solutions with $b_0=2/5$. The red, dashed curves correspond to the lowest-entropy solution that we were able to construct. The solid, blue curves show the horizonless, supersymmetric singular solution. These two solutions only differ in a tiny IR region. The dotted, black curves correspond to the horizonless, regular solution.}
		\label{fig.ZeroEntropyLimit}
	\end{center}
\end{figure}
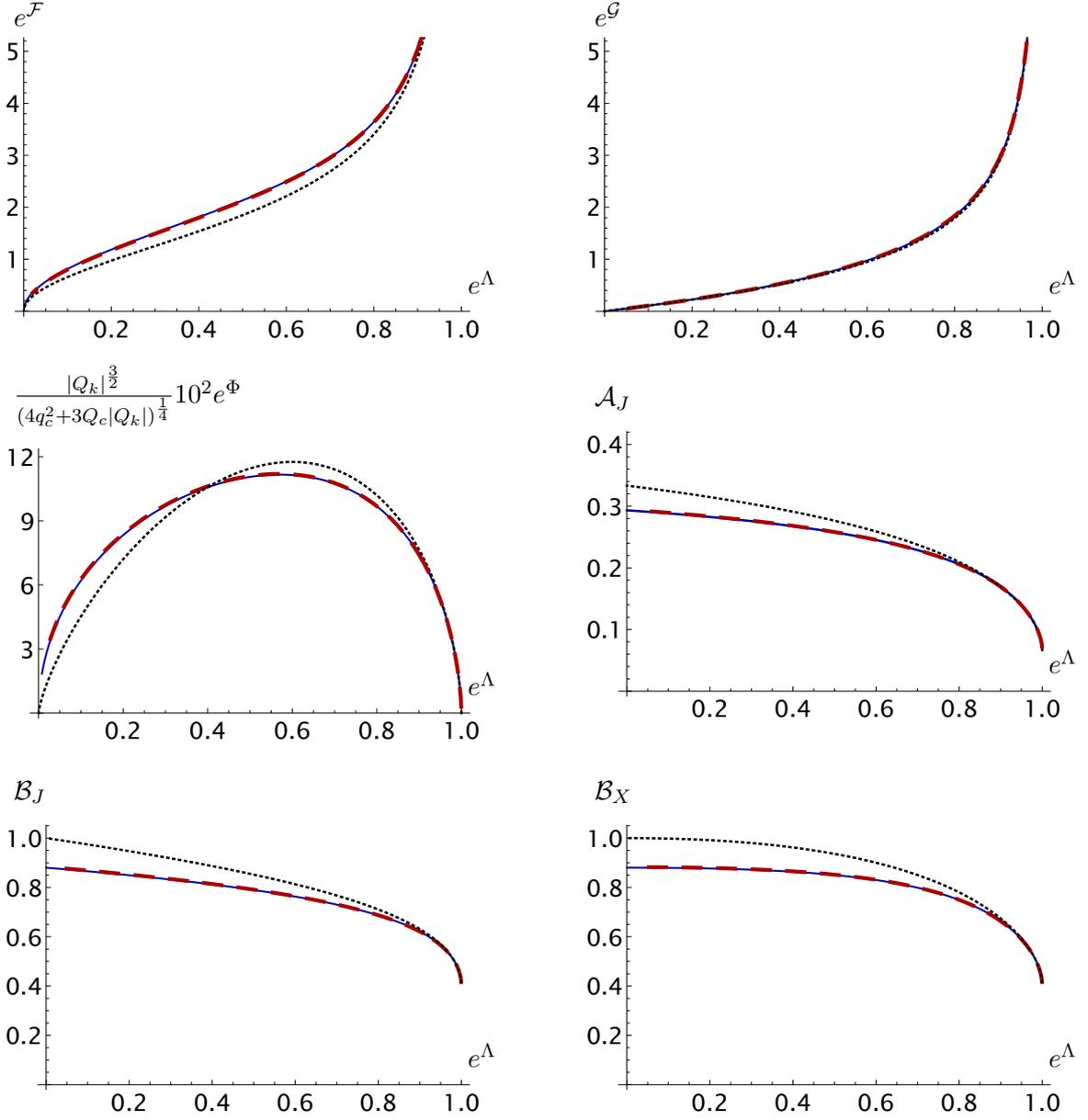

One indication comes from the fact that the eleven-dimensional curvature at the horizon diverges as we remove it. This suggests that a naked singularity is uncovered in this limit. This is illustrated  in Fig.~\ref{fig:Ricci}, where we show the Ricci scalar in terms of the horizon value of $e^\Lambda$,  $\lambda_h$, in order to obtain a manifestly gauge-invariant plot.  It can be seen that the scalar curvature at the horizon diverges as $R\sim \lambda_h^{-2/3}$ in the limit $\lambda_h\to0$, that is, as $u_h\to u_N$.
This behaviour is of course inherited from that of the metric functions. Indeed, it is possible to infer their dependence on $\lambda_h$ near zero from the numerics. For the internal components we obtain
\begin{equation}
e^\FF \sim \lambda_h^{1/2}\,,\qquad \qquad\qquad e^\GG \sim \lambda_h\,. 
\end{equation}
It turns out that this is precisely their behaviour in the IR of the $\mathbb{B}_8$ family, which is generically $\mathbb{R}^4\times{\rm S}^4$ (see Fig.~\ref{fig:triangle}). The transverse geometries in eleven dimensions\footnote{As given by Eq.~\eqref{11dtrans}, with the metric functions found numerically.} obtained by removing the horizon are thus perfectly regular, capping off smoothly at a certain $u_N$. Furthermore, the coincidence with a $\mathbb{B}_8$ metric goes beyond the IR and extends to the entire solution. In Fig.~\ref{fig.ZeroEntropyLimit} (top) we compare the metric functions that describe the transverse metric for the solution with lowest entropy that we constructed for $b_0=2/5$ with those of the $\mathbb{B}_8$ metric that ends at the same value of $u_N$. They coincide within numerical precision except in a tiny IR region due to the small horizon that still exists in the thermal solution. We conclude that, as far as the eight-dimensional transverse geometry is concerned, the zero-entropy limit $u_h\to u_N$ describes a regular $\mathbb{B}_8$ metric with the end-of-space at $u_N$ instead of the $u_s$ of the corresponding ground state.

The singularity in the full eleven-dimensional metric must then come from the warp factor. As we approach the zero-entropy limit it goes as
\begin{equation}\label{divh}
 \mathbf{h}\sim \lambda_h^{-2}\,.
\end{equation}
Again, this behaviour can be identified with that of a known solution. As explained in Section~\ref{LowT}, for the horizonless solutions to be regular it is necessary to include internal fluxes (describing fractional branes) with fine-tuned values. If these values are not the correct ones the warp factor is IR-singular, with a divergence given exactly by Eq.~\eqref{divh}. We thus conclude that the zero-entropy limit of our black-brane solutions is a supersymmetric solution with fluxes, but the limiting values of these fluxes are not the correct ones to render the solution regular in the IR.  This can be seen in Fig.~\ref{fig.ZeroEntropyLimit}, where we compare, for $b_0=2/5$, the lowest-entropy black brane solution that we constructed  (i) with the supersymmetric solution to the BPS equations in \cite{Faedo:2017fbv} with the end-of-space at $u_N(2/5)$, and  (ii) with the regular supersymmetric ground state. The first two backgrounds agree within numerical precision except in a tiny IR region due to the impossibility of reaching $u_h\to u_N$ numerically, while the regular one is clearly distinct.  
 
In summary, in the limit in which we remove the horizon and the entropy vanishes we find a supersymmetric solution which is nevertheless singular, since the subleading coefficients that we recover are not the ones that ensure regularity of the ground state. The singularity is good in the sense of \cite{Gubser:2000nd} because it can be hidden behind a horizon. It is interesting that this limit of zero entropy is reached at a finite temperature, as we will see  in detail in the next section.

\section{Thermodynamics and the phase diagram}
\label{sec:thermodynamics}

In this section we will discuss the main thermodynamic properties of the black branes we have constructed and confirm the presence of phase transitions between themselves and between them and the low-temperature phases described in Section~\ref{LowT}. In this way we will construct   the phase diagram in the $\left(b_0,T\right)$-plane for the entire family of theories. 
 
\subsection{Thermodynamic quantities}

The renormalized four-dimensional bulk action $S_{\mbox{\footnotesize ren}}$ describing the system is obtained in Section~\ref{sec:4Deffectivetheory}. From this it is possible to extract the different thermodynamic quantities. The free energy density is given by
\begin{equation}
F= -\frac{S_{\mbox{\footnotesize ren}}}{\beta V_2}\,,
\end{equation}
with $V_2$ the (infinite) volume in the spatial directions and $\beta$ the period of the compact Euclidean time. This period, which is related to the temperature through Eq.~\eqref{beta}, is fixed in the black-brane  geometries by requiring the absence of conical singularities. As usual the Bekenstein--Hawking entropy is given by the area of the horizon. Using the known UV and horizon expansions, Eqs.~\eqref{UVexpansions} and \eqref{Horizon_expansions}, to evaluate these expressions, we obtain the following dimensionless values for the free energy, entropy and temperature 
\begin{eqnarray}\label{themo_quantities}
\overline{F}&=&\frac{2\kappa_4^2 }{|Q_k|^5}\ F =-\frac{411}{2}-6 f_4-2 f_5+\frac{3}{2}  \mathsf{b}_5\,,\nonumber\\ [2mm]
\overline{S}&=&\frac{2\kappa_4^2}{|Q_k|^3(4q_c ^2+3Q_c|Q_k|)^{1/2}} \ S =\frac{64 \pi  f_h^4 g_h^2 \sqrt{h_h}}{\lambda _h^2}\,,\\ [2mm]
\overline{T}&=&\frac{(4q_c^2 + 3Q_c|Q_k|)^{1/2}}{|Q_k|^2}\ T =- \frac{1}{4\pi} \frac{\mathsf{b}_hu_h^2}{\sqrt{h_h}}\,,\nonumber
\end{eqnarray}
where $\mathsf{b}_5$ is understood to be given by Eq.~\eqref{b5_conserved} in terms of horizon data. Similarly, the internal energy density $E$  and pressure $P$, computed from the energy-momentum tensor in Eq.~\eqref{EMtensor}, are given by
\begin{equation}
	\frac{2\kappa_4^2 P}{|Q_k|^5} = \frac{411}{2} +6 f_4+2 f_5-\frac{3}{2}  \mathsf{b}_5\,, \qquad \qquad
	\frac{2\kappa_4^2 E}{|Q_k|^5} =-\frac{411}{2}-6 f_4-2f_5-\frac{7}{2}  \mathsf{b}_5\,,
\end{equation}
which fulfil the expected thermodynamic relations $F=-P$ and $F=E-TS$. Finally, it must be verified that 
\begin{equation}
S= - \frac{\dd F}{\dd T}\,,
\end{equation}
which can be used as a check of the numerical results. Other quantities like the specific heat and the speed of sound can be straightforwardly computed from these thermodynamic potentials.

\subsection{Phase diagram}
\label{sec:phase_diagrams}

For each value of  $b_0$, namely for each gauge theory in the family, we must determine the preferred solution at each temperature. The first type of solutions that compete are those of Section~\ref{LowT}, which are obtained from the horizonless, supersymmetric, regular solutions simply by compactifying the Euclidean time. The period of this thermal circle is arbitrary, so these geometries exist for any  temperature and, in our renormalization scheme, they have vanishing  free energy and entropy. The second type of solutions are the black branes found in Section~\ref{HighT}. Their free energy and entropy are given in terms of UV and horizon data by Eq.~\eqref{themo_quantities} and have a non-trivial dependence on the temperature. If several solutions exist at a given temperature, the one with lower free energy will be thermodynamically preferred. Since the free energy of the first type of solutions vanishes, the black branes will dominate if their free energy is negative. At points where the free energy of black branes crosses zero there is a phase transition to the gapped state.   The nature of the phase transition depends on the value of $b_0$. We identified two special values of $b_0$ at which the qualitative features change: 
\begin{equation}
\label{below}
b_0^\textrm{critical}\approx 0.6815 \, ,\qquad b_0^\textrm{triple}\approx 0.6847 \,.
\end{equation}
Note that these are very close to one another. We will come back to this in Section~\ref{sec:conclusions}. These two values give rise to the following three regions:  
\begin{description}
\item[{\textbf{Case A:}} Phase transitions for \mbox{$b_0 \in (b_0^\textrm{triple},1]$}.]
In this range the situation is similar to the typical Hawking--Page phase transition. The free energy crosses the horizontal axis at some critical temperature $T_c$. The preferred phase above $T_c$ is the black brane and below $T_c$ it is the horizonless, regular solution. Therefore at $T_c$ there is a phase transition between a gapped and an ungapped phase. At the critical temperature the entropy density jumps between some finite value and zero, signaling that the transition is first-order. In Fig.~\ref{fig.TypeI_PT} we show the free energy and the entropy densities as a function of temperature for a representative case with  $b_0=0.7902$.
	\begin{figure}[t]
		\begin{center}
			\begin{subfigure}{.45\textwidth}
				\includegraphics[width=\textwidth]{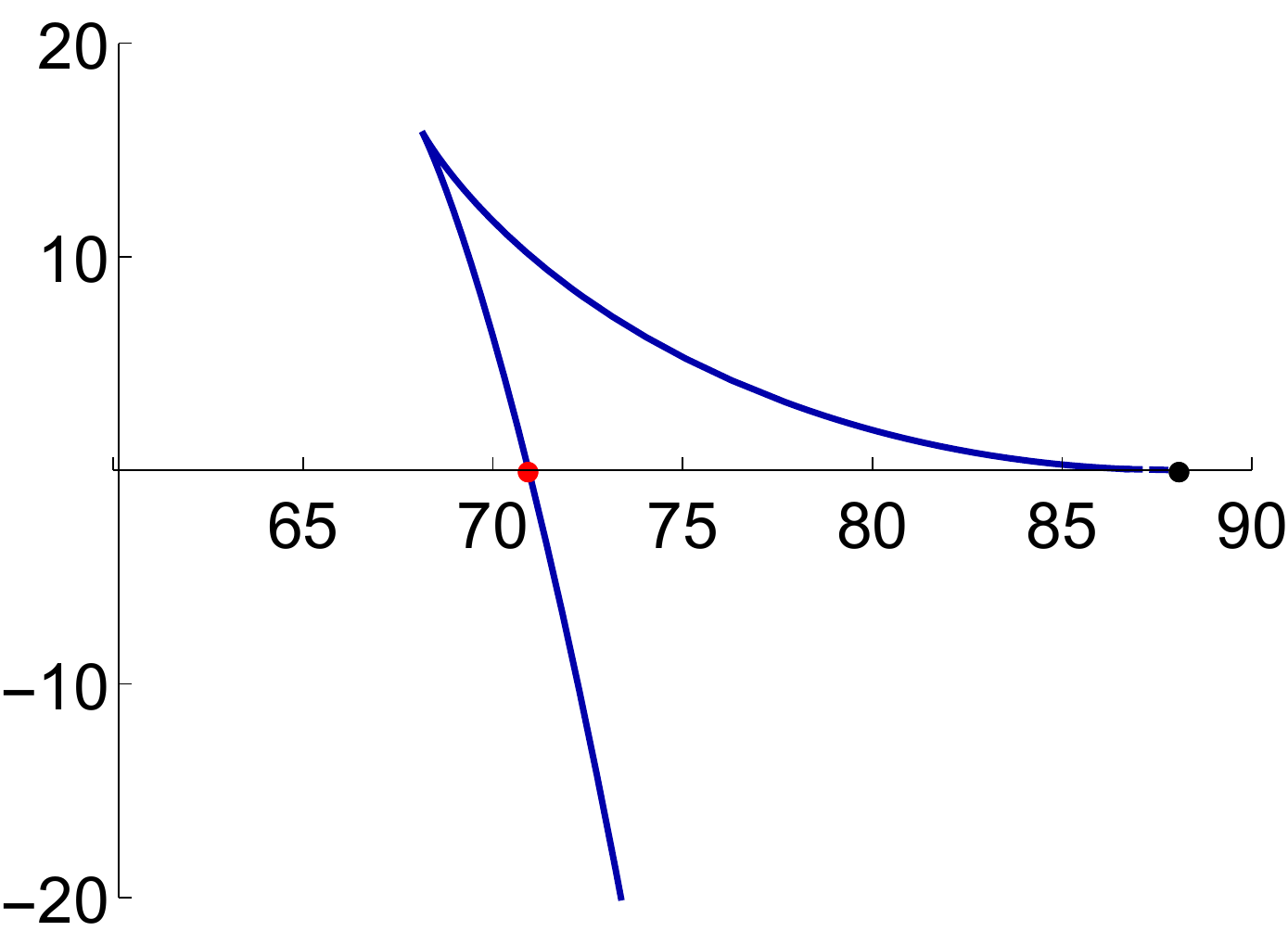} 
				\put(-190,150){$10^{-4}\ \overline F$}
				\put(-10,80){$\overline T$}
			\end{subfigure}\hfill
			\begin{subfigure}{.45\textwidth}
				\includegraphics[width=\textwidth]{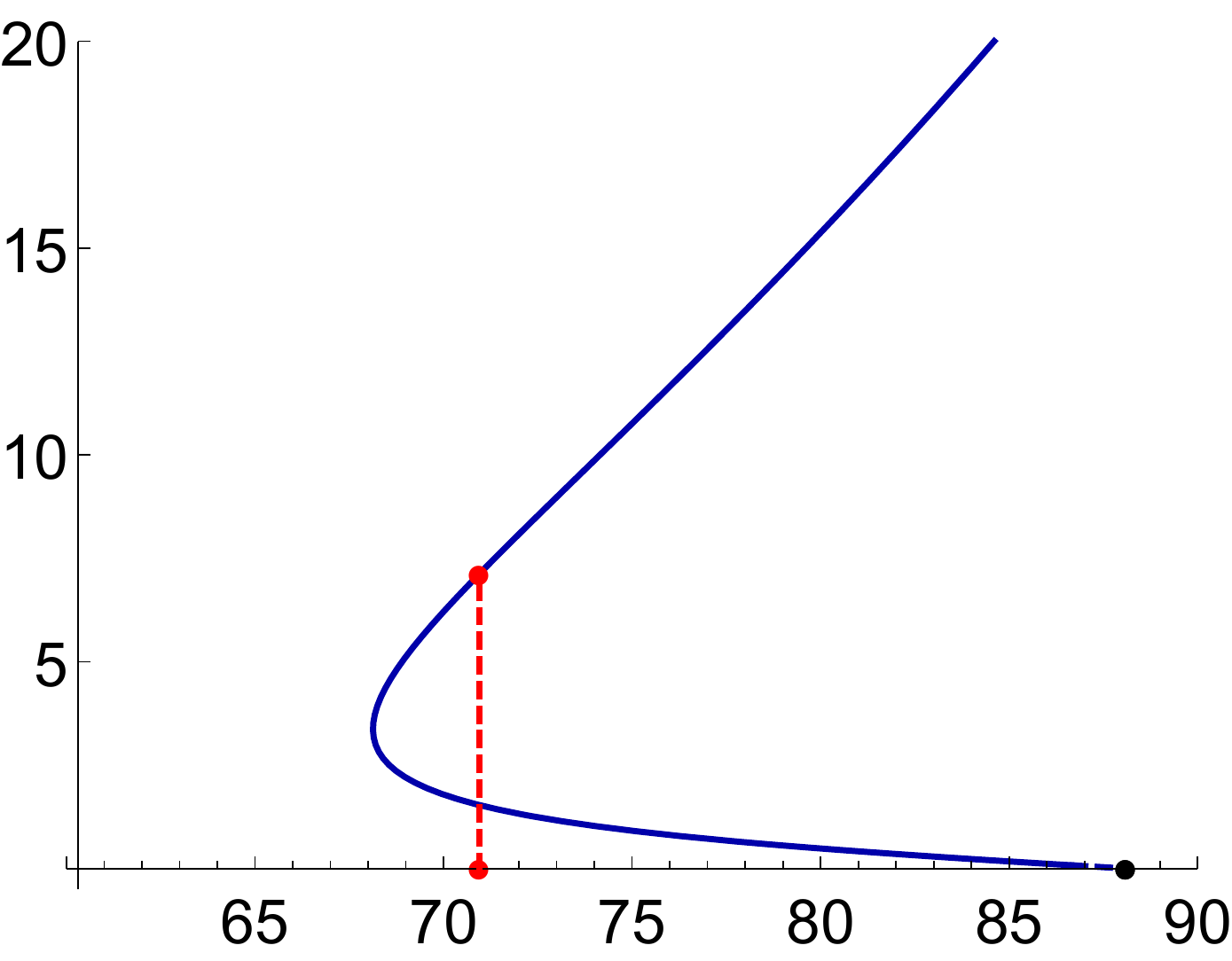} 
				\put(-200,155){$10^{-4}\  \overline S$}
				\put(-10,20){$\overline T$}
			\end{subfigure}
			\caption{\small Free energy density (left) and entropy density (right) as a function of the temperature for black branes in the theory with $b_0=0.7902$ (Case A). When the free energy crosses the axis, there is a first-order phase transition at which the entropy changes discontinuously. The dashed, red line indicates  the critical temperature. The region close to the horizontal axis of the blue curves, shown with dashes, is the result of an extrapolation to zero entropy.}\label{fig.TypeI_PT}
		\end{center}
	\end{figure}
\item[{\textbf{Case B:}} Phase transitions for  
\mbox{$b_0 \in (b_0^\textrm{critical},b_0^\textrm{triple})$}.]
In this small range the theories exhibit two phase transitions. This is illustrated in Fig.~\ref{fig.TypeII_PT}, where we show the free energy and the entropy of the black branes for the theory with $b_0=0.6835$. Decreasing the temperature from asymptotically high values we first find a first-order phase transition between two black brane geometries at which the entropy density changes discontinuously. Therefore this transition takes place between two ungapped phases and is indicated in Fig.~\ref{fig.TypeII_PT} by the dashed, red line. Decreasing the temperature further along the low-temperature black-brane branch we find that the entropy and the free energy densities vanish at a finite value of the temperature. In this zero-entropy limit we recover the horizonless, singular solution described in Section~\ref{sec:zeroEntropyLimit}. Strictly speaking, since this solution is singular, its temperature is undetermined. However, the limit along the regular branch of black brane solutions results in a fixed, non-zero value of the temperature, indicated by the black dot in Fig.~\ref{fig.TypeII_PT}. At this temperature we expect a transition between the supersymmetric,  horizonless, singular solution and the supersymmetric,  horizonless, regular solution with the same value of $b_0$. This transition is analogous to that in Case A  except for the fact that the entropy density is continuous across the transition. Nevertheless, the transition is still first-order because the solution changes discontinuously, as illustrated in Fig.~\ref{fig.ZeroEntropyLimit}. In the gauge theory this would be reflected in a discontinuity in observable quantities such as $n$-point functions. Note that, strictly speaking, the supergravity description breaks down sufficiently close to the transition since the curvature diverges at that point, as shown in Fig.~\ref{fig:Ricci}. 
	\begin{figure}[t]
	\begin{center}
		\begin{subfigure}{.45\textwidth}
			\includegraphics[width=\textwidth]{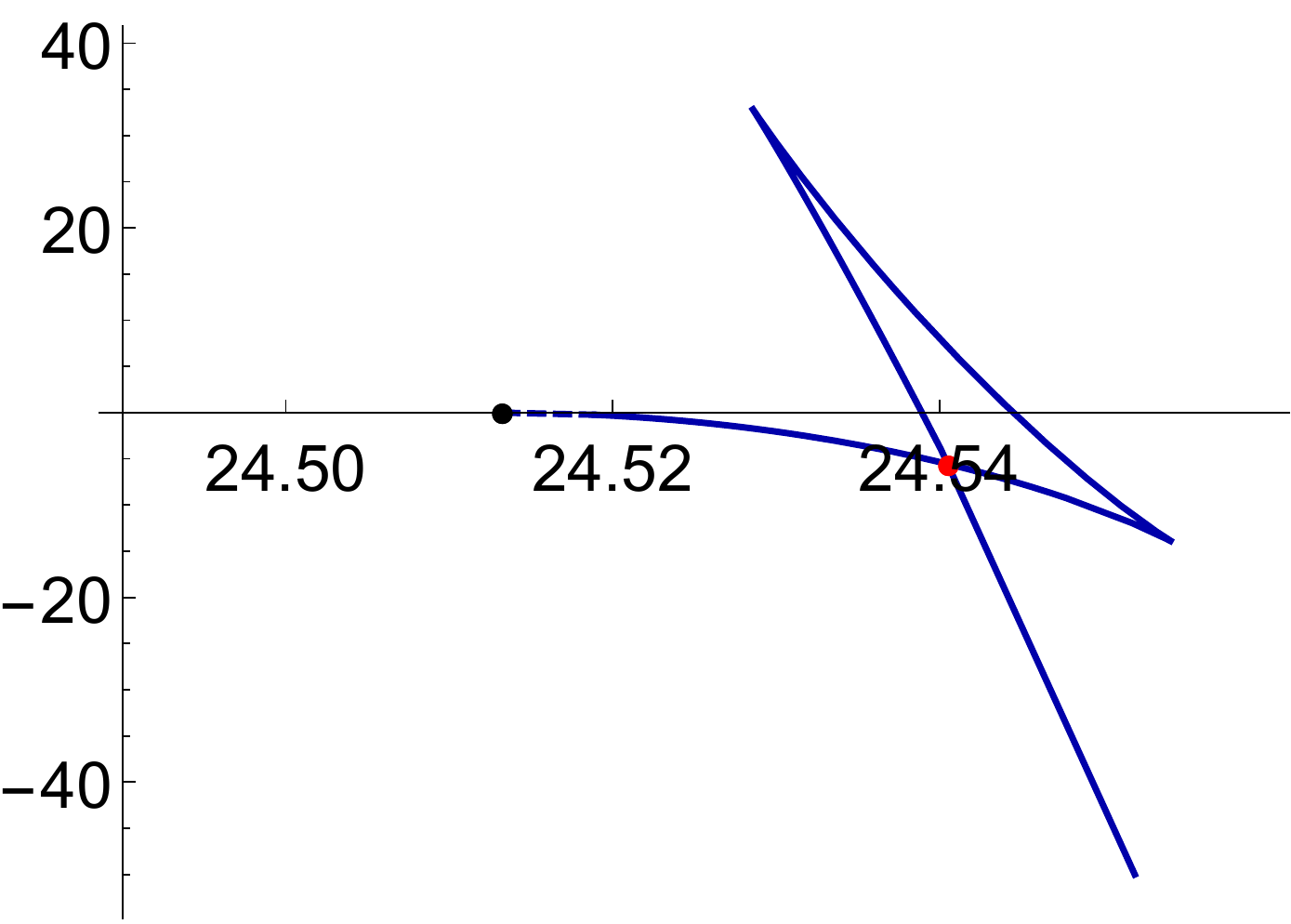} 
			\put(-170,130){$\overline F$}
			\put(-15,90){$\overline T$}
		\end{subfigure}\hfill
		\begin{subfigure}{.45\textwidth}
			\includegraphics[width=\textwidth]{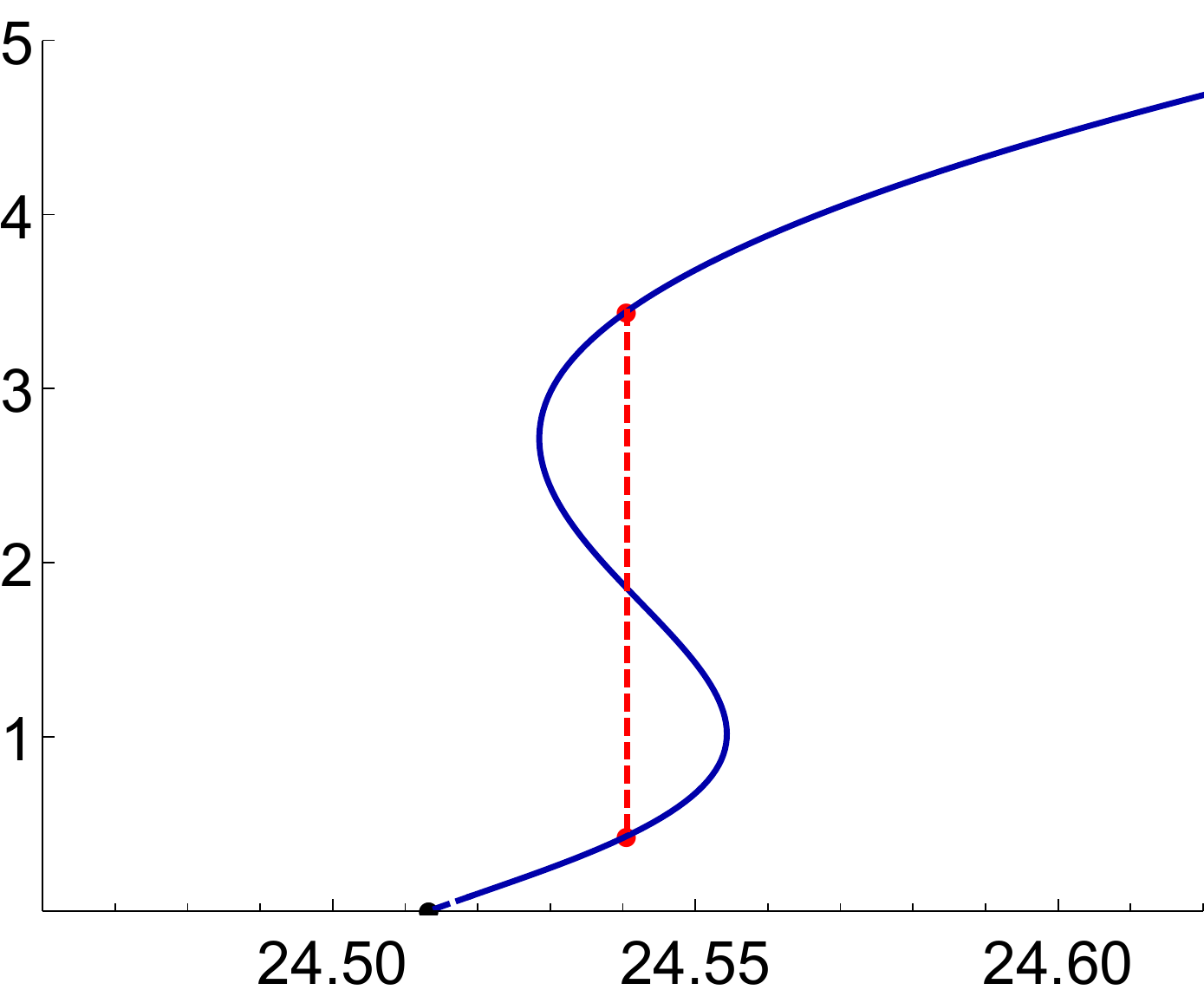} 
			\put(-180,150){$ 10^{-3}\ \overline S$}
			\put(-10,20){$\overline T$}
		\end{subfigure}
		\caption{\small Free energy (left) and entropy (right) as a function of the temperature for black branes in the theory with $b_0=0.6835$ (Case B). As the temperature decreases, there is a first order phase transition between two branches of black branes. At that point the entropy changes from some finite value to another finite value, represented by the dashed red line. Decreasing further the temperature, one would find another phase transition from the second black brane phase to the regular phase without a horizon. This happens at some finite value of the temperature and vanishing entropy. The region close to the horizontal axis of the blue curve, shown with dashes, is the result of an extrapolation to zero entropy.}\label{fig.TypeII_PT}
	\end{center}
\end{figure}

\item[{\textbf{Case C:}} Phase transitions for  \mbox{$b_0 \in (0, b_0^\textrm{critical})$}.]
In this range of values the phase transition between black branes disappears. The qualitative behaviour of the free energy and entropy densities  is similar to that in Fig.~\ref{fig.TypeIII_PT}, where we show the results for $b_0 = 0.5750$. For the theories in this range there is a unique phase transition between the ungapped black brane phase and the regular gapped solution. Again, this happens at zero entropy but finite temperature, as in Case B. 
\begin{figure}[t]
	\begin{center}
		\begin{subfigure}{.45\textwidth}
			\includegraphics[width=\textwidth]{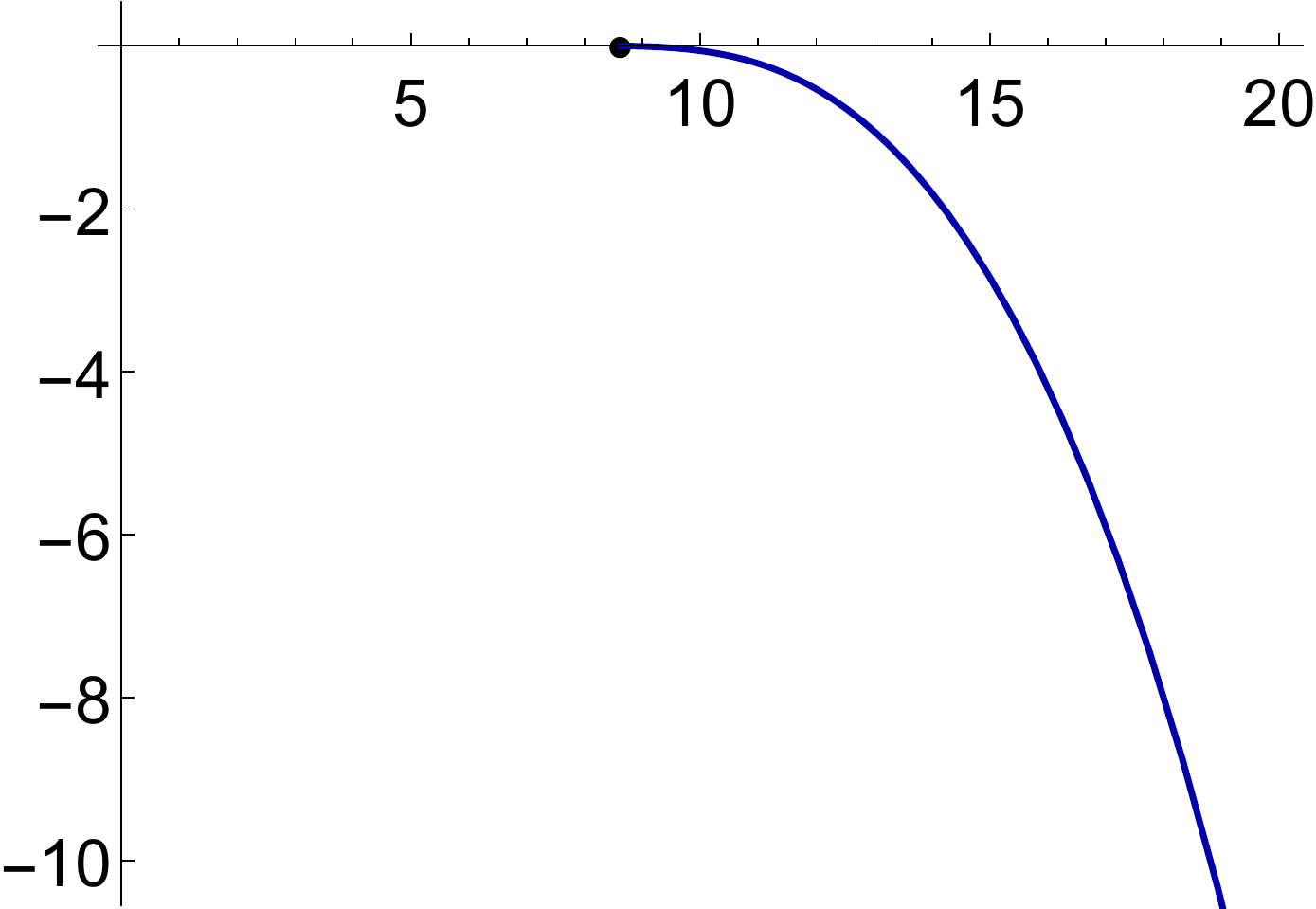} 
			\put(-180,140){$10^{-4}\ \overline F$}
			\put(-5,100){$\overline T$}
		\end{subfigure}\hfill
		\begin{subfigure}{.45\textwidth}
			\includegraphics[width=\textwidth]{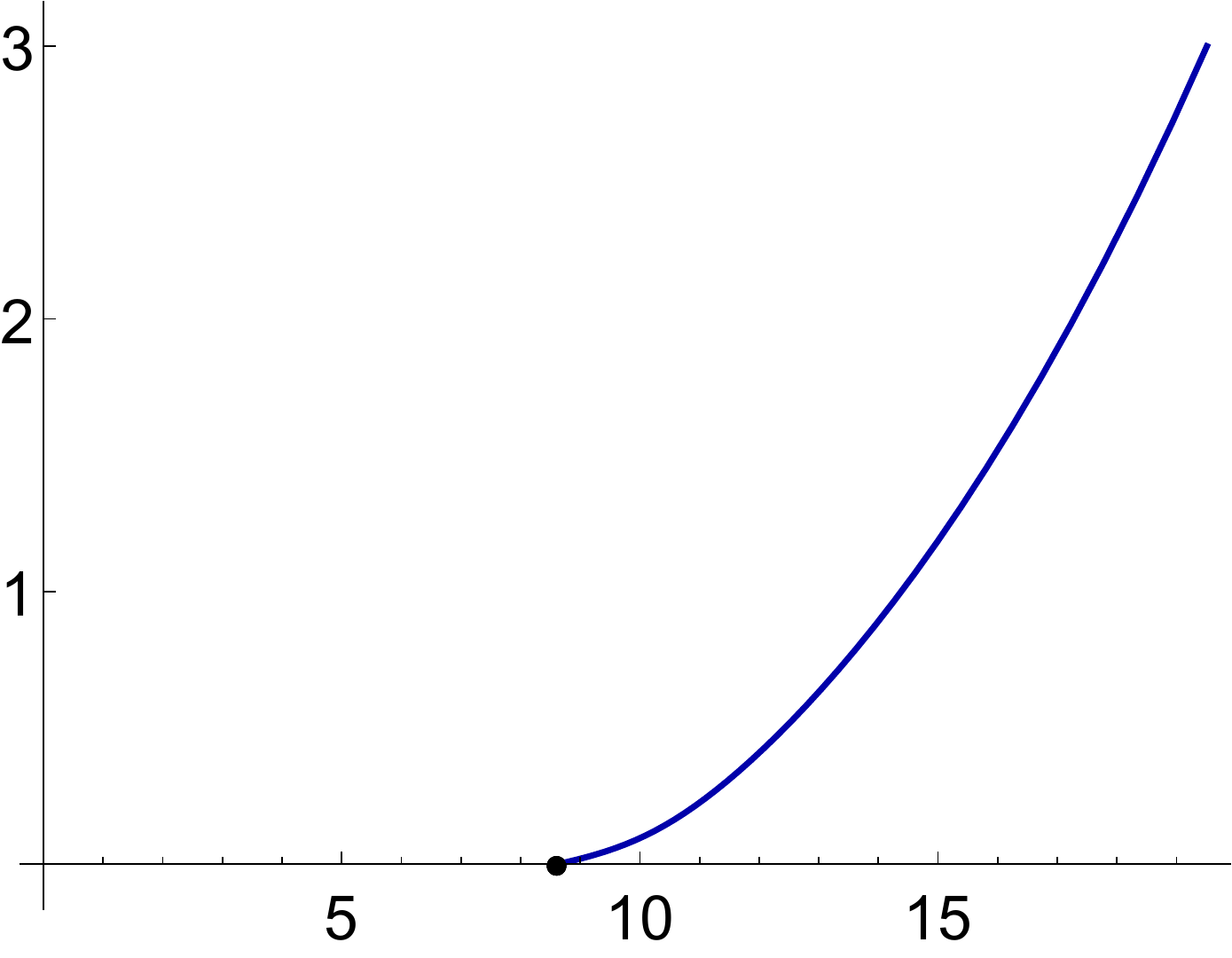} 
			\put(-180,140){$10^{-4} \ \overline S$}
			\put(-10,20){$\overline T$}
		\end{subfigure}
		\caption{\small Free energy density (left) and entropy density (right) as a function of the temperature for black branes in the theory with $b_0 = 0.5750$ (Case C). In this example, there is a single phase transition that takes place at finite temperature and zero entropy. The region close to the horizontal axis of the blue curve, shown with dashes, is the result of an extrapolation to zero entropy.}\label{fig.TypeIII_PT}
	\end{center}
\end{figure}
\end{description}

\begin{figure}[t]
	\begin{center}
		\begin{subfigure}{.90\textwidth}
			\includegraphics[width=\textwidth]{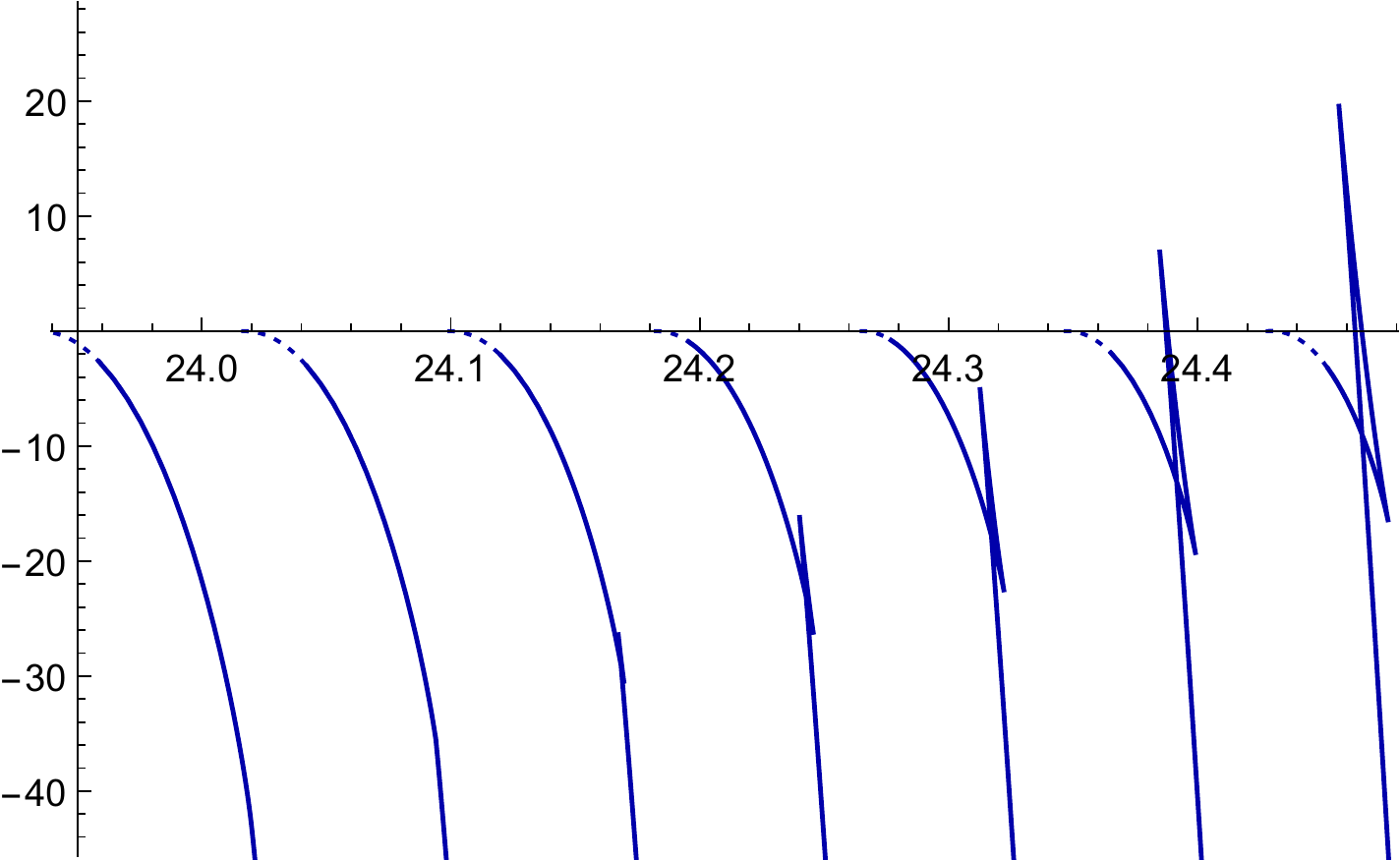} 
			\put(-360,240){$\overline F$}
			\put(5,160){$\overline T$}
		\end{subfigure}\vspace{9mm}
		\begin{subfigure}{.90\textwidth}
			\includegraphics[width=\textwidth]{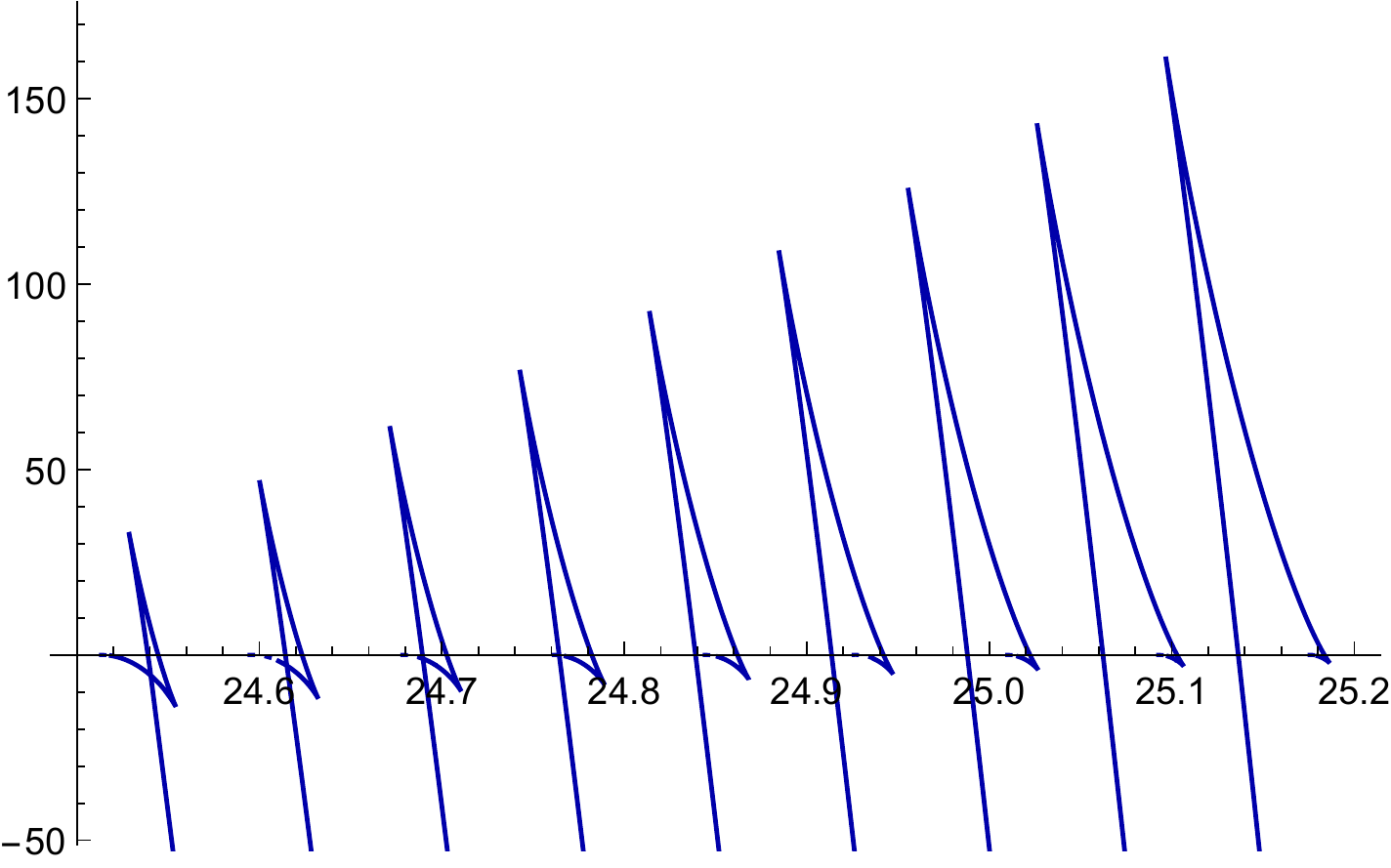} 
			\put(-360,240){$\overline F$}
			\put(5,70){$\overline T$}
		\end{subfigure}
		\caption{\small Free energy density  for different theories with $b_0$ varying from $b_0=0.6813 \lesssim b_0^\textrm{critical}$ to $b_0=0.6861\gtrsim b_0^\textrm{triple}$ from left to right and from top to bottom, with an approximate separation of $\Delta b_0 \approx 0.0003$ between adjacent curves. A triangle develops as we decrease $b_0$ (bottom panel). If $b_0$ is decreased further then the triangle disappears (top panel). The region close to the horizontal axis of the curve, shown with dashes, is the result of an extrapolation to zero entropy.}
		\label{fig:FreeEnergyCrossOver}
	\end{center}
\end{figure}

\begin{figure}[t]
	\begin{center}
		\includegraphics[width=.70\textwidth]{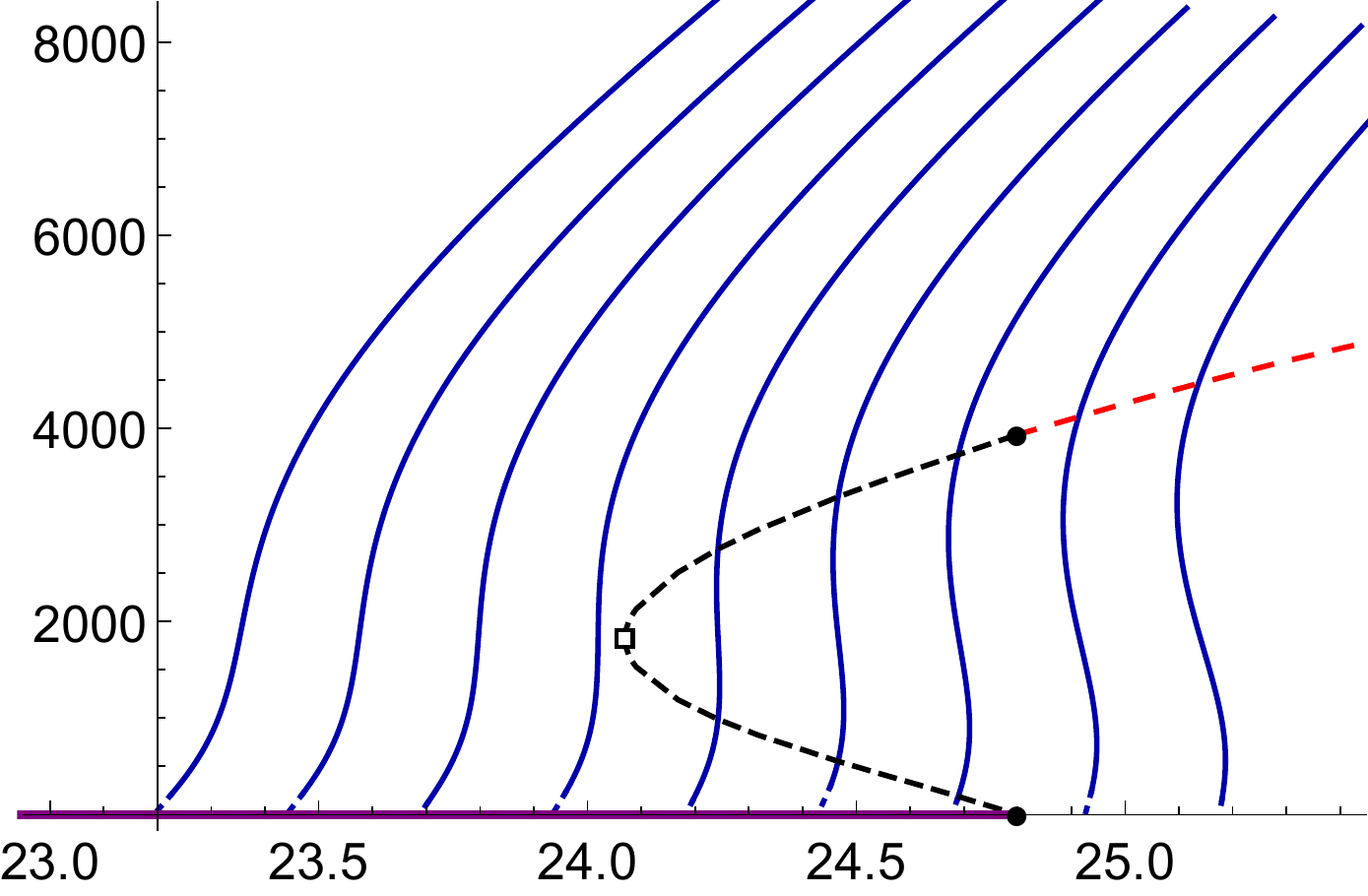} 
		\put(-260,195){\rotatebox{0}{$\overline S $}}
		\put(0,25){$\overline T$}
		\caption{\small Entropy density for different theories with $b_0$ varying from $b_0=0.6783 \lesssim b_0^\textrm{critical}$ to \mbox{$b_0=0.6861\gtrsim b_0^\textrm{triple}$} from left to right, with an approximate separation of $\Delta b_0 \approx 0.001$ between adjacent curves. The dashed, red curve indicates first-order phase transitions between a black brane phase and a regular horizonless phase with a discontinuous jump in the entropy density. The dashed, black curve indicates first-order phase transitions between two black branes. The turning point of this line, indicated with a square, corresponds to the second-order phase transition at the critical point. The solid, purple line on the axis indicates first-order transitions between the zero-entropy limit of a black brane branch and a regular horizonless solution. The entropy density is continuous across these transitions. The phase transition corresponding to the triple point is indicated with two black dots. The region close to the horizontal axis of the blue curves, shown with dashes, is the result of an extrapolation to zero entropy.}
		\label{fig:EntropyCrossOver}
	\end{center}
\end{figure}

The change from one behaviour to another happens smoothly as we vary $b_0$. Essentially, Case B is an intermediate scenario between Cases A and C. This can be seen by plotting the free energy and entropy densities for different values of the parameter in a small region that contains that of Case B entirely. The result is shown in Figs.~\ref{fig:FreeEnergyCrossOver} and \ref{fig:EntropyCrossOver}, respectively. The largest value of the parameter in the bottom panel of Fig.~\ref{fig:FreeEnergyCrossOver} is $b_0\approx0.6861 \gtrsim b_0^\textrm{triple}$ and corresponds to the rightmost curve. As we decrease $b_0$, a small triangle, ending on the axis, starts to grow below the unstable branch. The lower side of this triangle is thermodynamically stable but not dominant. Decreasing further the parameter $b_0$, the lower side of the triangle ends up crossing the previous stable branch and becomes favoured. This explains the change from Case A to Case B. At this point the system goes from having one first-order transition to having two of them. Therefore this is a triple point at which three phases can coexist. 

As the size of the second stable branch increases, the size of the triangle decreases. This explains the change from Case B to Case C: by reducing $b_0$ the triangle eventually disappears. This can be seen in the top panel of Fig.~\ref{fig:FreeEnergyCrossOver}, where the leftmost curve corresponds to $b_0\approx0.6813 \lesssim b_0^\textrm{critical}$.  The  point where the triangle disappears is a critical point, namely the end of a line of first-order transitions. At this point the transition becomes second-order, as can be seen in the plot of the entropy density of Fig.~\ref{fig:EntropyCrossOver}. 
	\begin{figure}[t]
	\begin{center}
		\begin{subfigure}{.48\textwidth}
			\includegraphics[width=\textwidth]{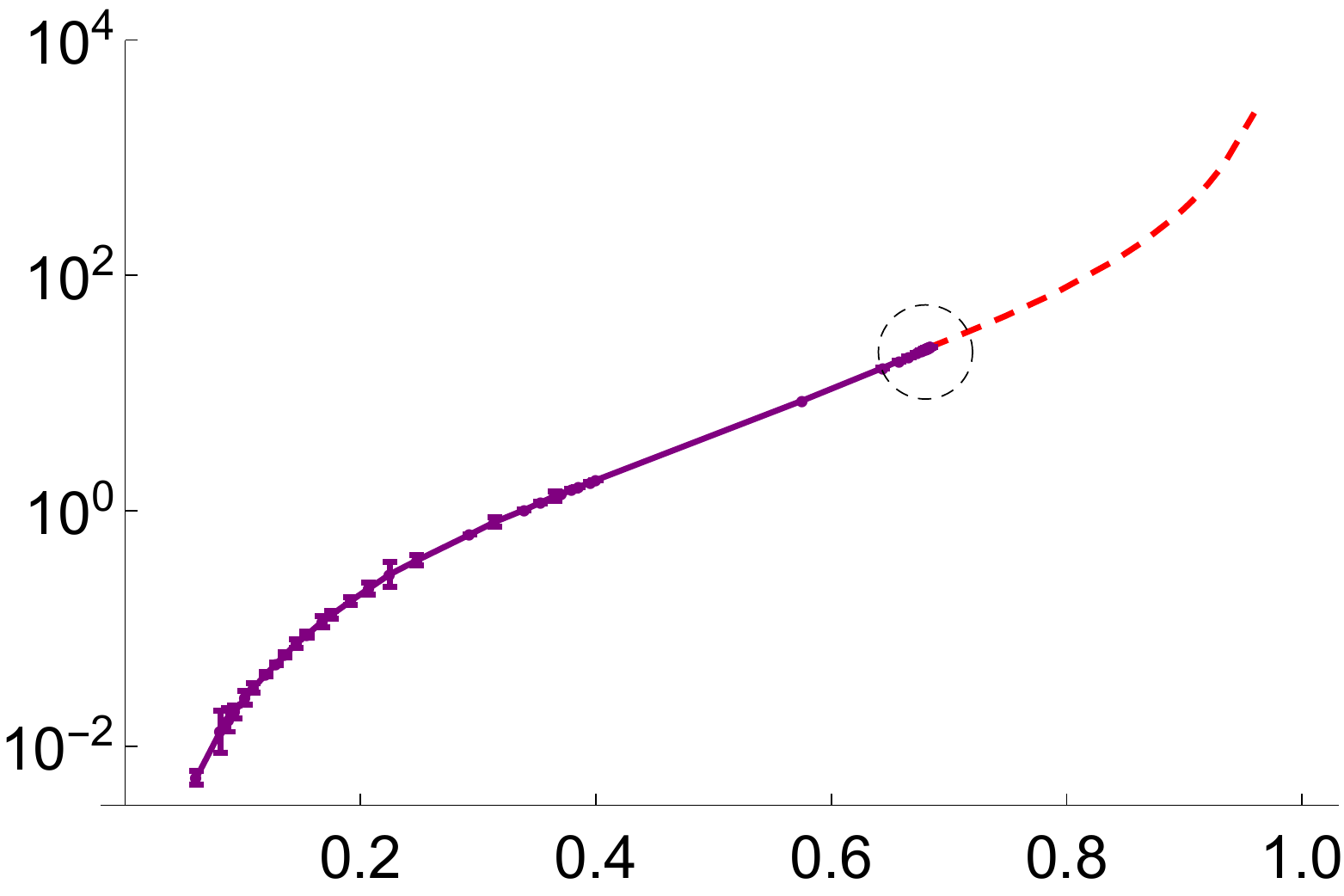} 
			\put(-180,130){\rotatebox{0}{$\overline T_c $}}
			\put(-5,20){$b_0$}
		\end{subfigure}\hfill
		\begin{subfigure}{.48\textwidth}
			\includegraphics[width=\textwidth]{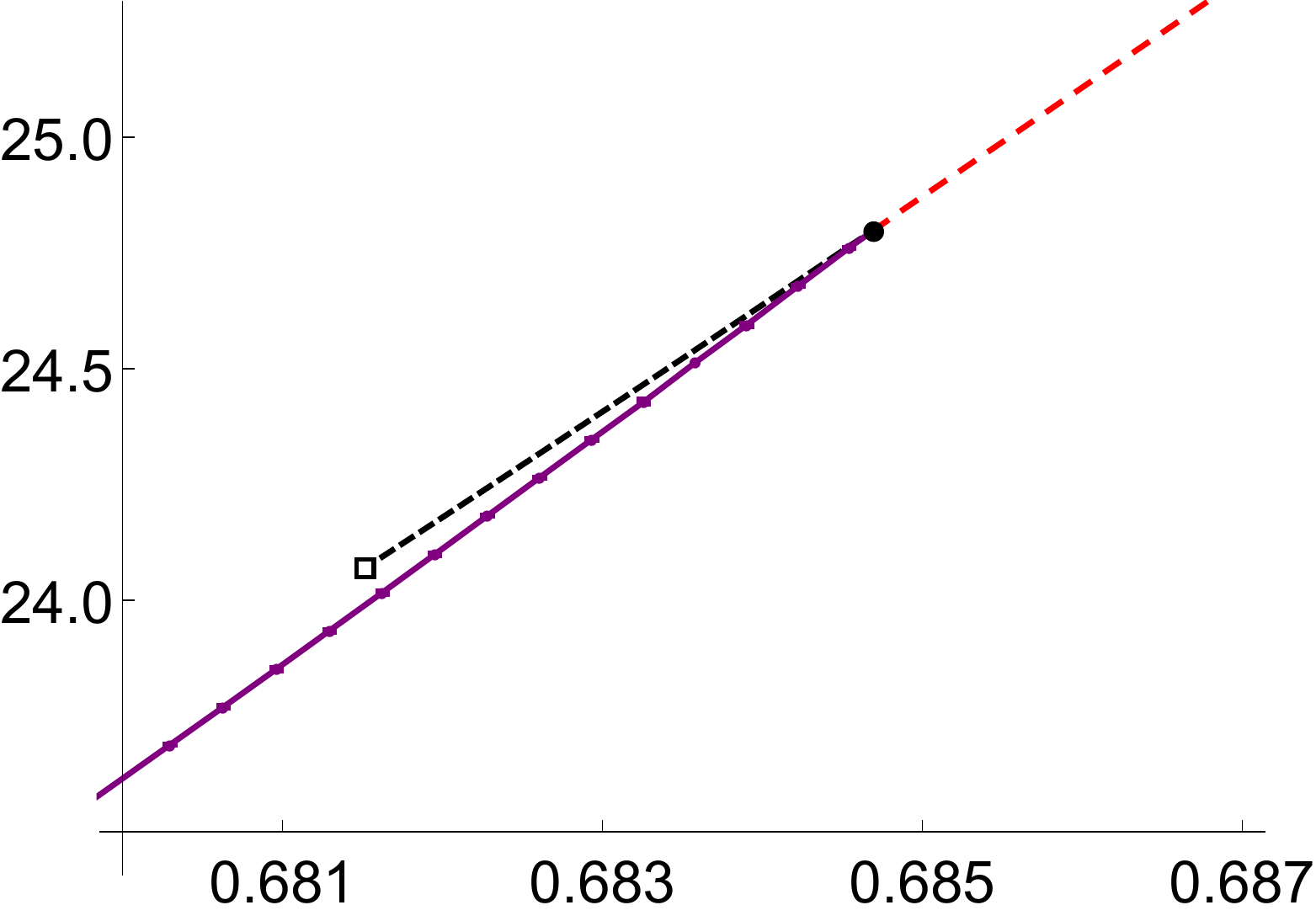} 
			\put(-180,130){\rotatebox{0}{$\overline T_c $}}
			\put(-5,20){$b_0$}
		\end{subfigure}
		\caption{\small Phase diagram for the entire family of gauge theories (left) and zoomed in version of the circle around $b_0\approx 0.68$. The dashed, red line represents  first-order transitions between a black brane and a regular horizonless solution with a discontinuity in the entropy density (Case A). The dashed, black line corresponds to phase transitions between two black brane solutions (Case B). The dotted, purple line indicates first-order transitions between the zero-entropy limit of a black brane branch and a regular horizonless solution. The entropy density is continuous across these transitions (Cases B and C). The purple curve requires an extrapolation to zero entropy. We have estimated the associated error bars, shown explicitly in the figure, by comparing a linear and a quadratic extrapolation.}
		\label{fig:PhaseDiagram}
	\end{center}
\end{figure}
With all this information we can draw the phase diagram of the entire family in the $\left(b_0,T\right)$-plane, which is shown in Fig.~\ref{fig:PhaseDiagram}. The region above the curves of critical temperatures in the left panel is dominated by black branes and corresponds to  ungapped phases in the dual gauge theories. Below these curves, the regular horizonless solutions are dominant and the preferred phases are gapped. The right panel zooms into the range of values that include Case B, where two transitions take place.  The triple point, where the three lines meet, is indicated with a black dot. The critical point, where the line of first-order phase transitions between black branes ends, is indicated with a square.

\begin{figure}[t]
	\begin{center}

	\end{center}
\end{figure}

\subsection{Quasi-conformal thermodynamics}

As described in Section~\ref{LowT}, there are two special values of the parameter with distinct IR dynamics. When $b_0=0$ the ground state RG flow, denoted by $\B_8^\infty$, ends on a fixed point, as indicated by the leftmost arrow in Fig.~\ref{fig:triangle}. The properties of this particular theory at finite temperature were already discussed around Eq.~\eqref{FreeECFT}. In this section we comment on the imprints that this fixed point leaves on the thermodynamics of flows passing nearby. 

\begin{figure}[t]
	\begin{center} 
		\includegraphics[width=.80\textwidth]{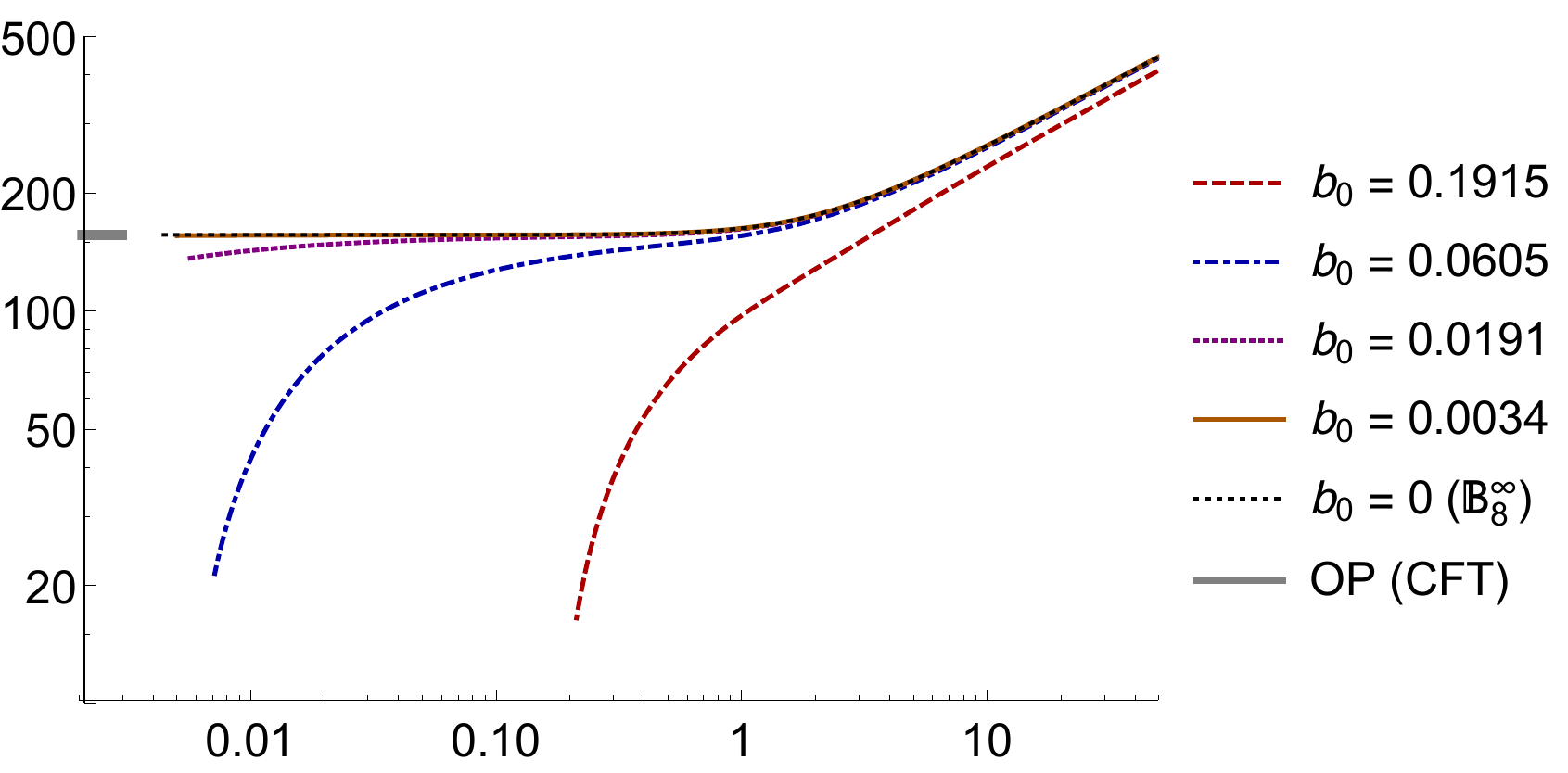} 
		\put(-310,160){$\overline S\  \overline{T}^{-2}$}
		\put(-70,10){$\overline T$}
		\caption{\small Dimensionless combination $S\,T^{-2}$ for different flows with $b_0\gtrsim0$. This quantity is constant in a CFT at finite temperature, as in the IR of $\mathbb{B}_8^{\infty}$.}
		\label{fig:quasi_conformal}
	\end{center}
\end{figure}

Conformal invariance dictates that when a three-dimensional CFT is heated up, the entropy grows as $S\propto T^2$. Flows with $b_0\gtrsim0$ approach the fixed point but eventually fail to reach it, so it is expected that the entropy should scale approximately as in a CFT for some range of temperatures. This behaviour is confirmed by the results plotted in Fig.~\ref{fig:quasi_conformal}. For $\B_8^\infty$, the dimensionless quantity $S T^{-2}$ becomes nearly independent of $T$ for all temperatures below Eq.~\eqref{Tconformal}. For black brane solutions with a small but non-zero value of $b_0$, a plateau develops before their zero-entropy limit is eventually reached at a finite temperature. The size and the steepness of the plateau, in which there is quasi-conformal behaviour, is controlled by the smallness of $b_0$. 

\subsection{Quasi-confining thermodynamics}

\begin{figure}[t]
	\begin{center}
		\begin{subfigure}{.47\textwidth}
			\includegraphics[width=\textwidth]{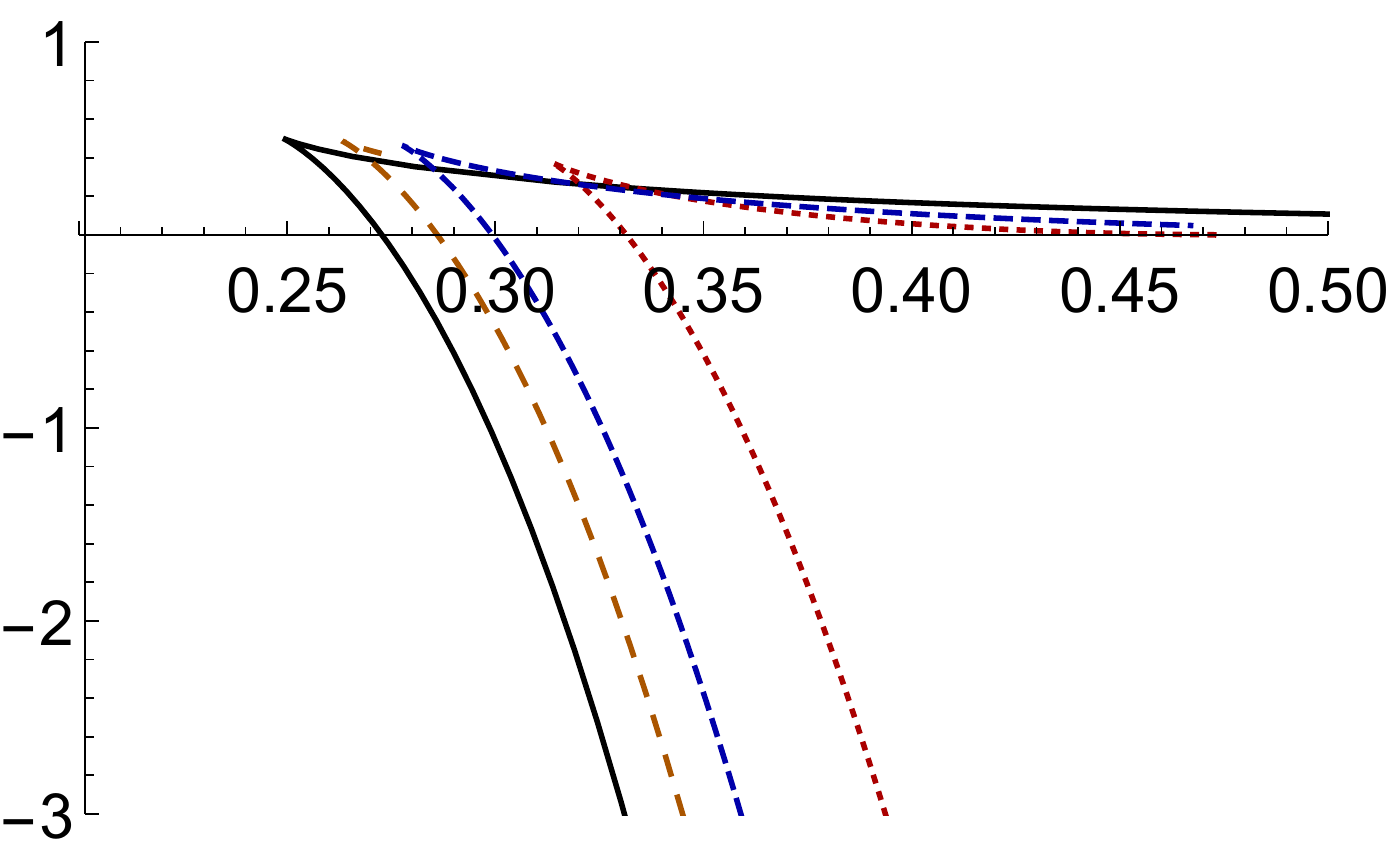} 
			\put(-180,120){$\overline F$}
			\put(-10,100){$\overline T$}
		\end{subfigure}\hfill
		\begin{subfigure}{.47\textwidth}
			\includegraphics[width=\textwidth]{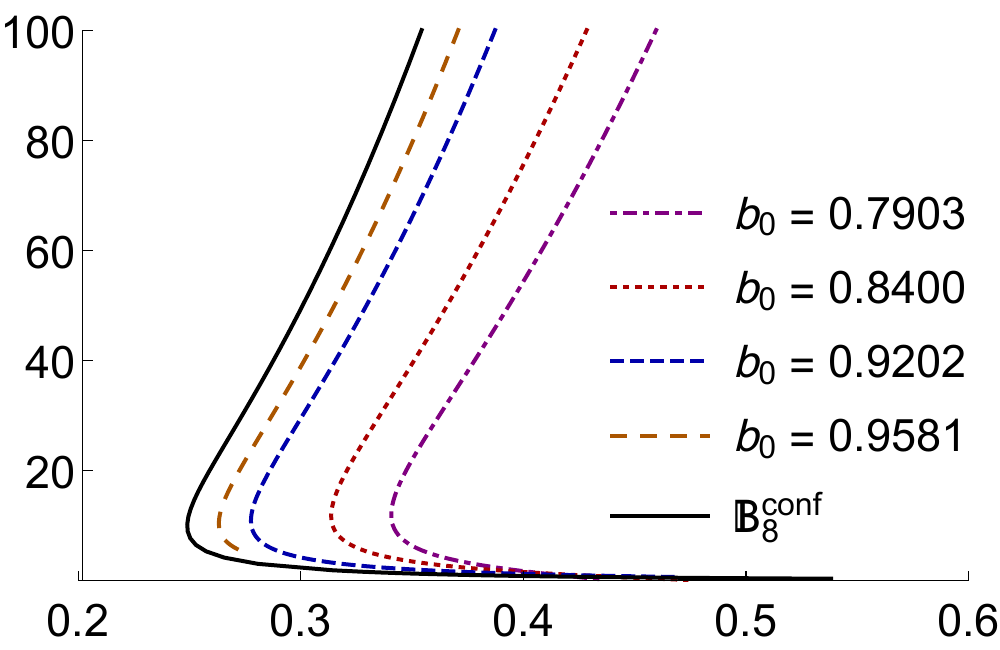} 
			\put(-180,120){$\overline S$}
			\put(0,10){$\overline T$}
		\end{subfigure}
		\caption{\small Free energy (left) and entropy (right) for different $\B_8$ metrics with $b_0 \lesssim 1$, with the appropriate scaling of the charges, in dashed and dotted lines. In the limit $b_0\to1$, we recover those of $\Bconf$, shown in solid black line.}
		\label{fig:QuasiConfining}
	\end{center}
\end{figure}

For $b_0=1$ the solution is $\Bconf$, at the right of Fig.~\ref{fig:triangle}. This theory is not only gapped but truly confining, in the sense that the quark-antiquark potential displays an area law (see also the arguments in \cite{Herzog}). In \cite{Faedo:2017fbv} this particularity was attributed to the vanishing of the CS level, $k=0$. Due to this, if we aim to understand the solution as a limit of the $\B_8$ family of metrics, we have to rescale the charge $Q_k$ --- related to the CS level by Eq.~\eqref{gaugeparam} --- before taking $k\to0$ (see \cite{Faedo:2017fbv,Elander:2018gte} for further details). Moreover, the radial coordinate in Eq.~\eqref{ucoord} is no longer valid, so we define a new one through
\begin{equation}
\dd r=-\frac{\rho_0}{\xi ^2 \sqrt{1-\xi ^4}}\ \dd\xi\,, \qquad\qquad \xi \in (0,1)\,,
\end{equation}
with $\rho_0$ a constant with dimensions of length. We factor out the charges from the fluxes, Eq.~\eqref{fluxesansatz}, to obtain dimensionless functions as
\begin{equation}
b_J\,=\,\frac{Q_c}{4 q_c}+\frac{2q_c}{3 \rho _0}\  \mathcal{B}_J\,,\quad b_X\,=\,-\frac{Q_c}{4 q_c}-\frac{2q_c}{3 \rho _0}\  \mathcal{B}_X\,,\quad a_J\,=\, -\frac{q_c}{2}-q_c \ \mathcal{A}_J\,,\quad h = \frac{128 q_c^2 }{9 \rho _0^6} \ \mathbf{h}\,.
\end{equation}
The free parameters in the expansions both at the UV and at the horizon are essentially the same as in the rest of the $\mathbb{B}_8$ family, since $\xi$ and $u$ coincide asymptotically. However, now we need to gauge fix $f_1=0$ to compare with the ground state solution. After the appropriate rescaling in order to compare the same dimensionless quantities, it can be seen that the free energies and entropies for different values of $b_0$ approach those of $\Bconf$ as $b_0\to1$. This is shown in Fig.~\ref{fig:QuasiConfining}. In the case of $\Bconf$, as we remove the horizon, that is, in the limit $S \to 0$, the temperature diverges, as in the case of small black branes in global AdS spacetimes. This branch is however unstable: before reaching it, there is a first-order phase transition at a critical temperature 
$\overline{T}_c\approx 0.27$. In the dual gauge theory this corresponds to a genuine confinement/deconfinement first-order phase transition.

\subsection{Regime of validity}
\label{sec:validity}

In this section we determine the range of validity of the supergravity solutions we have studied. For the ground state, as well as for the low temperature solutions considered in Section~\ref{LowT}, there are two requisites, already discussed in \cite{Faedo:2017fbv}. In this reference it was shown that the type IIA picture is trustable for energies in the gauge theory 
\begin{equation}\label{val1}
U \ll \lambda \left(1+\frac{\overline{M}^2}{2 N|k| }\right)(1-b_0^2)\,,
\end{equation}
above which the perturbative regime is adequate.\footnote{Recall that the three-dimensional gauge theories are superrenormalizable and asymptotically free.} On the other hand, the IR is regular in eleven dimensions. In order for the curvature to be small it is required that
\begin{equation}\label{val2}
\frac{\overline{M}^2}{2}+N|k| \gg 1\,.
\end{equation}
On top of this, we have seen for instance in Fig.~\ref{fig:Ricci} that the curvature at the horizon grows unbounded as we remove it. This is important for phase transitions in Cases B and C, since they occur precisely in this limit. As we approach the critical temperature, the eleven-dimensional Ricci scalar in Planck units grows as
\begin{equation}
\ell_p^2 R\sim \left(\frac{\overline{M}^2}{2} +N|k|\right)^{-1/3}(\overline T-\overline T_c)^{-2/3}\,.
\end{equation}
Therefore, there is always an interval of temperatures
\begin{equation}
\overline T\in \left(\overline T_c,\overline T_c+\left(\frac{\overline{M}^2}{2} +N|k|\right)^{-1/2}\right)\,,
\end{equation}
where curvatures are large and higher-order curvature corrections must be considered. Nevertheless, this region is small provided Eq.~\eqref{val2} is satisfied.

\section{Conclusions and discussion}
\label{sec:conclusions}

In this work we have studied the finite-temperature physics of a family of gauge theories in three dimensions by means of holography. The UV, which is weakly coupled, is expected to be described by a two-sites quiver Yang--Mills theory deformed by CS-terms, as discussed around Eq.~\eqref{quiver}. The parameter $b_0$, labelling the different solutions within the family, controls the asymptotic difference between the microscopic couplings of each of the two factors in the quiver. The IR of the ground state is generically gapped except for a particular value of the parameter that drives the RG flow to a fixed point. Despite the gap, the models have no linear quark-antiquark potential at large separations unless the CS level is zero \cite{Faedo:2017fbv}.  

The gap is lost at some non-zero critical temperature, as we have shown by constructing the black brane geometries dual to the ungapped phase and demonstrating their dominance above $T_c$. The nature of the phase transition depends on the value of $b_0$ and there are three distinct regions. For large values of the parameter there is a single degapping first-order transition. At intermediate values, there is a small range in which two transitions take place: as the temperature is decreased from high values, there is a first-order transition between two different black branes, corresponding to two ungapped states, while at a lower critical temperature there is a degapping first-order transition at which the entropy is continuous. Finally, at even lower values of the parameter there is only a phase transition of this second type. This is summarised in the phase diagram of Fig.~\ref{fig:PhaseDiagram}, where a triple point and a critical point are indicated by a black dot and by a square, respectively.

As noted below Eq.~\eqref{below}, the size of the intermediate region where two phase transitions take place is very small compared to the range of the $b_0$ parameter. In other words, \mbox{$b_0^\textrm{triple}-b_0^\textrm{critical} =0.0032 \ll 1$}. Nevertheless, we have verified that the existence of this intermediate region is not a numerical artifact by integrating the equations with two different integrators and by varying the control parameters in each of them. It would be interesting to understand in detail how such a small region arises dynamically in our model. A dynamically-generated small parameter  was also encountered in \cite{Basu:2011yg} in the phase diagram of a holographic color superconductor. 

The black branes that we have found reach a finite temperature in the limit in which the horizon is removed and the entropy vanishes. As a result, the transition for low values of $b_0$ seems to take place at zero entropy but finite temperature. However, the geometries recovered in this limit, which were identified in Section~\ref{sec:zeroEntropyLimit}, are singular. This means that,  for temperatures slightly above the critical one,  the curvature is large and the supergravity approximation is unreliable in a small region near the horizon, as detailed in Section~\ref{sec:validity}. The singularity is a good one according to the classification of \cite{Gubser:2000nd}, since by construction it can be cloaked behind a horizon. It would be interesting to discern how string theory resolves the singularity and how the phase transition manifests itself in that picture.  

An analogous phase transition was observed in \cite{Dias}, where black branes in a similar system were obtained. In that case the ground state is conformal in the UV and preserves $\mathcal{N}=2$, being a higher-supersymmetric version of the RG flow denoted $\mathbb{B}_8^{\textrm{\tiny OP}}$ at the left of Fig.~\ref{fig:triangle}. Our results can also be compared to those in \cite{Aharony,Buchel}, where the Klebanov--Tseytlin and Klebanov--Strassler theories at finite temperature were studied. The phase transition shown to happen in those models is comparable to the one encountered here for large values of $b_0$. Indeed, when the CS-level vanishes, which happens for the flow called $\mathbb{B}_8^{\rm{conf}}$ at the right of Fig.~\ref{fig:triangle},\footnote{A black brane solution in this background, perturbative in the number of fractional branes, was constructed in \cite{Giecold}.} the first-order phase transition is genuinely a confinement/deconfinement transition, as in \cite{Aharony,Buchel}. Yet, the UV of our models, which are asymptotically free, is simpler than the infinite cascade of the four-dimensional relatives.  

Both \cite{Dias} and \cite{Buchel} discuss the possibility of hiding the singularity caused by including antibranes in the ground state geometries behind a horizon. This would signal that the singularity is physical and give support to KKLT-type constructions of de Sitter vacua \cite{Kachru}. As argued in \cite{Bena}, for this to be realised, the charge measured at the horizon of the black branes must have the opposite sign to the asymptotic one. We have computed the relevant Maxwell charges and verified that they do not change sign along the RG flow up to the horizon.

\vspace{1.0cm}
\begin{acknowledgments}
We thank David Pravos for collaboration in an early stage of the project. We were supported by grants FPA2016-76005-C2-1-P, FPA2016-76005-C2-2-P, SGR-2017-754, MDM-2014-0369 of ICCUB, and ERC Starting Grant HoloLHC-306605. JGS acknowledges support from the FPU program, fellowship FPU15/02551. DE was supported by the OCEVU Labex (ANR-11-LABX-0060) and the A*MIDEX project (ANR-11-IDEX-0001-02) funded by the ``Investissements d'Aveni'' French government program managed by the ANR. AF is supported by the ``Beatriz Galindo'' program, as well as grant PGC2018-096894-B-100. 
\end{acknowledgments}

\appendix

\section{Ansatz and equations of motion}

In this appendix we give the details needed to reproduce the equations of motion that we solve in the bulk of the paper, both from the ten- and four-dimensional points of view.

\subsection{Ten-dimensional ansatz}
\label{sec:10Dansatz}

Although the ground state solutions are regular only in eleven dimensions, we will give the details of the ten-dimensional setup, since the geometric picture is clearer. In string frame, the equations of motion for the forms of type IIA supergravity read
\begin{eqnarray}\label{formeqs}
\dd *F_4 + H\wedge F_4 &=&0\,,	\nonumber\\[2mm]
\dd  *F_2 + H\wedge* F_4 &=&0\,,\\[2mm]
\dd\left(e^{-2\Phi}  * H\right)-F_2\wedge *F_4 -\frac{1}{2} F_4\wedge F_4 &=&0\,.\nonumber
\end{eqnarray}
The field strengths are such that the following Bianchi identities must be satisfied
\begin{equation}
\dd H=0\,, \qquad\qquad \dd F_2 =0\,,  \qquad\qquad \dd F_4 = H\wedge F_2\,.
\end{equation}
On top of that, we have the equations governing  the dilaton
\begin{eqnarray}
R+4\nabla_M\nabla^M \Phi - 4 \nabla^M\Phi \nabla_M \Phi-\frac{1}{12} H^2 = 0\,,
\end{eqnarray}
and the metric
\begin{equation}
R_{MN} + 2 \nabla_M\nabla_N \Phi -\frac{1}{4} H_{MN}^2 = e^{2\Phi}\left[
\frac{1}{2} (F_2^2)_{MN} + \frac{1}{12}(F_4^2)_{MN} - \frac{1}{4}g_{MN}\left(
\frac{1}{2}F_2^2 +\frac{1}{24}F_4^2\right)\right]\,,
\end{equation}
in a self-explanatory notation. 

All the solutions considered are based on the $\mathbb{CP}^3$ as internal manifold and asymptote to a stack of coincident D2-branes in the UV. We then take as ansatz for the metric and dilaton
\begin{eqnarray}
\dd s_{\rm st}^2 &=&h^{-\frac12}\left(-\mathsf{b}\ \dd t^2 + \dd x_1^2 + \dd x_ 2^2\right)+h^{\frac12} \left(\frac{\dd r^2}{\mathsf{b}}+e^{2f}\dd\Omega_4^2+e^{2g}\left[\left(E^1\right)^2+\left(E^2\right)^2\right] \right)\,,\nonumber\\[2mm]
e^\Phi&=&h^{\frac14} \, e^\Lambda \,,
\end{eqnarray}
where the functions $f$, $g$, $h$, $\mathsf{b}$ and $\Lambda$ all depend on the radial coordinate. In this picture the complex projective plane is seen as the coset ${\rm Sp}(2)/{\rm U}(2)$. This is a ${\rm S}^2$, described by the vielbeins $E^1$ and $E^2$, fibered over ${\rm S}^4$, with metric $\dd\Omega_4^2$. A suitable choice of coordinates is the following. Let $\omega^i$ be a set of left invariant forms on the three-sphere. Then the metric of the unit-radius four-sphere can be written as
\begin{equation}
\dd\Omega_4^2\,=\,\frac{4}{\left(1+\xi^2\right)^2}\left[\dd \xi^2+\frac{\xi^2}{4}\omega^i\omega^i\right]\,,
\end{equation}
with $\xi$ a non-compact coordinate. If we parametrise the two-sphere by the angles $\theta$ and $\varphi$, then the non-trivial fibration is described by the vielbeins 
\begin{eqnarray}
\label{introduced}
E^1&=&\dd \theta+\frac{\xi^2}{1+\xi^2}\left(\sin\varphi\,\omega^1-\cos\varphi\,\omega^2\right)\,,\nonumber\\[2mm]
E^2&=&\sin\theta\left(\dd\varphi-\frac{\xi^2}{1+\xi^2}\omega^3\right)+\frac{\xi^2}{1+\xi^2}\cos\theta\left(\cos\varphi\,\omega^1+\sin\varphi\,\omega^2\right)\,.
\end{eqnarray}
For our purposes, it is convenient to consider a rotated version of the vielbeins on the four-sphere that read:
\begin{eqnarray}
\mathcal{S}^1&=&\frac{\xi}{1+\xi^2}\left[\sin\varphi\,\omega^1-\cos\varphi\,\omega^2\right]\,,\nonumber\\
\mathcal{S}^2&=&\frac{\xi}{1+\xi^2}\left[\sin\theta\,\omega^3-\cos\theta\left(\cos\varphi\,\omega^1+\sin\varphi\,\omega^2\right)\right]\,,\nonumber\\
\mathcal{S}^3&=&\frac{\xi}{1+\xi^2}\left[\cos\theta\,\omega^3+\sin\theta\left(\cos\varphi\,\omega^1+\sin\varphi\,\omega^2\right)\right]\,,\nonumber\\
\mathcal{S}^4&=&\frac{2}{1+\xi^2}\,\dd\xi\,.
\end{eqnarray}
Despite their explicit dependence on the angles of the two-sphere, it can be easily checked that $\mathcal{S}^n\mathcal{S}^n=\dd \Omega_4^2$. Using these vielbeins we can construct a set of left-invariant forms on the coset. This set contains the two-forms
\begin{equation}
X_2\,=\,E^1\wedge E^2\,,\qquad\qquad\qquad J_2\,=\,\mathcal{S}^1\wedge\mathcal{S}^2+\mathcal{S}^3\wedge\mathcal{S}^4\,,
\end{equation}
as well as the three-forms 
\begin{eqnarray}
X_3&=&E^1\wedge\left(\mathcal{S}^1\wedge\mathcal{S}^3-\mathcal{S}^2\wedge\mathcal{S}^4\right)-E^2\wedge\left(\mathcal{S}^1\wedge\mathcal{S}^4+\mathcal{S}^2\wedge\mathcal{S}^3\right)\,,
\nonumber\\[2mm]
J_3&=&-E^1\wedge\left(\mathcal{S}^1\wedge\mathcal{S}^4+\mathcal{S}^2\wedge\mathcal{S}^3\right)-E^2\wedge\left(\mathcal{S}^1\wedge\mathcal{S}^3-\mathcal{S}^2\wedge\mathcal{S}^4\right)\,.
\end{eqnarray}
These are related by exterior differentiation as  
\begin{equation}
\dd X_2\,=\,\dd J_2\,=\,X_3\,,\qquad\qquad\qquad \dd J_3\,=\,2\left(X_2\wedge J_2+J_2\wedge J_2\right)\,.
\end{equation}
Higher forms constructed by wedging of these are also left invariant, so we have the two four-forms $X_2\wedge J_2$ and $J_2\wedge J_2$ appearing in the equation above together with the volume form $\Omega_6 = - (E_1 \wedge E_2)\wedge (\mathcal{S}^1\wedge\mathcal{S}^2\wedge\mathcal{S}^3\wedge\mathcal{S}^4)$. There are no adequate one- or five-forms. We will write the fluxes in terms of these left-invariant forms, since this symmetry ensures the consistency of the ansatz, meaning that all the internal angles will drop from the resulting equations, which will depend just on the radial coordinate. Thus, we take the following forms 
\begin{equation}\label{fluxesansatz}
\begin{array}{rclcrcl}
F_4 &=&  \mathbf{f}_4 *\Omega_6 + G_4 + B_2\wedge F_2 \,,&\qquad\qquad\qquad& F_2& =& Q_k (X_2 - J_2)\,,\\[2mm]
 G_4 &=& \dd(a_J J_3) + q_c\left(J_2\wedge J_2 - X_2\wedge J_2\right)\,,&\qquad \qquad\qquad& B_2& =& b_X X_2 + b_J J_2\,,
 \end{array}
\end{equation}
where we have defined the quantity 
\begin{equation}
\mathbf{f}_4= Q_k b_J^2 +2(q_c+2a_J)b_X - 2 b_J(q_c-2a_J+Q_kb_X) +Q_c\,.
\end{equation}
The parameters $Q_c$, $q_c$ and $Q_k$ are constants, related to gauge theory parameters as stated around Eq.~\eqref{gaugeparam}. On the other hand, $b_J$, $b_X$ and $a_J$ depend on the radial coordinate, their dynamics dictated by the form equations \eqref{formeqs}.

The ground state is supersymmetric and has $\mathsf{b}=1$. The explicit form of the rest of the functions is not important for our purposes and can be found in \cite{Faedo:2017fbv}. Finally, in eleven dimensions the backgrounds correspond to M2-branes 
\begin{equation}
\label{M2metric}
\dd s_{11}^2\,=\,H^{-2/3}\,\dd x_{1,2}^2+H^{1/3}\,\dd s_8^2\,,
\end{equation}
with eight-dimensional transverse space
\begin{equation}\label{11dtrans}
\dd s_8^2\,=\,e^{-\Lambda}\,\Bigg[ \dd r^2+e^{2f}\,\dd\Omega_4^2+e^{2g}\,\left[\left(E^1\right)^2+\left(E^2\right)^2\right] \Bigg]+e^\Lambda\,Q_k^2\,\left(\dd\psi-\cos\theta\,\dd \varphi+\xi\,\mathcal{S}^3\right)^2\,.
\end{equation}

\subsection{Four-dimensional reduction}
\label{sec:4Deffectivetheory}

To study aspects like spectra \cite{Elander:2018gte} or thermodynamic properties it is convenient to work with a four-dimensional consistent truncation. This truncation was presented in \cite{Faedo:2017fbv}. In order to reduce from ten to four dimensions, we take the following metric
\begin{equation}\label{different_ansatz}
\dd s_{\rm st}^2\,=\,e^{\Phi/2}\left(e^{-2U-4V}\dd s_4^2+e^{2U}\,\frac14\left[\left(E^1\right)^2+\left(E^2\right)^2\right]+e^{2V}\,\frac12\dd\Omega_4^2\right)\,.
\end{equation}
The forms read exactly as in Eq.~\eqref{fluxesansatz}. Assuming that all the functions depend just on the coordinates in $\dd s_4^2$, the resulting equations of motion can be obtained from the action 
\begin{eqnarray}\label{reduced_Action}
S_4&=&\frac{1}{2\kappa_4^2}\,\int\,\left[R*1+\LL_{\mbox{\footnotesize kin}}-\mathcal{V}*1\right]\nonumber\\[2mm]
&=&\frac{1}{2\kappa_4^2}\,\int\,\left[R*1-\frac12\left(\dd\Phi\right)^2-4\left(\dd U\right)^2-12\left(\dd V\right)^2-8\dd U\cdot\dd V-4e^{-4V-\Phi}\left(\dd b_J\right)^2\right.\nonumber\\[2mm]
&&\qquad\qquad-8e^{-4U-\Phi}\left(\dd b_X\right)^2-32e^{-2U-4V+\Phi/2}\left(\dd a_J\right)^2-\mathcal{V}*1\bigg]\,,
\end{eqnarray}
with the potential
\begin{eqnarray}\label{potential}
\mathcal{V}&=&128\,e^{- 6 U -12 V-\Phi/2} \left[Q_c + 4 a_J \left(b_J + b_X\right) +Q_kb_J\left(b_J-2b_X\right)+2 q_c \left(b_X - b_J\right)\right]^2 \nonumber\\[2mm]
&+&  32 \left(b_J + b_X\right)^2 e^{-4 U - 8 V - \Phi} + 64 \left[2 a_J +Q_k\left(b_J-b_X\right) -  q_c\right]^2 e^{-6 U - 8 V + \Phi/2} \nonumber\\[2mm]
&+& 32 \left(2 a_J -Q_kb_J + q_c\right)^2 e^{-2 U - 12 V + \Phi/2} +4Q_k^2 e^{-2 U - 8 V + 3\Phi/2 } \nonumber\\[2mm]
&+&8Q_k^2 e^{-6 U - 4 V + 3\Phi/2}- 24 e^{-2 U - 6 V} -  8 e^{-4 U - 4 V} + 2 e^{-8 V} \,. 
\end{eqnarray}
Thus, the action contains six scalars: $U$ and $V$ from the metric, $b_X$, $b_J$ and $a_J$ from the fluxes (RR and NS forms) plus the dilaton $\Phi$. By comparing Eq.~\eqref{different_ansatz} with Eq.~\eqref{D2brane} it is possible to translate into the variables used in the bulk of the paper. Additionally, we have three constants, $Q_c$, $Q_k$ and $q_c$, whose relation with gauge-theory parameters was reported in Eq.~\eqref{gaugeparam}. 
The potential can be recovered from the superpotential  
\begin{eqnarray}\label{superpotential}
\mathcal{W}&=&e^{-4 V} + 2 e^{-2U -2 V} + Q_k\,e^{-3 U - 2 V + 3\Phi/4} - Q_k\,e^{-U - 4 V + 3\Phi/4} \\[2mm]
&-& 4 e^{-3 U - 6 V - \Phi/4} \left[Q_c + 4 a_J \left( b_J + b_X\right) +Q_k b_J \left(b_J-2b_X\right)+2q_c \left(b_X - b_J\right) \right]\nonumber
\end{eqnarray}
through the usual relation
\begin{equation}
\mathcal{V}\,=\,4\,G^{ij}\partial_i\mathcal{W}\partial_j\mathcal{W}-6\,\mathcal{W}^2\,.
\end{equation}
We now perform standard holographic renormalization on this action. As a first step, we cut off the radial coordinate at some $\mu_{\mbox{\tiny UV}}$. The action diverges in the limit $\mu_{\mbox{\tiny UV}}\to \infty$, so we have to find the counterterms $S_{\mbox{\footnotesize ct}}$ that regularize it. If these are correctly chosen, adding them to the action and removing the regulator gives a finite quantity. Additionally, we have to include the Gibbons--Hawking term \cite{Emparan:1999}. The regularized action is thus
\begin{eqnarray}\label{Sren}
S_{\mbox{\footnotesize reg}}=\frac{1}{2\kappa_4^2}\,\int_{\MM}\,\Big[R*1+\LL_{\mbox{\footnotesize kin}}-\mathcal{V}*1\Big] +S_{\mbox{\tiny GH}} + S_{\mbox{\footnotesize ct}},
\end{eqnarray}
where $\MM$ is the whole four dimensional spacetime, $S_{\mbox{\footnotesize ct}}$ is the counterterm piece and $S_{\mbox{\tiny GH}}$ is the Gibbons--Hawking term
\begin{eqnarray}
S_{\mbox{\tiny GH}}&=&\frac{1}{\kappa_4^2}\,\int_{\partial \MM}\,K*1,
\end{eqnarray}
being $K$ the extrinsic curvature induced on the boundary $\partial \MM$.

Introducing a horizon in the gravitational description and thus considering the gauge theory at finite temperature is an infrared deformation. On the other hand, it is known that the superpotential renormalizes the supersymmetric solution. If the black brane enjoys the same UV asymptotic behaviour as the ground state, the UV divergences will be cancelled out in the same way. Therefore, the counterterm we consider is
\begin{eqnarray}
S_{\mbox{\footnotesize ct}}&=&-\frac{2}{\kappa_4^2}\,\int_{\partial \MM}\, \mathcal{W}*1,
\end{eqnarray}
where $\mathcal{W}$ is the superpotential, Eq.~\eqref{superpotential}. The renormalized on-shell action used to extract the thermodynamic quantities is obtained through the usual manipulations as 
\begin{eqnarray}\label{Full_Sren}
S_{\mbox{\footnotesize ren}}&=&\lim_{\mu_{\mbox{\tiny UV}}\to\infty}S_{\mbox{\footnotesize reg}}\,=\, - \frac{\beta V_2}{2\kappa_4^2}\,\lim_{\mu_{\mbox{\tiny UV}}\to\infty}\left[\sqrt{\gamma}\ (2{K^t}_t - 2K + 4\mathcal{W})\big|_{\mu_{\mbox{\tiny UV}}} -2\sqrt{\gamma} {K^t}_t\big|_{H}\right]\nonumber\\
\end{eqnarray}
where the last term is to be evaluated at the horizon. Here $\gamma$ is the determinant of the boundary metric,  $\beta$ is the period of the Euclidean time and $V_{2}$ is the volume of $\mathbb{R}^2$. From this renormalized action we can compute the energy-momentum tensor of the dual gauge theory by varying with respect to the induced metric, evaluated at the boundary
\begin{equation}\label{EMtensor}
{T^i}_j = - \frac{1}{ \kappa_4^2} \lim_{\mu_{\mbox{\tiny UV}}\to\infty}\left\{\sqrt{\gamma}\ ({K^i}_j - {\delta^i}_j (K-2\mathcal{W}))\big|_{\mu_{\mbox{\tiny UV}}}\right\} = \mbox{diag}(-E,P,P)\,.
\end{equation}

There is a final observation which is useful in the numerical computations. Given a four-dimensional metric of the form
\begin{eqnarray}\label{domain_wall_ansatz}
	\dd s_4^2&=&-g_{tt}\ \dd t^2+g_{xx}\ \dd x_1^2 +g_{xx}\ \dd x_2^2+g_{rr}\ \dd r^2\,,
\end{eqnarray}
it can be seen that the action is invariant under
\begin{equation}
g_{tt}\mapsto \mu^2 g_{tt}\qquad \qquad g_{xx}\mapsto \mu^{-1} g_{xx}\,.
\end{equation}
Using Noether's theorem there must be a conserved current associated to this symmetry, which is
\begin{equation}
j^\mu =\frac{g_{tt}g_{xx}'-g_{tt}'g_{xx}}{\sqrt{g_{tt}g_{rr}}}\,\delta^\mu_r\,,
\end{equation}
satisfying
\begin{equation}\label{conserved_current}
\partial_\mu j^\mu = \partial_r \left(\frac{g_{tt}g_{xx}'-g_{tt}'g_{xx}}{\sqrt{g_{tt}g_{rr}}}\right) = 0,\qquad \Rightarrow\qquad \frac{g_{tt}g_{xx}'-g_{tt}'g_{xx}}{\sqrt{g_{tt}g_{rr}}}=\mbox{ constant}\,.
\end{equation}
This equation relates a combination of UV parameters with horizon data as in Eq.~\eqref{b5_conserved}. This can be employed either as a check of the numerical coefficients obtained while solving the equations or, using it as an input, to reduce the number of unknown constants in the shooting problem.  

\section{UV expansions}\label{ap:UVexp}

In this appendix we present some details of the UV expansions. The undetermined parameters are those in Eq.~\eqref{UVexpansions}, while the remaining ones, shown here, are given in terms of them. In the following expressions we have already fixed $f_1=-1$ as explained in the bulk of the paper. Although we only show a few terms, we indicate in the sums the order up to which we solved the equations --- for example, up to order $u^{22}$ in the case of $e^\FF$. 

For the metric function $e^\FF$ we expanded around $u=0 $ as
$$e^\FF = \frac{1}{u\sqrt{2}} \left[1+\sum_{n=1}^{22}  f_n \ u^n + \OO(u^{23})  \right]\,.$$
The first few coefficients read
\begin{eqnarray}
f_2&=&-\frac{5}{2}\,,\qquad\qquad\qquad f_3\,=\,-\frac{13}{2}\,,\nonumber\\ [2mm]
f_6&=&\frac{1}{120 \left(b_0^2-1\right)}\Big[3 b_0^2 \left(52 \mathsf{b}_5-12436 f_4+224 f_5-33671\right)-96 \left(196 b_4-5 b_6\right) b_0+ \nonumber\\ &&\quad  +4 \mathsf{b}_5-3 \left(364 f_4+224 f_5+11705\right)\Big]\,,\nonumber\\ [2mm]
f_7 &=& \frac{1}{2520 \left(b_0^2-1\right)} 
\Big[4 \left(5321 b_0^2-141\right) \mathsf{b}_5- 21 \Big(b_0^2 \left(229456 f_4-3308 f_5+599069\right)+\nonumber\\ &&\quad  +592 \left(196 b_4-5 b_6\right) b_0+7344 f_4+3308 f_5+240387\Big)\Big]\,.
\end{eqnarray}
Similarly, the function $e^\GG$ enjoys an expansion
$$e^\GG = \frac{1}{{2u}} \left[1+\sum_{n=1}^{22}  g_n \ u^n + \OO(u^{23})  \right]\,,$$
with the first coefficients being
\begin{eqnarray}
g_1&=&-2\,,\qquad g_2\,=\,-4\,,\qquad g_3\,=\,-8\,,\qquad g_4 \,=\, -2 f_4-\frac{205}{4}\,,\qquad g_5 \,=\,3 f_4+f_5+\frac{283}{4}\,,\nonumber\\ [2mm]
g_6&=& -\frac{b_0^2 \left(-36 \mathsf{b}_5+8160 f_4-72 f_5+22917\right)+21 \left(196 b_4-5 b_6\right) b_0+\mathsf{b}_5+240 f_4+72 f_5+6861}{15 \left(b_0^2-1\right)}\,,\nonumber\\ [2mm]
g_7 &=& \frac{1}{315 \left(b_0^2-1\right)}
\Big[\left(4624 b_0^2-354\right) \mathsf{b}_5 - \nonumber\\ 
&&\quad -21 \left(b_0^2 \left(47780 f_4-308 f_5+141577\right)+122 \left(196 b_4-5 b_6\right) b_0+1020 f_4+308 f_5+31419\right)\Big]\,.\nonumber\\ [2mm]
\end{eqnarray}

Analogously, for the function $e^\Lambda$ we have
$$e^\Lambda  = 1+\sum_{n=1}^{22}  \lambda_n \ u^n + \OO(u^{23}) \,, $$
where
\begin{eqnarray}
\lambda_1 &=& 0\,,\qquad \lambda_2\, =\, -4\,,\qquad \lambda_3 \,=\, -16\,,\qquad \lambda_4 \,=\, -48\,,\qquad \lambda_5 \,=\, \frac{6 f_4}{5}+2 f_5-\frac{71}{10}\,,\nonumber\\ [2mm]
\lambda_6 &=& \frac{2 \left(b_0^2 \left(5 \mathsf{b}_5-1164 f_4+20 f_5-3102\right)+\left(15 b_6-588 b_4\right) b_0-4 \left(9 f_4+5 f_5+288\right)\right)}{3 \left(b_0^2-1\right)}\,,\nonumber\\ [2mm]
\lambda_7 &=& \frac{4}{105 \left(b_0^2-1\right)}
\Big[ b_0^2 \left(680 \mathsf{b}_5-151788 f_4+1740 f_5-412161\right)-  \nonumber\\ 
&&\quad -3 \left(10 \mathsf{b}_5+1404 f_4+580 f_5+46953\right)-390 \left(196 b_4-5 b_6\right) b_0\Big]\,.
\end{eqnarray}

The first non-trivial coefficient in the expansion of the blackening factor appears at order $u^5$ due to the D2-brane asymptotics imposed and is undetermined. The expansion can be written as
$$\mathsf{b} = 1 +  \sum_{n=5}^{23}  \ \mathsf{b}_n \ u^n + \OO(u^{24}) \,, $$
with the parameters 
\begin{eqnarray}
\mathsf{b}_6&=& \frac{20 \mathsf{b}_5}{3}\,,\qquad \mathsf{b}_7 \,=\, \frac{240 \mathsf{b}_5}{7}\,,\qquad\mathsf{b}_8 \,=\, 160 \mathsf{b}_5\,,\qquad \mathsf{b}_9 \,=\, \frac{6400 \mathsf{b}_5}{9}\,,\nonumber\\ [2mm]
\mathsf{b}_{10} &=& \frac{1}{20} \mathsf{b}_5 \left(4 f_4-20 f_5+60513\right)\,,\nonumber\\ [2mm]
\mathsf{b}_{11} &=& -\frac{2 \mathsf{b}_5}{33 \left(b_0^2-1\right)}\Big[b_0^2 \left(25 \mathsf{b}_5-5916 f_4+220 f_5-223857\right)  + \left(75 b_6-2940 b_4\right) b_0-\nonumber\\ 
&&\quad - 84 f_4-220 f_5+202587\Big]\,.
\end{eqnarray}

Likewise, the leading order in the expansion of the warp factor is $u^5$ so that the D2-brane asymptotics is maintained
$$\mathbf{h} = \sum_{n=5}^{27}  \ h_n \ u^n + \OO(u^{28}) \,. $$
The first few coefficients are
\begin{eqnarray}
h_5 &=& -\frac{16}{15} \left(b_0^2-1\right)\,,\qquad h_6 \,=\, -\frac{64}{9} \left(2 b_0^2-1\right)\,,\qquad h_7 \,=\, -\frac{64}{315} \left(581 b_0^2-201\right) \,, \nonumber\\ [2mm]
h_8&=& -\frac{512}{45} \left(69 b_0^2-19\right)\,,\qquad\qquad h_9 \,=\, -\frac{1024}{945} \left(4263 b_0^2-1003\right)\,,\nonumber\\ [2mm]
h_{10} &=& \frac{4 \left(7 b_0^2 \left(252 f_4+180 f_5-1387925\right)+784 b_4 b_0-5 \left(84 f_4+252 f_5-401183\right)\right)}{1575}\,, \nonumber\\ [2mm]
h_{11} &=& \frac{32}{10395}  \Big[21 b_0^2 \left(85 \mathsf{b}_5-15904 f_4+1000 f_5-1896225\right)-21 \left(8092 b_4-225 b_6\right) b_0- \nonumber\\ 
&& \qquad-4 \left(3213 f_4+3465 f_5-1794298\right)\Big]\,.
\end{eqnarray}

Finally, the fluxes are expanded as
\begin{equation}
\mathcal{B}_J = \sum_{n=0}^{23} \ \mathcal{B}_{J,n} u^n + \OO (u^{24})\nonumber,\qquad 
\mathcal{B}_X= \sum_{n=0}^{23} \ \mathcal{B}_{X,n} u^n + \OO (u^{24})\nonumber,\qquad
\mathcal{A}_J = \sum_{n=0}^{22} \ \mathcal{B}_{J,n} u^n +  \OO(u^{23}) \nonumber.\end{equation}
For convenience, we keep the undetermined parameters in the same function, namely $\BB_J$. The parameter $b_0$ labelling the representative of the family is the leading order of both $\BB_J$ and $\BB_X$. The following coefficients in the expansion of $\BB_J$ are
\begin{eqnarray}
\mathcal{B}_{J,1} &=& 4 b_0\,,\qquad \mathcal{B}_{J,2} \,=\, 8 b_0\,,\qquad \mathcal{B}_{J,3} \,=\, -16 b_0\,,\qquad \mathcal{B}_{J,5} \,=\, \frac{8}{5} b_0 \left(8 f_4+45\right)+\frac{34 b_4}{5}\,,\nonumber \\ [2mm]
\mathcal{B}_{J,7} &=& -\frac{4}{525 \left(b_0^2-1\right)}\Big[ b_0^3 \left(108 b_5+21108 f_4-244 f_5+246963\right)-  \nonumber\\ 
&& \qquad -b_0 \left(178 b_5+4308 f_4-244 f_5+187407\right) + 42 \left(139 b_4-20 b_6\right) b_0^2+126 \left(19 b_4+5 b_6\right)\Big]\,,\nonumber\\
\end{eqnarray}
while for $\BB_X$ we have
\begin{eqnarray}
\mathcal{B}_{X,1} &=& 4 b_0\,,\qquad\quad \mathcal{B}_{X,2} \,=\, 12 b_0\,,\qquad\quad \mathcal{B}_{X,3} \,=\, 32 b_0\,,\qquad\quad\mathcal{B}_{X,4} \,=\, -64 b_0-\frac{b_4}{2}\,,\nonumber\\ [2mm]
\mathcal{B}_{X,5} &=& \frac{1}{5} \left(-b_0 \left(56 f_4+1755\right)-6 b_4\right)\,,\qquad\mathcal{B}_{X,6} \,=\, -16 b_0 \left(8 f_4+93\right)-42 b_4+b_6\,,\nonumber\\ [2mm]
\mathcal{B}_{X,7} &=& \frac{2}{525 \left(b_0^2-1\right)} \Big[4 b_0^3 \left(72 b_5-87918 f_4-46 f_5-511593\right)+  \nonumber\\ 
&& \qquad +4 b_0 \left(173 b_5+29118 f_4+46 f_5+303147\right)-21 \left(7332 b_4-185 b_6\right) b_0^2+38724 b_4-945 b_6\Big]\,.\nonumber\\ [2mm]
\end{eqnarray}
The remaining flux, $\AAA_J$, has coefficients 
\begin{eqnarray}
\mathcal{A}_{J,0}&=&\frac{b_0}{6}\,,\qquad\quad \mathcal{A}_{J,1} \,=\, \frac{4 b_0}{3}\,,\qquad\quad \mathcal{A}_{J,2} \,=\, 8 b_0\,,\qquad\quad \mathcal{A}_{J,3} \,=\, \frac{1}{6} \left(-32 b_0-b_4\right)\,,\nonumber\\ [2mm]
\mathcal{A}_{J,4} &=& b_0 \left(-2 f_4-\frac{167}{12}\right)-\frac{2 b_4}{3}\,,\qquad \quad\mathcal{A}_{J,5} \,=\, -\frac{16}{15} \left[b_0 \left(8 f_4+45\right)+3 b_4\right]\,, \nonumber\\ [2mm]
\mathcal{A}_{J,6} &=& \frac{1}{225 \left(b_0^2-1\right)}\Big[b_0^3 \left(4 b_5-3 \left(5732 f_4+44 f_5+22325\right)\right)+\nonumber\\ 
&& \qquad +3 b_0 \left(12 b_5+2532 f_4+44 f_5+10981\right)-24 \left(336 b_4-5 b_6\right) b_0^2+3360 b_4\Big] \,,\nonumber\\ [2mm]
\mathcal{A}_{J,7} &=& -\frac{4}{17325 \left(b_0^2-1\right)} \Big[b_0^3 \left(7594 b_5+1537164 f_4+39748 f_5+5893635\right)- \nonumber\\ 
&& \qquad  - b_0 \left(12004 b_5+478764 f_4+39748 f_5+2141607\right)+84 \left(9204 b_4-145 b_6\right) b_0^2 - 
\nonumber\\ 
&& \qquad -210 \left(1212 b_4+5 b_6\right)\Big]\,.\nonumber\\ [2mm]
\end{eqnarray}

\newpage

\section{Expansions near the horizon} \label{ap:horizonexp}

In this appendix we consider the expansion around the horizon, Eq.~\eqref{Horizon_expansions}. Although we have solved the equations to sixth order, we only show the first term for the different functions. They read

\begin{eqnarray}
f^h_1 &=& \frac{\lambda _h^2 \left(9 h_h \left(f_h^4+g_h^4\right)+\left(-6 \alpha _h+\xi _h+\chi
	_h\right){}^2\right)-2 g_h^2 \left(9 h_h g_h^2 \left(g_h^2-3 f_h^2\right)+\left(\xi
	_h-\chi _h\right){}^2\right)}{18 h_h \mathsf{b}_h f_h^3 g_h^4 u_h^4}\,,\nonumber\\[2mm]
g^h_1&=& \frac{g_h^2 \lambda _h^2 \left(9 h_h f_h^4+2 \left(6 \alpha _h+\xi
	_h\right){}^2\right)+f_h^4 \left(9 h_h \left(f_h^4+g_h^4\right)-\left(\xi _h-\chi
	_h\right){}^2\right)}{9 h_h \mathsf{b}_h f_h^8 g_h u_h^4}\,,\nonumber\\[2mm]
\lambda_1^h&=& \frac{\lambda _h}{9
	h_h \mathsf{b}_h f_h^8 g_h^4 u_h^4} \Big[f_h^4 \left(18 h_h g_h^4 \lambda _h^2-2 g_h^2 \left(\xi _h-\chi
	_h\right){}^2+\lambda _h^2 \left(-6 \alpha _h+\xi _h+\chi _h\right){}^2\right)+ \nonumber \\ && \quad +9 h_h
	f_h^8 \lambda _h^2+2 g_h^4 \lambda _h^2 \left(6 \alpha _h+\xi _h\right){}^2\Big]\,,\nonumber\\[2mm]
h_1^h&=&-\frac{\lambda _h^2}{81 h_h \mathsf{b}_h f_h^8 g_h^4 u_h^4} \Big[18 h_h f_h^4 \left(9 h_h g_h^4+\left(-6 \alpha _h+\xi _h+\chi
	_h\right){}^2\right)+81 h_h^2 f_h^8+\nonumber \\ && \quad  +36 h_h g_h^4 \left(6 \alpha _h+\xi
	_h\right){}^2 + \left(2 \chi _h \left(6 \alpha _h+\xi _h\right)-12 \alpha _h \xi _h+\xi
	_h^2-3\right){}^2\Big]\,,\nonumber\\[2mm]
\mathsf{b}_1^h&=& \frac{9 h_h f_h^4 \left(-\mathsf{b}_h g_h^2 u_h^3-1\right)-54 h_h f_h^2 g_h^2+9 h_h
	g_h^4+\left(\xi _h-\chi _h\right){}^2}{9 h_h f_h^4 g_h^2 u_h^4}\,,\nonumber\\[2mm]
\BB_{J,1}^h &=&
\frac{1}{9 h_h \mathsf{b}_h f_h^4 g_h^4 u_h^4}\Big[9 h_h f_h^4 \left(2 g_h^2 \left(\xi _h-\chi _h\right)+\lambda _h^2 \left(-6 \alpha
	_h+\xi _h+\chi _h\right)\right)+ \nonumber \\ &&  \qquad +\lambda _h^2 \Big(18 h_h g_h^4 \left(6 \alpha _h+\xi
	_h\right)+\left(6 \alpha _h-\xi _h-\chi _h\right) \left(12 \alpha _h \left(\xi
	_h-\chi _h\right)-\xi _h \left(\xi _h+2 \chi _h\right)\right)-\nonumber \\ && \qquad\qquad +- 3 \left(\xi _h+\chi
	_h\right)\Big)+18 \alpha _h \lambda _h^2\Big]\,,\nonumber\\[2mm]
\BB_{X,1}^h &=&
\frac{1}{9 h_h \mathsf{b}_h f_h^8 u_h^4}\Big[18 h_h f_h^4 \left(2 g_h^2 \left(\chi _h-\xi _h\right)+\lambda _h^2 \left(-6
	\alpha _h+\xi _h+\chi _h\right)\right)-\nonumber \\ && \qquad\qquad\qquad-2 \lambda _h^2 \left(6 \alpha _h+\xi
	_h\right) \left(12 \alpha _h \left(\xi _h-\chi _h\right)-\xi _h \left(\xi _h+2 \chi
	_h\right)+3\right)\Big]\,,
\nonumber\\[2mm]
\AAA_{J,1}^h&=& -\frac{1}{27 h_h \mathsf{b}_h f_h^4 g_h^2 u_h^4} \Big[9 h_h f_h^4\left(-6 \alpha _h+\xi _h+\chi _h\right)-18 h_h g_h^4 \left(6 \alpha
	_h+\xi _h\right)+\nonumber \\ && \qquad\ +\left(\xi _h-\chi _h\right) \left(-12 \alpha _h \left(\xi _h-\chi
	_h\right)+2 \xi _h \chi _h+\xi _h^2-3\right)\Big]\,.\nonumber\\[2mm]
\end{eqnarray}

\section{Numerical output}\label{ap:solutions}
In this appendix we show  some of the data obtained in our numerical computation. 
\begin{figure}[t]
	\begin{center}
		\begin{subfigure}{0.45\textwidth}
			\includegraphics[width=\textwidth]{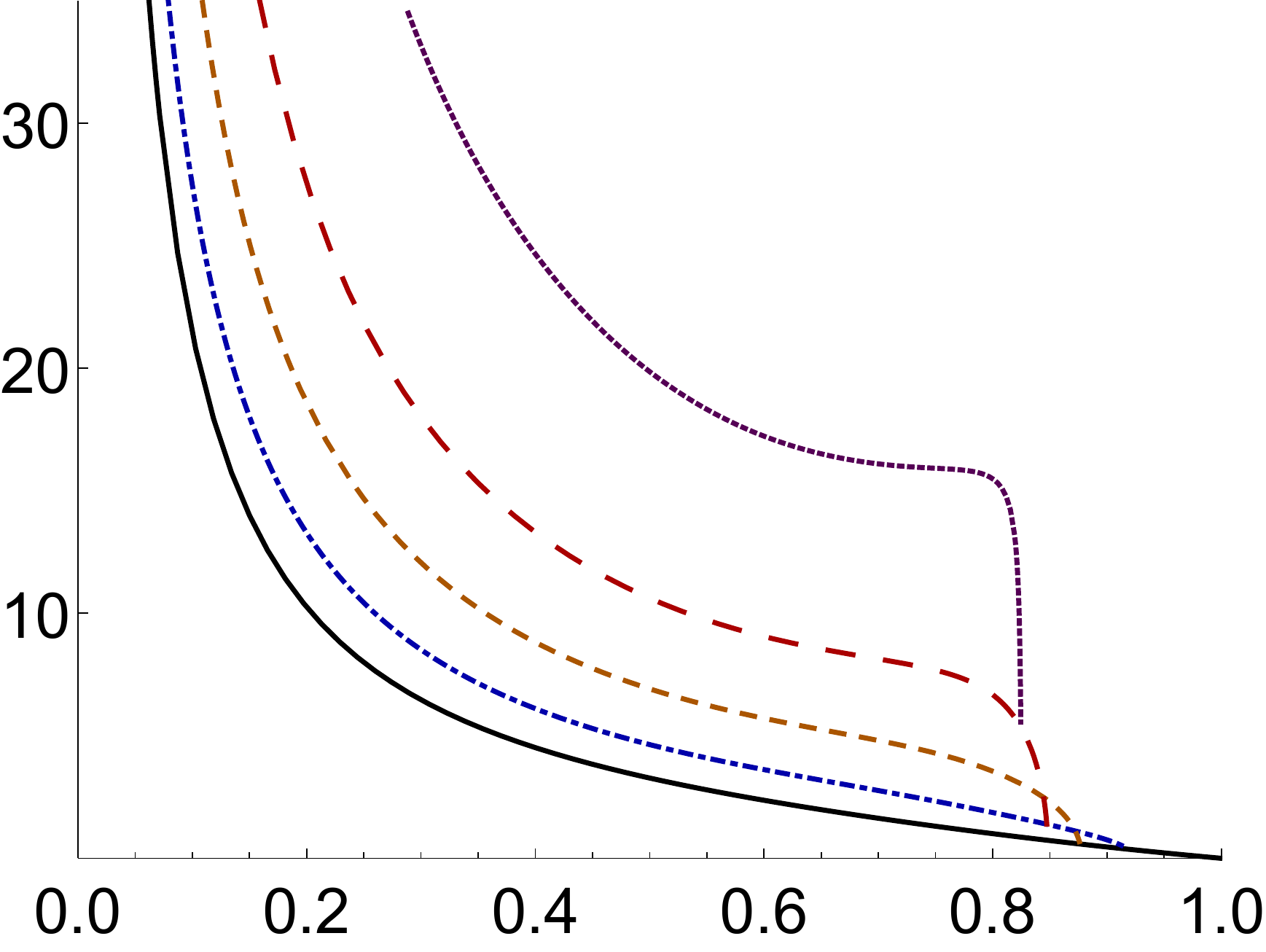} 
			\put(-200,150){$f_h$}
			\put(-20,20){$u_h/u_s$}
		\end{subfigure}\hfill
		\begin{subfigure}{0.45\textwidth}
	\includegraphics[width=\textwidth]{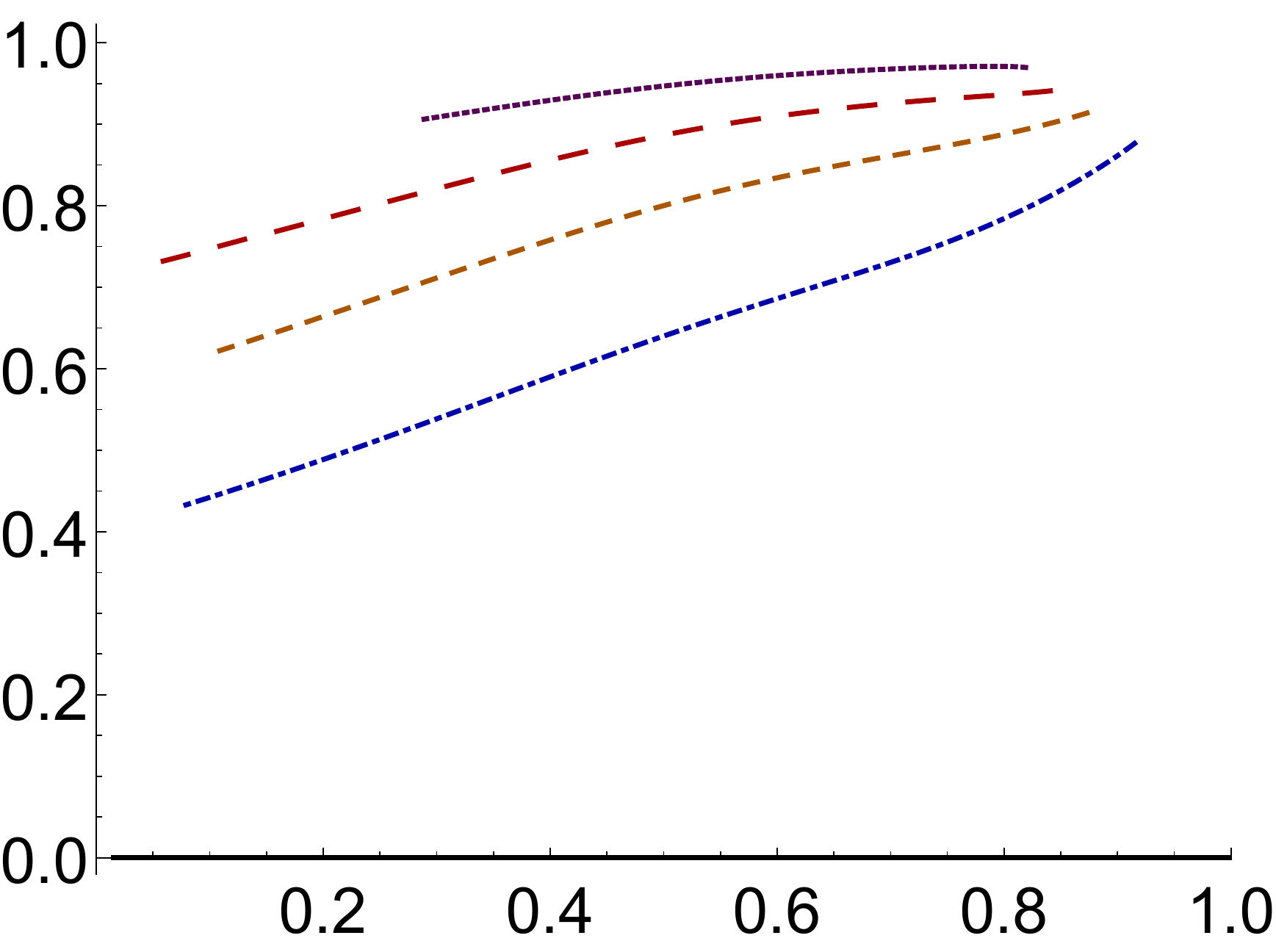} 
	\put(-200,150){$\xi_h$}
	\put(-20,20){$u_h/u_s$}
\end{subfigure} \vspace{3mm}
		\begin{subfigure}{.45\textwidth}
			\includegraphics[width=\textwidth]{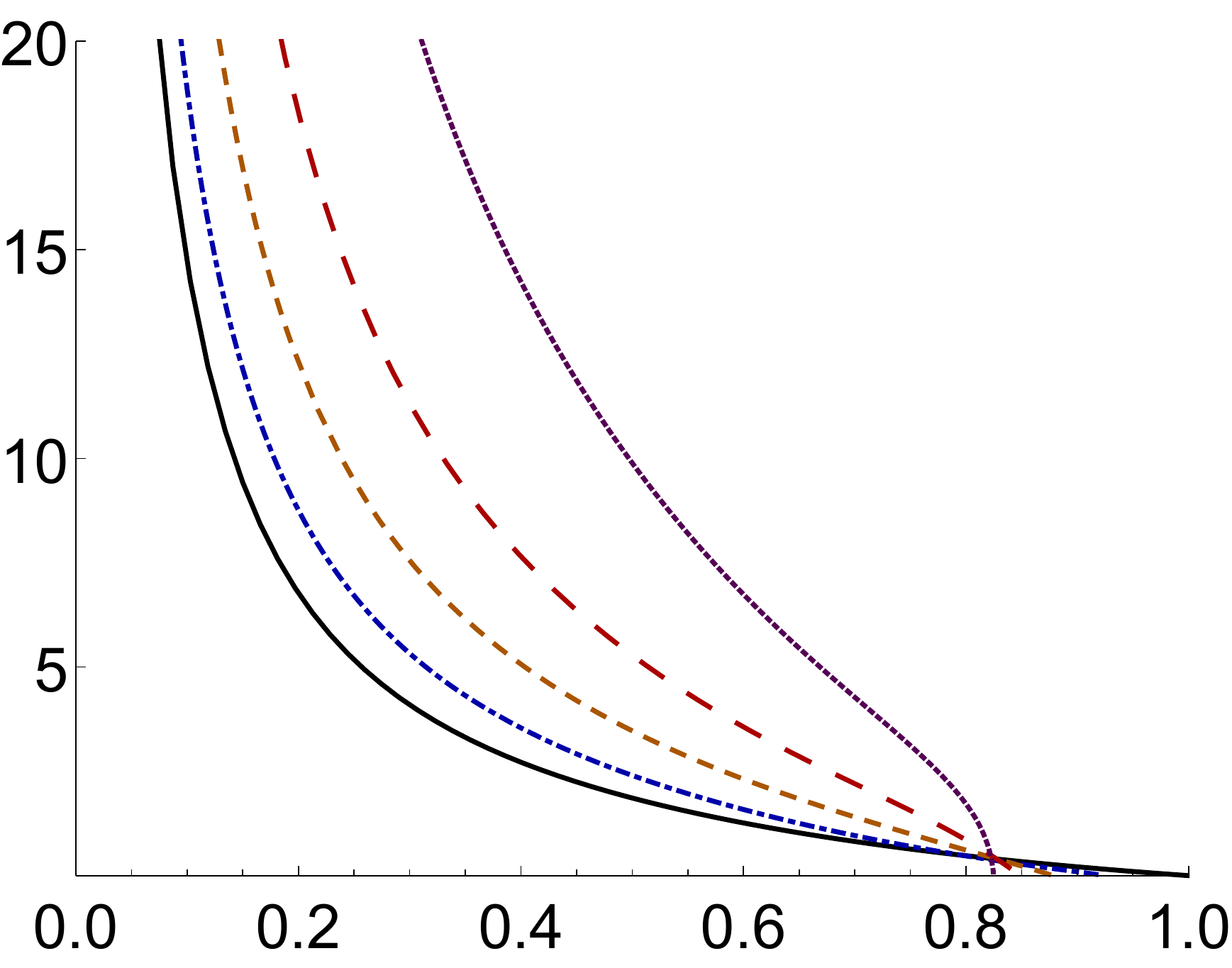} 
			\put(-200,175){$ $}
			\put(-200,160){$g_h$}
	\put(-20,20){$u_h/u_s$}
		\end{subfigure}\hfill
		\begin{subfigure}{.45\textwidth}
	\includegraphics[width=\textwidth]{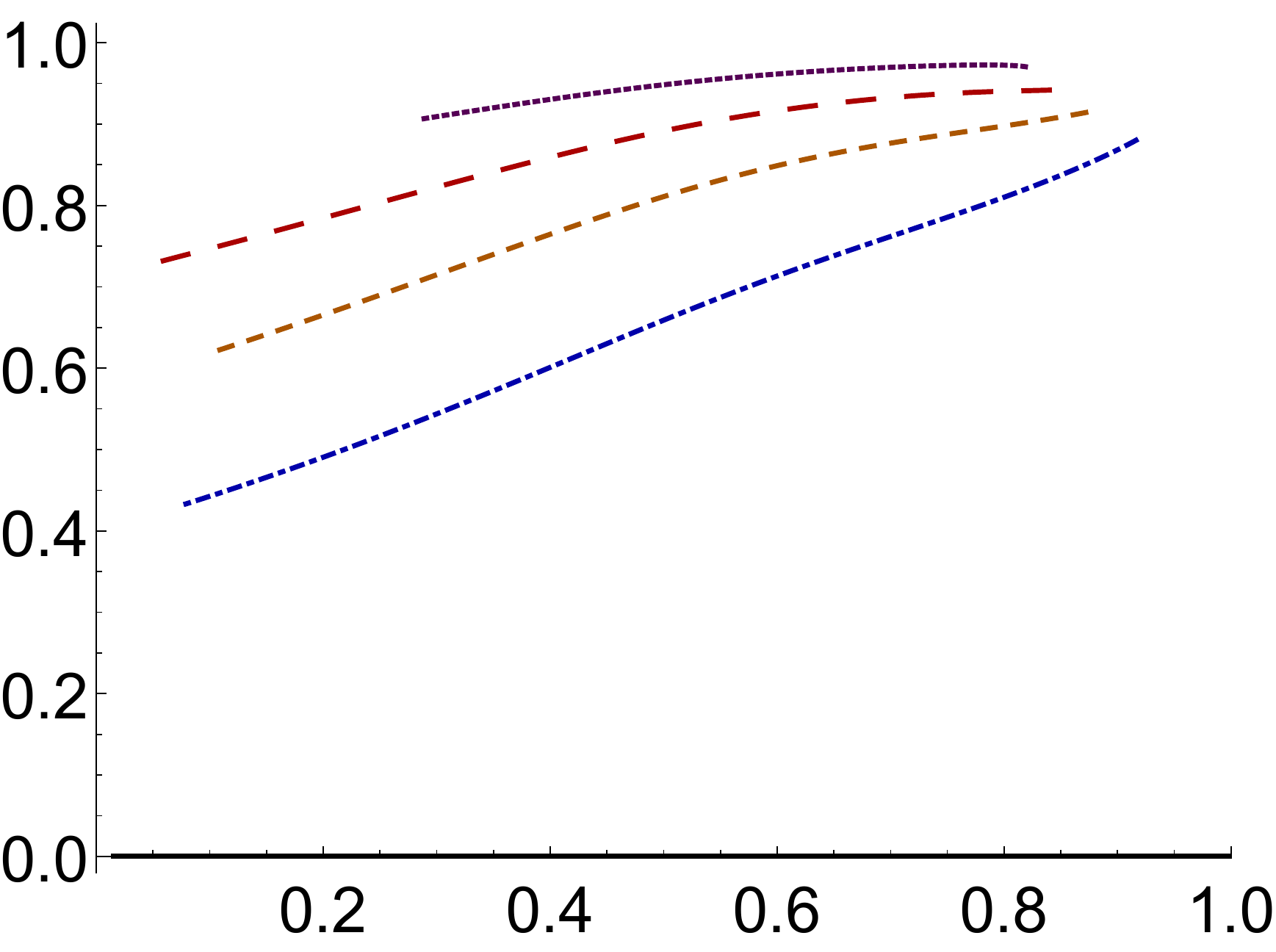} 
	\put(-200,160){$\chi_h$}
	\put(-20,20){$u_h/u_s$}
	\end{subfigure}
		\begin{subfigure}{.45\textwidth}
			\includegraphics[width=\textwidth]{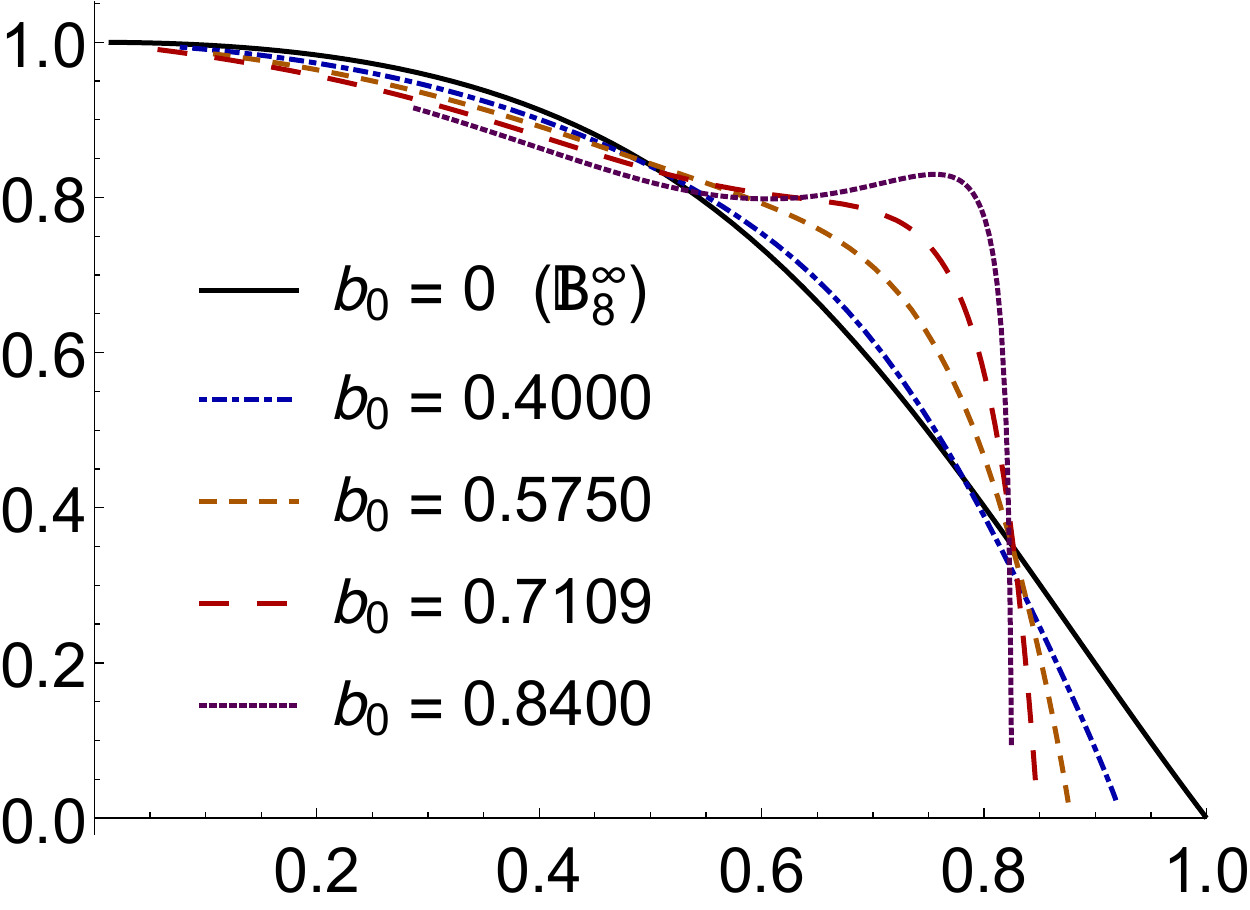} 
	\put(-200,150){$\lambda_h$}
	\put(-13,25){$u_h/u_s$}
		\end{subfigure}\hfill
		\begin{subfigure}{.45\textwidth}
	\includegraphics[width=\textwidth]{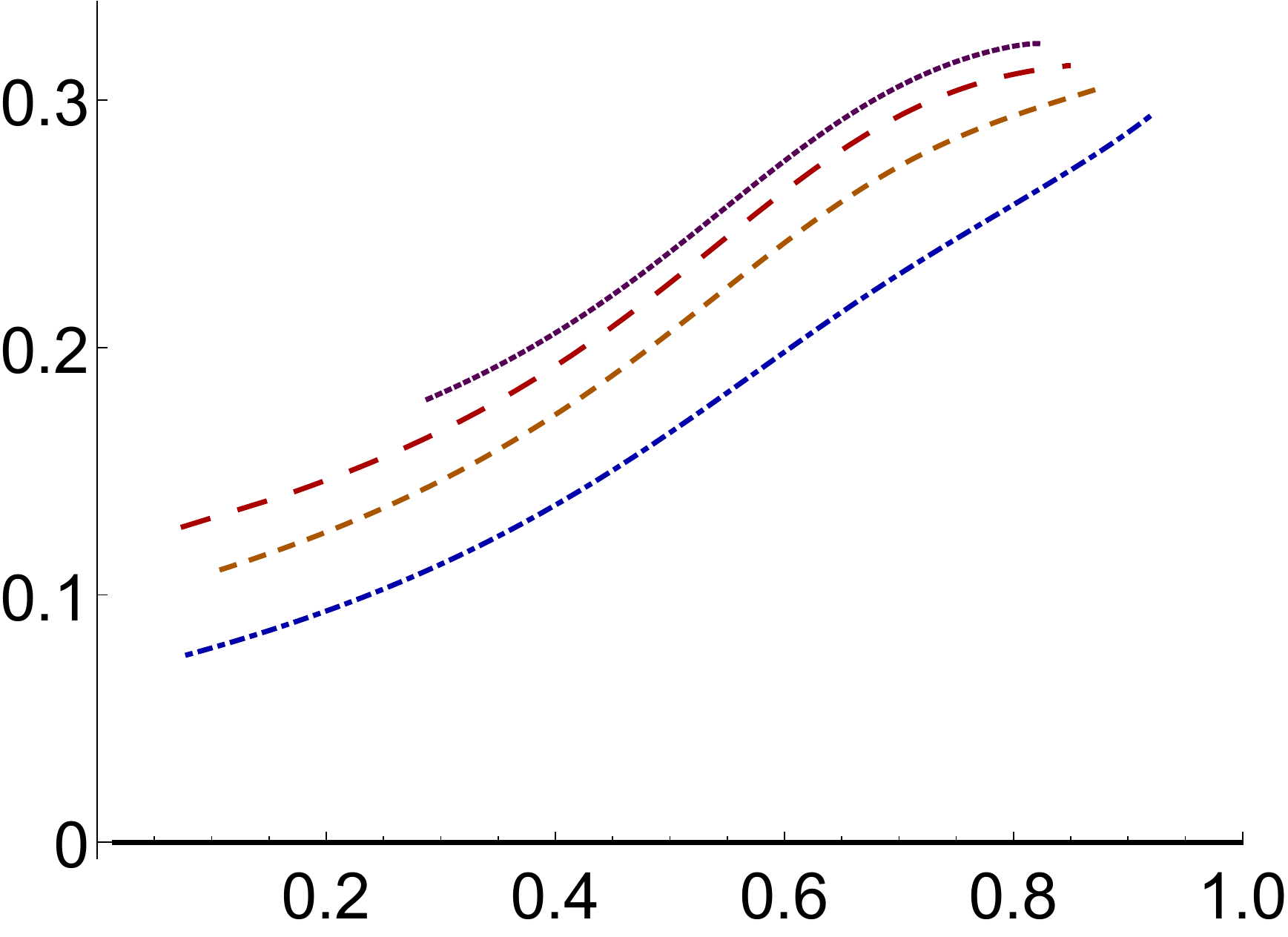} 
	\put(-180,150){$\alpha_h$}
	\put(-13,25){$u_h/u_s$}
	\end{subfigure}
		\caption{\small Value of the functions $e^\FF$, $e^\GG$ and $e^\Lambda$ at the horizon as a function of the position of the horizon normalized to $u_s$ for the metrics showed in the legend (left). Value of the fluxes at the horizon as a function of the position of the horizon normalized to $u_s$ for the metrics showed in the legend (right).}\label{fig.FunctionsBXBJAJ_Horizon}
	\end{center}
\end{figure}
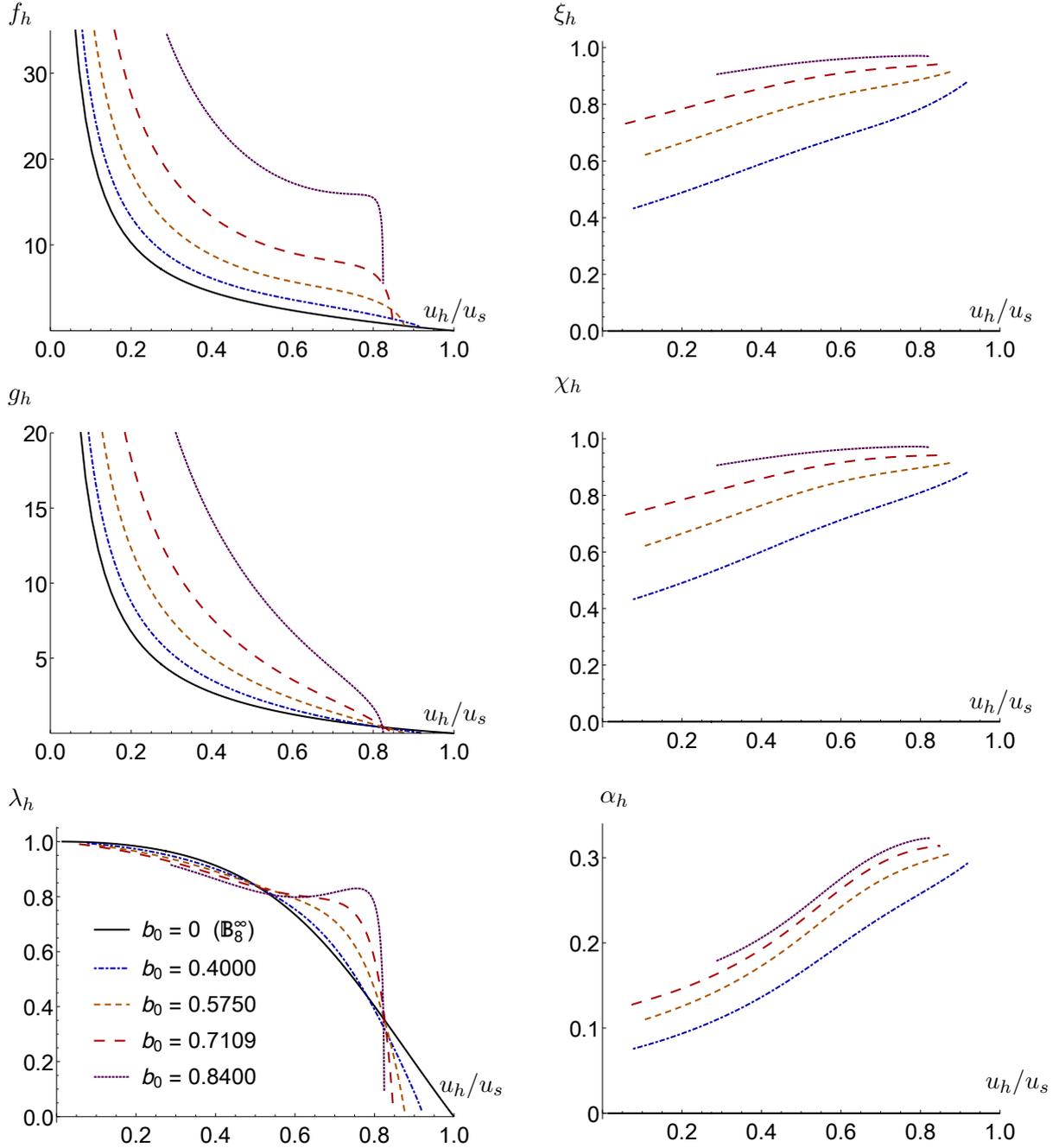
First of all, in Fig.~\ref{fig.FunctionsBXBJAJ_Horizon} (left) we show the value of the functions $e^\FF$, $e^\GG$ and $e^\Lambda$ for some representative values of $b_0$, as a function of the position of the horizon normalized to the point where the regular solution ends. Recall that we already argued that there was a maximum $u_N<u_s$ above which we cannot place the horizon. This can be clearly seen in Fig.~\ref{fig.FunctionsBXBJAJ_Horizon} (left), since the values at the horizon go to zero at some point $u_N/u_s<1$. Moreover, in these plots we see that as $b_0$ increases, the ratio between $u_N$ and $u_s$ decreases. For completeness, in Fig.~\ref{fig.FunctionsBXBJAJ_Horizon} (right) we show the value of the fluxes at the horizon. 

It is often the case that, when ``heating up'' a geometry, the value of the scalar fields at the horizon is similar to those in the zero-temperature solution at the same value of the radial coordinate (this statement is meaningful if the radial gauge is completely fixed, and in the same way, in both geometries). In our case this does not happen, as can be seen, for example, in the plot of $e^\Lambda$: whilst for the supersymmetric solutions (both regular and irregular) the function $e^\Lambda$ is monotonically decreasing, its horizon values develop a minimum and a maximum, resulting in a manifest difference in the dilaton behaviour.

\begin{figure}[t]
	\begin{center}
		\begin{subfigure}{0.45\textwidth}
			\includegraphics[width=\textwidth]{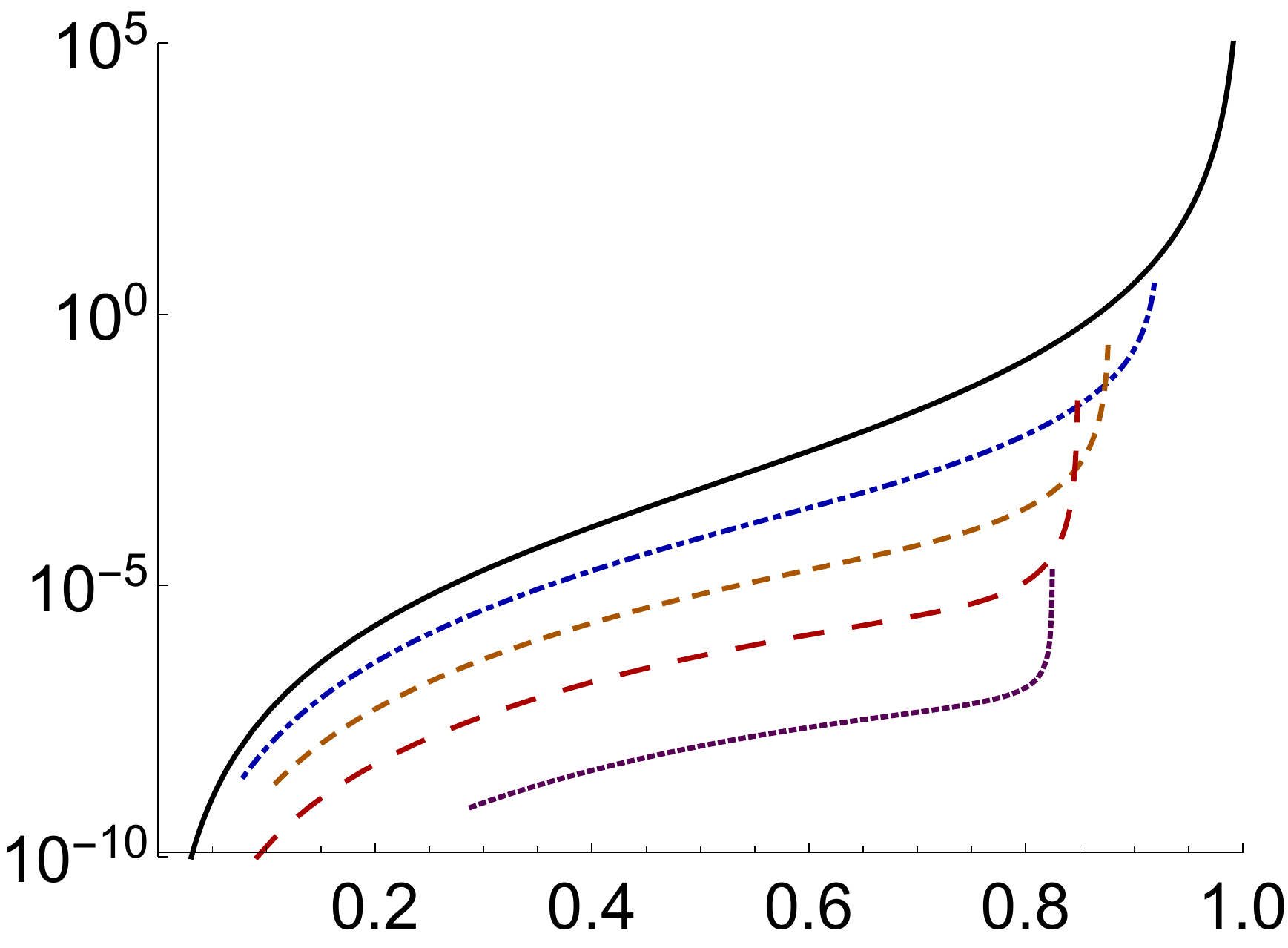} 
			\put(-200,150){$h_h$}
			\put(-5,20){$u_h/u_s$}
		\end{subfigure}\hfill
		\begin{subfigure}{.45\textwidth}
			\includegraphics[width=\textwidth]{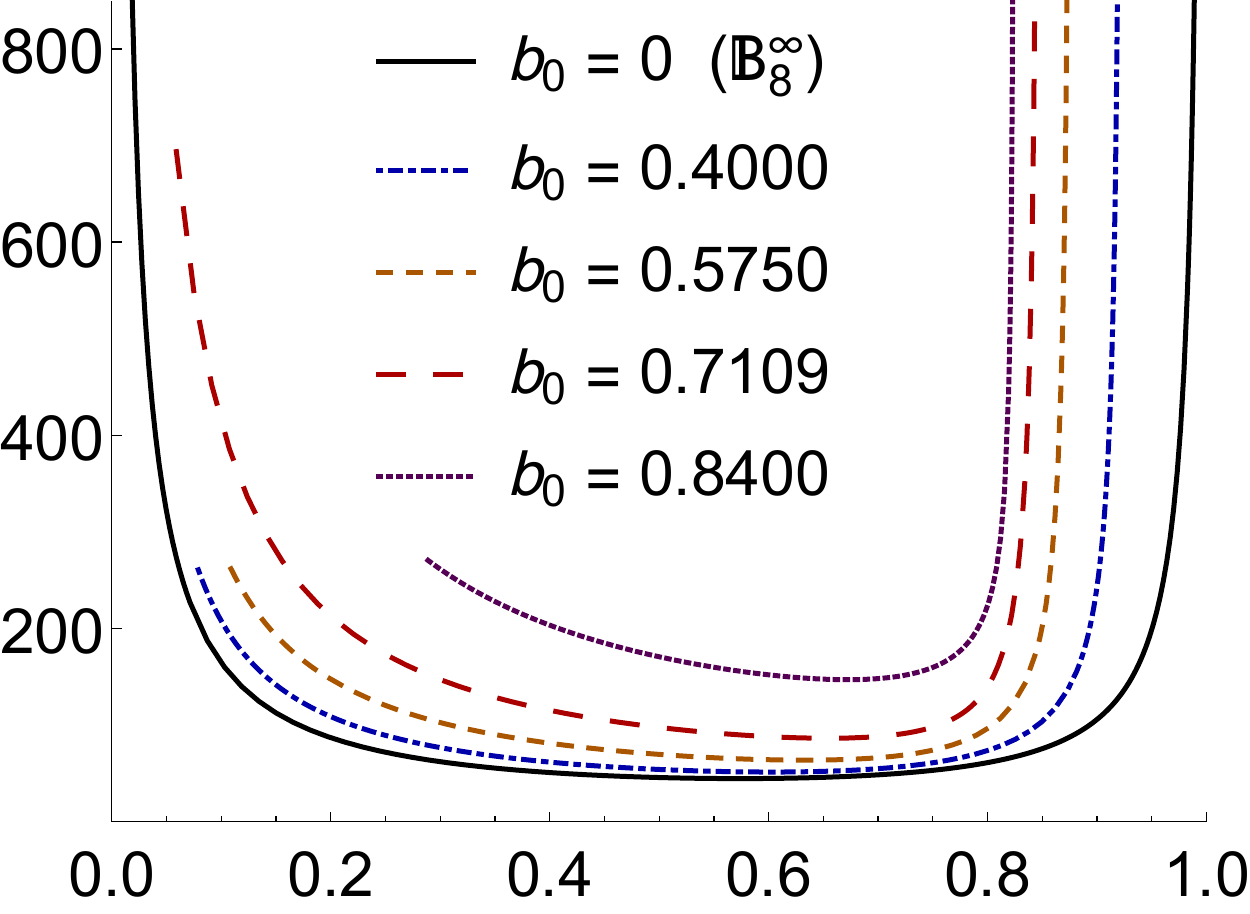} 
			\put(-200,150){$-\textsf{b}_h$}
			\put(-5,20){$u_h/u_s$}
		\end{subfigure}
		\caption{\small  Value of the warp factor at the horizon (left) and value of the derivative of the blackening factor at the horizon (right) as a function of the position of the horizon normalized to $u_s$. }\label{fig.H_B_Horizon}
	\end{center}
\end{figure}

In Fig.~\ref{fig.H_B_Horizon} we show the value of the warp factor at the horizon and the value of $\mathsf{b}_h$, which is the derivative of the blackening factor at the horizon. Notice that both magnitudes diverge as $u_h$ approach $u_N$: it is the contribution of both divergences (one in the denominator and the other in the numerator) that gives a finite temperature in the zero-entropy limit.

\end{document}